\documentclass{jfm_arxiv}

\usepackage[utf8]{inputenc}
\usepackage[T1]{fontenc}
\usepackage[english]{babel}

\usepackage{graphicx}
\usepackage{newtxtext}
\usepackage{newtxmath}
\usepackage{dsfont}
\usepackage{floatrow}
\usepackage{subcaption}
\RawFloats[figure,table]
\newfloatcommand{ftabbox}{table}[\nocapbeside]

\usepackage[separate-uncertainty=true]{siunitx}

\usepackage{booktabs}
\usepackage{colortbl}
\usepackage{multirow}

\usepackage{natbib}
\usepackage[colorlinks,urlcolor=blue,citecolor=black]{hyperref}
\usepackage{cleveref}

\newcommand*{\paragraph}[1]{\textit{#1}\quad}

\usepackage{thmtools}
\declaretheorem{lemma}
\declaretheorem{theorem}
\declaretheorem{proposition}
\declaretheorem[style=definition]{definition}

\usepackage{algorithm,algpseudocode}

\DeclareMathOperator{\adj}{adj}
\DeclareMathOperator{\erf}{erf}
\DeclareMathOperator{\erfc}{erfc}
\DeclareMathOperator{\supp}{supp}
\DeclareMathOperator{\tr}{tr}
\newcommand*{\Reals}{\mathbb{R}}
\newcommand*{\Complex}{\mathbb{C}}

\newcommand*{\Integer}{\mathbb{Z}}

\newcommand*{\Sphere}{\mathbb{S}^2}
\newcommand*{\E}{\mathrm{e}}
\newcommand*{\I}{\mathrm{i}}
\newcommand*{\areafrac}{\alpha_\mathrm{f}}
\newcommand*{\bigO}{\mathcal{O}}
\newcommand*{\Ca}{\mbox{\textit{Ca}}}
\newcommand*{\Con}{\mbox{\textit{Cn}}}
\newcommand*{\centerv}{\bvec{c}}
\newcommand*{\centerc}{c}
\newcommand*{\centervinit}{\centerv_\mathrm{init}}
\newcommand*{\centercyinit}{\centerc_{2,\mathrm{init}}}
\newcommand*{\dcenterc}{\dot{\centerc}}
\newcommand*{\dcentercxsteady}{\dcenterc_{1,\mathrm{steady}}}
\newcommand*{\centerlat}{\centerv_{\mathrm{L}}}
\newcommand*{\centerlatsteady}{\centerv_{\mathrm{L,steady}}}
\newcommand*{\dcenterlat}{\dot{\centerv}_{\mathrm{L}}}
\newcommand*{\centerch}{\hat{c}}
\newcommand*{\dcenterch}{\dot{\centerch}}
\newcommand*{\centerchyinit}{\centerch_{2,\mathrm{init}}}
\newcommand*{\driftfactor}{K}
\newcommand*{\driftfactormat}{\bmat{\driftfactor}}
\newcommand*{\Es}{E_\mathrm{s}}
\newcommand*{\Ed}{E_\mathrm{d}}
\newcommand*{\Estilde}{\tilde{E}_\mathrm{s}}
\newcommand*{\Edtilde}{\tilde{E}_\mathrm{d}}
\newcommand*{\Fs}{\bmat{F}_\mathrm{S}}
\newcommand*{\forceop}{\bvec{\mathcal{F}}}
\newcommand*{\forceoplin}{\widetilde{\forceop}}
\newcommand*{\forceops}{\forceop_\mathrm{s}}
\newcommand*{\forceopd}{\forceop_\mathrm{d}}
\newcommand*{\forceopslin}{\widetilde{\forceop}_\mathrm{s}}
\newcommand*{\forceopdlin}{\widetilde{\forceop}_\mathrm{d}}
\newcommand*{\gammar}{\gamma_\mathrm{r}}
\newcommand*{\nr}{\bvec{n}_\mathrm{r}}
\newcommand*{\gradgamma}{\bnabla_\gamma}

\newcommand*{\gradsphere}{\bnabla_{\Sphere}}

\newcommand*{\Deltasphere}{\Delta_{\Sphere}}
\newcommand*{\GammaL}{\Gamma_\mathrm{L}}
\newcommand*{\OmegaL}{\Omega_\mathrm{L}}
\newcommand*{\DeltaL}{\Delta_\mathrm{L}}
\newcommand*{\Lch}{L}
\newcommand*{\Lper}{L_\mathrm{per}}
\newcommand*{\numpatches}{n_\text{p}}
\newcommand*{\odetol}{\tau_\mathrm{ode}}
\newcommand*{\rc}{r_\mathrm{c}}
\newcommand*{\rv}{\nu}
\newcommand*{\rvr}{\rv_\mathrm{r}}
\newcommand*{\sph}{Y}
\newcommand*{\spha}{r}
\newcommand*{\sphb}{s}
\newcommand*{\vsphv}{\bvec{V}}
\newcommand*{\vsphw}{\bvec{W}}
\newcommand*{\vsphx}{\bvec{X}}
\newcommand*{\strain}{\mathcal{E}}
\newcommand*{\Tch}{T}
\newcommand*{\Ts}{T_\mathrm{s}}
\newcommand*{\Td}{T_\mathrm{d}}
\newcommand*{\Tbg}{T_\infty}
\newcommand*{\Ubg}{U_\infty}
\newcommand*{\ubgmax}{u_{\infty,\mathrm{max}}}
\newcommand*{\ubgavg}{u_{\infty,\mathrm{avg}}}
\newcommand*{\bvec}[1]{\boldsymbol{#1}}
\newcommand*{\bmat}[1]{\mathsfbi{#1}}
\newcommand*{\T}{\mathrm{T}}
\newcommand*{\CT}{\mathrm{H}}
\newcommand*{\bLambda}{\boldsymbol{\mathsf{\Lambda}}}
\newcommand*{\gridspace}{h}
\newcommand*{\coordspace}{h_0}

\newcommand*{\coordspaceupipe}{h_{u}}
\newcommand*{\coordspacevpipe}{h_{v}}
\newcommand*{\mupipe}{m_{u}}
\newcommand*{\mvpipe}{m_{v}}

\newcommand*{\qed}{\hfill\begin{picture}(7,7)\put(0,0){\framebox(7,7){}}\end{picture}}

\newcommand*{\iconAsmall}{\raisebox{-1pt}{\includegraphics[width=3mm,trim=14pt 14pt 14pt 14pt,clip]{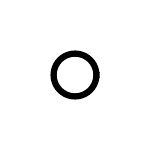}}}
\newcommand*{\iconBsmall}{\raisebox{-1pt}{\includegraphics[width=3mm,trim=14pt 14pt 14pt 14pt,clip]{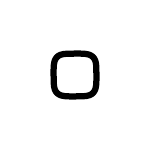}}}
\newcommand*{\iconCsmall}{\raisebox{-1pt}{\includegraphics[width=3mm,trim=14pt 14pt 14pt 14pt,clip]{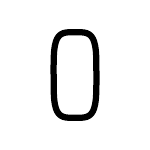}}}
\newcommand*{\iconDsmall}{\raisebox{-1pt}{\includegraphics[width=3mm,trim=14pt 14pt 14pt 14pt,clip]{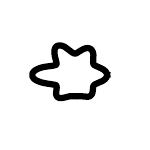}}}
\newcommand*{\iconEsmall}{\raisebox{-1pt}{\includegraphics[width=3mm,trim=14pt 14pt 14pt 14pt,clip]{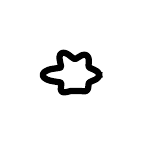}}}

\title{A boundary integral equation solver for capsule suspensions}
\author{Joar Bagge\corresp{\email{joar.bagge@gmail.com}}, Ziyu Du, \and George Biros}
\affiliation{Oden Institute, University of Texas at Austin, Austin, TX, 78712, United States}

\begin{document}
\maketitle

\begin{abstract}
Deformable capsules serve as models for biological cells, such as red blood cells, as well as for synthetic capsules in biomedical and industrial applications. In this work, we introduce a framework for fast and robust simulation of deformable capsules in three-dimensional Stokes flow, both in unbounded domains and in pipe-confined geometries. We solve the Stokes equations using a boundary integral formulation, employing a partition-of-unity discretization for the capsule membrane and the trapezoidal rule with regularization to handle the singular quadrature. Dense periodic interactions are evaluated via a fast Ewald summation scheme, achieving $\bigO(N \log N)$ complexity for $N$ grid points. The interfacial forces on the capsule membrane are modeled using the Skalak constitutive law. We carry out a spectral analysis of this model for small perturbations of a spherical capsule, which indicates that the relaxation dynamics are intrinsically stable. However, the nonlinear character of the Skalak model induces aliasing errors that can compromise numerical stability. To address this when needed, we perform de-aliasing through upsampling and spectral filtering, and we derive guidelines for choosing appropriate filter parameters, thereby stabilizing the method. We investigate the stability and convergence behavior of the resulting scheme and conduct simulations of long-time dynamics, focusing on the relaxation of unconfined capsules and the lateral migration of pipe-confined capsules under various flow conditions.
\end{abstract}

\tableofcontents
\section{Introduction}
\label{sec:introduction}

We investigate elastic capsules in three-dimensional Stokes flow, considering both unbounded domains and flow through cylindrical pipes. In our formulation, a capsule consists of an extensible membrane that encloses and is surrounded by a Newtonian fluid and that resists surface shear and tension but not bending. Synthetic capsules are central to numerous biomedical and industrial applications \citep{Barthes-Biesel1991,Lac2004,Barthes-Biesel2016}. They are also widely employed as simplified models of liquid-filled biological membranes, such as red blood cells (RBCs) and various other cell types.  Several constitutive relations have been proposed for connecting strains to elastic stresses; here, we use the Skalak model \citep{Skalak1973}. Some capsule models additionally possess bending rigidity, which is frequently included in mechanical models of RBCs, for example. In this work, however, we restrict our attention to capsules without bending resistance and assume that the viscosities of the internal and external fluids are identical, although our numerical framework can be extended to account for bending stiffness and viscosity contrast.

Capsules confined within pipes and channels play a significant role in many contexts, such as blood microcirculation and microfluidic devices \citep{Barthes-Biesel2016}. The presence of bounding walls gives rise to interesting effects; for instance, it is well known that RBCs tend to migrate away from the wall and toward the vessel centerline, which leads to a reduction in the apparent viscosity of blood, a behavior referred to as the F\aa hr\ae us--Lindqvist effect \citep{FahraeusLindqvist1931,DoddiBagchi2009,Zhao2010}. As a result, the volumetric flow rate associated with RBCs exceeds that of the bulk suspension. This effect is sensitive to the vessel diameter and becomes especially significant when the confinement ratio (the capsule-to-vessel size ratio) $L/b$ is greater than about 0.01 \citep{Pozrikidis2005}. Similar cross-stream migration has been examined numerically for capsules in Poiseuille (pipe-confined) flows \citep{DoddiBagchi2008,Koolivand2020,Ebrahimi2021} as well as in shear flows \citep{Singh2013,Nix2014}. A further widely used model for RBCs is the \emph{vesicle}, which has an inextensible membrane that does not resist shear deformations \citep{Barthes-Biesel2016,Veerapaneni2011} but does resist bending. In some cases, omitting bending effects is well justified~\citep{Barthes-Biesel2016}; whether bending must be included to accurately describe RBCs is beyond the scope of this work.

The Stokes flow equations are linear, which makes them well suited to boundary integral representations. We employ a single-layer potential integral formulation on the capsule membrane, while on the rigid walls we use the Hebeker formulation \citep{Hebeker1986,Zinchenko2006} (\cref{sec:formulation}). The resulting layer potentials are computed using a fast Ewald summation scheme that achieves $\bigO(N \log N)$ complexity for $N$ grid points. We use a standard Runge-Kutta 23 adaptive time stepping scheme. Below we discuss the novel aspects of the paper. 
\subsection{Contributions}

In this paper, we introduce a method for simulating capsules both in unbounded flows and in pipe flows, where the pipe may have an arbitrary but constant cross-section, and periodic boundary conditions are applied along the pipe axis. We focus on the case of a single capsule; extensions to systems with multiple interacting capsules will be presented in a future work. Our main contributions are:
\begin{itemize}
  \item 
    We perform a new stability analysis of the Skalak membrane force for a spherical capsule using spectral methods (\cref{sec:stability-analysis}), analogous in spirit to the spectral stability analysis of spherical vesicles in \cite{Veerapaneni2011}. This analysis characterizes the stability, stiffness, and null space of the force operator for small perturbations of the spherical configuration.
  \item
    We  evaluate surface derivatives on an upsampled grid to reduce aliasing.
    We develop a new analysis of the spectral filtering strategy used to stabilize the scheme and provide practical guidelines for choosing filter parameters based on a study of the interpolation error (\cref{sec:spectral_filtering}). This filtering procedure is shown to be essential for the stable simulation of non-spherical capsules. The proposed filter is not making a marginal improvement; it qualitatively changes the robustness of the \citep{Agarwal2024} solver.
\item 
    We generalize the partition-of-unity (POU) discretization for capsules in unbounded domains from \cite{Agarwal2024} to handle confined geometries. This capsule discretization is coupled to a representation of the pipe, and periodicity in the axial direction is enforced using a fast, spectrally accurate Ewald summation method \citep{Bagge2023} (\cref{sec:method}). The overall scheme is stabilized via a combination of grid-point blending and spectral filtering.

  \item  A natural question is whether the filtering changes the behavior of the capsule. To empirically address this question, we verify the stability and convergence of our method for capsules with a range of reduced volumes (including spherical and biconcave shapes) in \cref{sec:results-stability}. We also present simulations of unbounded shear flow and capsule relaxation (\cref{sec:results-unconfined}), as well as pipe flow for several cross-sectional geometries (\cref{sec:results-pipe-flow}). Collectively, these results show that our approach can robustly resolve capsule dynamics over long time horizons, capturing phenomena such as the lateral migration of capsules toward the pipe centerline. 
\end{itemize}

\subsection{Related work}

We focus the discussion on related work regarding aliasing errors and stability of capsules and vesicles, and we also briefly mention work concerning the choice of capsule representation and the lateral migration of capsules. For a more detailed discussion of the relation between the basic numerical scheme and the state of the art for vesicles and capsules, see \citep{Agarwal2024}. For a general review on elastic capsules see~\citep{Barthes-Biesel2016}.

\paragraph{Aliasing and stability.} The nonlinearities in the Skalak force model are known to generate aliasing errors, which in turn cause spurious amplification of high-frequency modes \citep{Zhao2010}. A standard remedy is to employ a dealiasing strategy in which the force is first computed on an upsampled mesh and then filtered. Implementing the filter is most straightforward in a spherical harmonics representation, as done, for example, in \cite{Zhao2010,Veerapaneni2011}. When a partition-of-unity (POU) discretization is used for the capsule surface, one must relate the POU resolution to an appropriate spherical harmonics truncation order to apply filtering. To the best of our knowledge, no prior work has established such a connection between POU resolution and spherical harmonics, and providing this relation is the purpose of \cref{sec:spectral_filtering}.

\cite{DodsonDimitrakopoulos2009} used a spectral boundary element method with non-overlapping patches and introduced a smoothing scheme to prevent discontinuities across patch boundaries from accumulating over time, thus stabilizing their method. In our method, blending across the overlapping patches (see \cref{sec:method-capsule,sec:method-time-stepping}) serves a similar purpose, together with the filtering mentioned above.

Several studies \citep{Lac2004,Pozrikidis2005,Zhao2010} have noted that when the ratio $\Ed/\Es$ of dilatational to shear modulus in the Skalak model is large (which is the case for RBCs), the model becomes prohibitively stiff. Time-step constraints of the CFL type are observed for explicit time-stepping methods \citep{DodsonDimitrakopoulos2009,Zhao2010}, which is also the case for droplets \citep{ZinchenkoDavis2005,Dimitrakopoulos2007}. Previous work has used moderate values for $\Ed/\Es$, such as 1 \citep{Lac2004,DodsonDimitrakopoulos2009}, 10 \citep{Lac2004} or 16 \citep{Zhao2010}. In the present work, we choose $\Ed/\Es = 10$, which is tractable for our standard explicit time-stepper.

\paragraph{Instabilities due to lack of bending resistance.} \cite{Lac2004} observed numerically that spherical capsules without bending resistance in shear flow attain a stable equilibrium only within a finite interval $[\Ca_\mathrm{L}, \Ca_\mathrm{H}]$ of capillary numbers. For $\Ca > \Ca_\mathrm{H}$, no steady state is found, whereas for $\Ca < \Ca_\mathrm{L}$, the membrane buckles and develops folds whose structure depends on the grid resolution. Although the grid spacing is a numerical parameter, such buckling has also been observed in experiments \citep{Barthes-Biesel2016}. To stabilize simulations once the capsule buckles, one must include bending resistance, either by introducing an explicit bending model, as in \cite{Zhao2010,Bryngelson2018,Lu2021}, or by employing a numerical technique that effectively imposes artificial bending stiffness, as in \cite{Walter2010}.

Note that, prior to stabilization, biconcave capsules with lower reduced volume are considerably more unstable than spherical ones, yet our filtering stabilizes all capsule shapes. (Another strategy to stabilize capsules without bending is to over-inflate, or pre-stress, the capsule, as in \cite{Kuriakosake2011,Barthes-Biesel2016}; this option is not pursued here.)
For vesicles, \cite{Veerapaneni2011} carried out a spectral analysis of small deformations of a spherical vesicle in a quiescent fluid to characterize stability and stiffness. We are not aware of any existing corresponding analysis for the Skalak force of capsules, and this is the contribution of \cref{sec:stability-analysis}. Our analysis shows that small deformations of a spherical capsule will vanish over time in a quiescent fluid, i.e., the spherical capsule is stable. This is not in conflict with the results by \cite{Lac2004}, since the negative principal tensions that cause the buckling go to zero as $\Ca \to 0$.

\paragraph{Spatial discretization of the capsule.}
A common way to represent the capsule membrane is through a spherical harmonics expansion \citep{Zhao2010,Veerapaneni2011,Zhu2015}, which is spectrally accurate and naturally suited for computing surface derivatives and applying spectral filtering. The standard approaches for handling singular integration on this surface—floating partitions of unity \citep{Bruno2001} and grid rotations \citep{Graham2002,Gimbutas2013}—have computational costs of $\bigO(N^{1.5})$ and $\bigO(N^2 \log N)$, respectively, for $N$ grid points. In contrast, the partition-of-unity formulation we employ (\cref{sec:method-capsule}), following \cite{Agarwal2024} and coupled with singular integration via regularization \citep{Tlupova2019}, achieves a lower complexity of $\bigO(N \log N)$, though it does not retain the spectral accuracy of \citep{Veerapaneni2011}.

\paragraph{Lateral migration.}
A number of studies have examined the lateral migration of capsules in pipes or channels, employing either the Skalak membrane model \citep{Nix2016,Koolivand2020,Ebrahimi2021} or a neo-Hookean membrane model \citep{DoddiBagchi2008,DoddiBagchi2009,Hu2012}. Lateral drift arises from two main mechanisms: a wall-induced effect that drives capsules away from the nearby wall and a deformation-induced effect (generated by the flow) that drives them toward regions of lower shear. Both mechanisms produce an asymmetric deformation of the capsule membrane, which in turn generates the lateral forces. In pipes with circular or square cross-sections, these effects lead capsules to migrate toward the pipe centerline. In curved vessels, an additional curvature-driven mechanism promotes migration toward the inner wall \citep{Ebrahimi2021}. Since migration requires particle deformability, it is not observed in rigid particle suspensions.   The purpose of this paper is not to provide a comprehensive analysis of lateral migration, but rather to present a few illustrative examples (in \cref{sec:results-pipe-flow}) that showcase the capabilities of our numerical method.

\subsection{Limitations}

Our method is restricted to low Reynolds numbers since inertial terms are ignored. We only consider capsules with spherical topology, without bending resistance and with no viscosity contrast between the inner and outer fluids.  

In a boundary integral formulation, evaluating the velocity field near the capsule or near the wall requires resolving nearly singular quadratures~\citep{Beale2024}. Such a nearly-singular quadrature scheme is required to resolve capsule-capsule and capsule-wall interactions.   In our numerical examples, the capsule typically does not approach the wall closely enough for such methods to be necessary. However, for dense suspensions involving multiple capsules, a near-singular integration scheme would be needed. Several existing approaches could be incorporated into our framework, including extrapolated regularization \citep{Beale2024} and line interpolation/extrapolation \citep{Morse2021}.

Our study is restricted to pressure-driven flow in pipes with a uniform cross-section. The methodology could be generalized to pipes with spatially varying cross-section by enforcing a prescribed pressure drop across the computational cell, following \cite{Bystricky2021}, and then connecting the average pressure gradient to the wall friction \citep{Hebeker1986}.

We also do not employ adaptive spatial discretizations, even though the partition-of-unity (POU) discretization, in principle, supports such an extension. For the POU function itself, we have observed that its steep gradients can be challenging to resolve on the capsule mesh, which may somewhat limit the attainable accuracy. Work on alternative POU functions is ongoing. For spectral filtering, we restrict attention to ideal (brick-wall) low-pass filters.

As already mentioned, we restrict our attention to the dynamics of a single capsule. Extending the approach to multiple capsules would require performance improvements geared toward high-performance computing and, for concentrated suspensions, the incorporation of a near-singular integration method.

\subsection{Outline of the paper}

The boundary integral formulation is presented in \cref{sec:formulation}, and the numerical methods for spatial and temporal discretization are discussed in \cref{sec:method}. The spectral stability analysis for small deformations of a spherical capsule is stated in \cref{sec:stability-analysis} (proofs are in \cref{app:stability-analysis}). \Cref{sec:spectral_filtering} contains the algorithm and analysis for the spectral filtering method that stabilizes the numerical method (with details found in \cref{appendix:DSHT,appendix:parameter_filtering}). We present results regarding stability, convergence, and physical quantities of interest, including lateral migration, in \cref{sec:results} (with some supplemental material in \cref{appendix:stable_conv_tbls,app:capsule-evolution}). Conclusions can be found in \cref{sec:conclusions}.

In addition to the appendices already mentioned above, we provide the parametrization of the biconcave capsule shape that we use in \cref{app:biconcave}, describe how the background flow for a pipe with an arbitrary constant cross-section is determined in \cref{app:pipes}, explain how the periodic length is selected in \cref{app:periodic-length}, and discuss the properties of the strain energy functional in \cref{appendix:strain_energy_functional}.

\section{Mathematical formulation}
\label{sec:formulation}

In this section, we describe the mathematical formulation in the
continuous setting, i.e., before discretization. We start with a
capsule in an unbounded flow in \cref{sec:formulation-unbounded},
and then introduce the additional machinery needed for
pipe-bounded flows in \cref{sec:formulation-pipe}. Finally,
characteristic quantities are introduced, and the equations are
written in dimensionless form, in
\cref{sec:formulation-nondimensionalization}.

\subsection{Unbounded flow}
\label{sec:formulation-unbounded}

Here, we give the formulation for a single capsule, but the
generalization to multiple capsules is straightforward. The
formulation here is the same as in \cite{Agarwal2024}; for a
comprehensive introduction, see e.g.\
\cite{Pozrikidis1992,Pozrikidis2001}. Let us
denote by $\gamma$ the capsule membrane, which is assumed to be a
compact, smooth, simply connected surface in $\Reals^3$.
The space outside $\gamma$, as well as the volume enclosed by
$\gamma$, is filled with a fluid governed by the (incompressible)
Stokes equations
\begin{gather}
  \label{eq:stokes-1}
  -\mu \Delta \bvec{u}(\bvec{x}) + \bnabla p(\bvec{x}) = \bvec{0},\  {\bf x} \text{ in } \mathbb{R}^3\backslash \gamma; \\
  \label{eq:stokes-2}
  \bvec{\nabla \cdot u}(\bvec{x}) = 0, \  {\bf x} \text{ in } \mathbb{R}^3\backslash \gamma; \\
  \llbracket p {\bf I } + \frac{\mu}{2} \left({\bf \nabla u + \nabla u^T}\right)\rrbracket(\bvec{x}) = {\bf f}(\bvec{x}), \ {\bf x} \text{~on~}  \gamma. \label{eq:jump}
\end{gather}
Here, $\bvec{u}(\bvec{x})$ is the fluid velocity, $p(\bvec{x})$ is the
pressure, and $\mu$ is the viscosity of the fluid (in principle
, the fluids inside and outside the capsule could have different
viscosities, but here we consider them to be the same). The fluid
velocity $\bvec{u}(\bvec{x})$ is continuous across $\gamma$, but the stress has a jump proportional to the membrane force density {\bf f}.  (An interface jump is denoted by $\llbracket\ \rrbracket$.) Once the velocity has been computed by solving \cref{eq:stokes-1}--\cref{eq:jump},  the capsule membrane moves along with this velocity, i.e.,
\begin{equation}
  \label{eq:time-evolution}
  \dot{\bvec{x}} = \bvec{u}(\bvec{x}), \qquad \bvec{x} \in \gamma,
\end{equation}
where the dot above $\bvec{x}$ denotes a time derivative.

Using a boundary integral equation formulation~\citep{Pozrikidis1992}, the fluid velocity $\bvec{u}(\bvec{x})$ that solves equations
\eqref{eq:stokes-1}--\eqref{eq:jump} is expressed as
\begin{equation}
  \label{eq:flow-field}
  \bvec{u}(\bvec{x}) = \bvec{u}_\infty(\bvec{x}) +
  \bvec{\mathcal{S}}[\bvec{f}](\bvec{x}),
\end{equation}
where the background flow $\bvec{u}_\infty(\bvec{x})$ is a given
function assumed to satisfy equations \eqref{eq:stokes-1}--\eqref{eq:jump} 
in all of $\Reals^3$ (for some unspecified pressure field
$p(\bvec{x})$), and the Stokes single-layer potential is given by
\begin{equation}
  \label{eq:single-layer}
  \bvec{\mathcal{S}}[\bvec{f}](\bvec{x})
  = \int_\gamma \bmat{G}(\bvec{x}-\bvec{y}) \bvec{f}(\bvec{y})
  \, \mathrm{d}S(\bvec{y}),
\end{equation}
where $\bvec{f}(\bvec{x})$ is a surface force density defined on
$\gamma$, and the Stokeslet tensor (the Green's function of the
Stokes equations) is given by
\begin{equation}
  \label{eq:stokeslet}
  \bmat{G}(\bvec{r}) = \frac{1}{8\pi\mu} \left(
  \frac{\bmat{I}}{\lVert \bvec{r} \rVert} + \frac{\bvec{r}
  \otimes \bvec{r}}{\lVert \bvec{r} \rVert^3}
  \right),
\end{equation}
where $\bmat{I}$ is the $3 \times 3$ identity matrix and
$\otimes$ denotes the outer product. We will also write
equation \eqref{eq:stokeslet} in index form as
\begin{equation}
  \label{eq:stokeslet-index}
  G_{jl}(\bvec{r}) = \frac{1}{8\pi\mu} \left(
  \frac{\delta_{jl}}{\lVert \bvec{r} \rVert} + \frac{r_j r_l}{\lVert \bvec{r} \rVert^3}
  \right),
  \qquad j, l \in \{1,2,3\},
\end{equation}
where $\delta_{jl}$ is the Kronecker delta.

For a deformable capsule, we consider the interfacial force density
$\bvec{f}(\bvec{x})$ in equations \eqref{eq:flow-field}--\eqref{eq:single-layer}
to be the elastic force generated by resistance to dilatation and
shear, according to a model introduced by \cite{Skalak1973}. In
this model, the capsule has a known relaxed reference
configuration $\gamma_\text{r}$, and the force is generated by
the surface deformation gradient $\bmat{F}_\text{S}(\bvec{x})$,
$\bvec{x} \in \gamma$, which can
be defined by the relations
\begin{align}
  \label{eq:surface-deformation-gradient-1}
  \bmat{F}_\text{S}(\bvec{x}) \bvec{a}^\text{r}_1(\bvec{x}_\text{r})
  &= \bvec{a}_1(\bvec{x}), \\
  \label{eq:surface-deformation-gradient-2}
  \bmat{F}_\text{S}(\bvec{x}) \bvec{a}^\text{r}_2(\bvec{x}_\text{r})
  &= \bvec{a}_2(\bvec{x}), \\
  \label{eq:surface-deformation-gradient-3}
  \bmat{F}_\text{S}(\bvec{x}) \bvec{n}^\text{r}(\bvec{x}_\text{r})
  &= \bvec{0},
\end{align}
where $\{\bvec{a}^\text{r}_1, \bvec{a}^\text{r}_2\}$ is a set of
tangent vectors%
\footnote{If $\gamma_\text{r}$ is locally parameterized by
$(u,v)$ around $\bvec{x}_\text{r}$, the tangent vectors are defined
by $\bvec{a}_1^\text{r} = \partial \bvec{x}_\text{r} / \partial u$
and $\bvec{a}_1^\text{r} = \partial \bvec{x}_\text{r} / \partial v$.}
at $\bvec{x}_\text{r} \in \gamma_\text{r}$, and 
$\{\bvec{a}_1, \bvec{a}_2\}$ is the corresponding set of tangent
vectors at $\bvec{x} \in \gamma$, the points being related by a
bijective map $\eta(\bvec{x}_\text{r}) = \bvec{x}$ between
$\gamma_\text{r}$ and $\gamma$ encoding the deformation.
Furthermore, $\bvec{n}^\text{r}$ is a normal to $\gamma_\text{r}$
at $\bvec{x}_\text{r}$. The interfacial force is then given by
\begin{equation}
  \label{eq:interfacial-force}
  \bvec{f}(\bvec{x}) = \gradgamma \bcdot
  \bLambda(\bvec{x}),
  \qquad \bvec{x} \in \gamma,
\end{equation}
with $(\gradgamma \bcdot)$ denoting the surface divergence,
where $\gradgamma = \bmat{P}(\bvec{x}) \bnabla$ and $\bmat{P}(\bvec{x})
= \bmat{I} - \bvec{n}(\bvec{x}) \otimes \bvec{n}(\bvec{x})$ is a
projection matrix for $\gamma$; here, $\bvec{n}(\bvec{x})$ is the
outward unit normal of $\gamma$. The in-plane elastic stress
tensor $\bLambda$ is a symmetric tensor given by
\begin{equation}
  \label{eq:stress-tensor}
  \bLambda(\bvec{x})
  = \frac{\Es}{2J(\bvec{x})} \Big(I_1(\bvec{x}) + 1\Big) \bmat{V}^2(\bvec{x}) +
  \frac{J(\bvec{x})}{2} \Big(\Ed I_2(\bvec{x}) - \Es\Big) \bmat{P}(\bvec{x}),
\end{equation}
where $\bmat{V}^2(\bvec{x}) =
\bmat{F}_\text{S}(\bvec{x})\bmat{F}_\text{S}^\T(\bvec{x})$.
Letting $\lambda_1^2$ and $\lambda_2^2$ denote the two nonzero
eigenvalues of $\bmat{V}^2$, we have the invariants
$I_1(\bvec{x}) = \lambda_1^2 + \lambda_2^2 - 2$ and
$I_2(\bvec{x}) = \lambda_1^2 \lambda_2^2 - 1$; finally  $J(\bvec{x}) =
(\lambda_1^2 \lambda_2^2)^{1/2}$ describes the relative change in surface
area. In equation \eqref{eq:stress-tensor}, $\Es$ is the elastic shear
modulus and $\Ed$ is the elastic dilatation modulus, both material
parameters of the membrane. The severe nonlinearities in
equation \eqref{eq:stress-tensor} have been cited as a source of
numerical instability by e.g.\ \cite{Lac2004,Pozrikidis2005,Zhao2010}.

To summarize, a capsule in unbounded flow is evolved in time by
computing its interfacial force due to deformation via
equations \eqref{eq:interfacial-force}--\eqref{eq:stress-tensor},
evaluating the fluid velocity for $\bvec{x} \in \gamma$ via
equations \eqref{eq:flow-field}--\eqref{eq:single-layer}, and evolving the surface $\gamma$ via equation \eqref{eq:time-evolution}.

\subsection{Pipe-bounded flow}
\label{sec:formulation-pipe}

\begin{figure}
  \centering
  \includegraphics[scale=0.8,trim=7mm 7mm 12mm 17mm,clip]{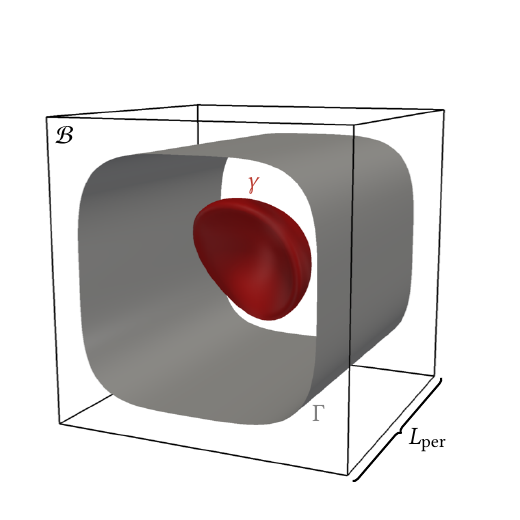}
  \caption{Flow setup with capsule $\gamma$ in pipe $\Gamma$. The primary simulation box $\mathcal{B}$ and periodic length $\Lper$ are also shown.}
  \label{fig:pipe-flow-setup}
\end{figure}

For pipe-confined flow, we consider a capsule suspended inside an infinitely long, fluid-filled cylindrical tube. In our simulations, this configuration is approximated by modeling only a finite portion of the pipe and imposing periodic boundary conditions in the axial direction, as illustrated in \cref{fig:pipe-flow-setup}. This approach is equivalent to repeating the modeled pipe segment and the capsule periodically, so that the capsule interacts with its own periodic images. We quantify the effect  of these periodic self-interactions in \cref{app:periodic-length}.

Let $\Gamma$ denote the surface of the pipe within the primary
simulation box $\mathcal{B}$, as shown in
\cref{fig:pipe-flow-setup}. A no-slip condition is enforced at
the pipe surface $\Gamma$, i.e., the fluid velocity must be zero
there. To ensure this, an extra term is added to
equation \eqref{eq:flow-field}, namely a correction flow generated by a
Hebeker formulation \citep{Hebeker1986,Zinchenko2006}
\begin{equation}
  \label{eq:hebeker-potential}
  \bvec{H}_\eta^\Gamma[\bvec{q}](\bvec{x})
  = \bvec{\mathcal{D}}^\Gamma[\bvec{q}](\bvec{x})
  + \eta
  \bvec{\mathcal{S}}^\Gamma[\bvec{q}](\bvec{x}),
\end{equation}
where $\eta > 0$ is a parameter with the physical dimension of
viscosity per length, $\bvec{\mathcal{S}}^\Gamma$ is defined as
in equation \eqref{eq:single-layer} but with $\Gamma$ in place of
$\gamma$, and the Stokes double-layer potential is
\begin{equation}
  \label{eq:double-layer}
  \mathcal{D}_j^\Gamma[\bvec{q}](\bvec{x})
  =
  \int_\Gamma T_{jlm}(\bvec{x} - \bvec{y})
  q_l (\bvec{y}) n_m (\bvec{y}) \, \mathrm{d}S(\bvec{y}),
\end{equation}
where Einstein's summation convention is used,
$\bvec{n}(\bvec{x})$ is the outward unit normal of $\Gamma$
(pointing away from the fluid inside the pipe), and the stresslet
tensor is given by
\begin{equation}
  T_{jlm}(\bvec{r}) = - \frac{6}{8\pi} \frac{r_j r_l r_m}{\lVert
  \bvec{r} \rVert^5}.
\end{equation}
The density $\bvec{q}$ that appears in
equations \eqref{eq:hebeker-potential}--\eqref{eq:double-layer} is
determined by enforcing the no-slip condition at $\Gamma$, which
will be stated explicitly in a moment. It can be noted that, in
principle, the correction flow enforcing no slip could be
generated by either a single-layer or double-layer potential
alone, but, in general, the double-layer operator alone has a
nontrivial nullspace, while the single-layer operator alone leads to an
ill-conditioned formulation. The Hebeker formulation
\eqref{eq:hebeker-potential} adds just enough of a single-layer
potential to give an invertible and well-conditioned formulation.

As mentioned, the periodic boundary conditions can be interpreted
as periodically replicating the objects in the primary simulation
box $\mathcal{B}$. This means that the equation for the fluid
velocity \eqref{eq:flow-field} is changed to read
\begin{equation}
  \label{eq:periodic-flow-field}
  \bvec{u}(\bvec{x}) = \bvec{u}_\infty(\bvec{x}) +
  \sum_{n=-\infty}^\infty \left(\bvec{\mathcal{S}}[\bvec{f}](\bvec{x} - \bvec{p}_n)
  + \bvec{H}^\Gamma_\eta[\bvec{q}](\bvec{x} - \bvec{p}_n)
  \right),
\end{equation}
where $\bvec{p}_n = (n \Lper, 0, 0)$, and $\Lper$ denotes the
periodic length, i.e., the length of the box $\mathcal{B}$ in the
axial direction (here assumed to be the first spatial dimension).
The background flow $\bvec{u}_\infty(\bvec{x})$ is assumed to be periodic
with period $\Lper$ in the axial direction. A way to select the
background flow given a pressure gradient when the pipe has
constant cross-section is described in \cref{app:pipe-background-flow}.

Taking $\bvec{x} \to \Gamma$ in equation \eqref{eq:periodic-flow-field}
and enforcing the no-slip condition, we end up with the integral
equation
\begin{equation}
  \label{eq:boundary-integral-equation}
  \frac{\bvec{q}(\bvec{x})}{2}
  +
  \sum_{n=-\infty}^\infty \left(
  \bvec{\mathcal{D}}^\Gamma[\bvec{q}](\bvec{x} - \bvec{p}_n)
  + \eta
  \bvec{\mathcal{S}}^\Gamma[\bvec{q}](\bvec{x} - \bvec{p}_n)
  \right) = -\bvec{u}_\infty(\bvec{x})
  - \sum_{n=-\infty}^\infty \bvec{\mathcal{S}}[\bvec{f}](\bvec{x} - \bvec{p}_n),
  \quad \bvec{x} \in \Gamma,
\end{equation}
which is an equation for $\bvec{q}(\bvec{x})$. The extra term
$\bvec{q}(\bvec{x})/2$ in this equation comes from the fact that
the double-layer potential $\bvec{\mathcal{D}}^\Gamma[\bvec{q}](\bvec{x})$
has a jump of size $-\bvec{q}(\bvec{x})/2$ when approaching the
pipe surface from within the pipe, and this must be corrected for
to get the limit from the inside.

In summary, for a capsule in pipe-bounded flow, the interfacial
force of the capsule is computed via
equations \eqref{eq:interfacial-force}--\eqref{eq:stress-tensor}
as in the unbounded case. Then, the boundary integral equation
\eqref{eq:boundary-integral-equation} is solved on the pipe
surface $\Gamma$, to enforce the no-slip condition. Finally, the
fluid velocity on the capsule membrane $\gamma$ is evaluated
using equation \eqref{eq:periodic-flow-field}, and the membrane is
evolved using equation \eqref{eq:time-evolution}. The periodic summation
that appears in equations \eqref{eq:periodic-flow-field} and
\eqref{eq:boundary-integral-equation} is evaluated using a fast
Ewald summation method which we describe in \cref{sec:method-pipe}.

\subsection{Nondimensionalization}
\label{sec:formulation-nondimensionalization}

In order to reduce the number of free parameters in the governing
equations, we write them in dimensionless form by dividing the
physical quantities by fixed characteristic quantities. For
example, given a characteristic length $\Lch$ and
characteristic time $\Tch$, the dimensionless spatial position
$\tilde{\bvec{x}}$ and time $\tilde{t}$ are defined as
\begin{equation}
  \tilde{\bvec{x}} = \frac{\bvec{x}}{L}, \qquad
  \tilde{t} = \frac{t}{T}.
\end{equation}
From the basic characteristic quantities $\Lch$ and $\Tch$, we
derive the characteristic speed $\Lch/\Tch$, area $\Lch^2$,
volume $\Lch^3$, pressure $\mu/\Tch$ and force $\mu \Lch^2 / \Tch$,
where $\mu$ is the fluid viscosity. Thus, the dimensionless
interfacial surface force density is
$\tilde{\bvec{f}}(\tilde{\bvec{x}}) = \bvec{f}(\bvec{x}) /
(\mu/\Tch)$, since $\bvec{f}$ has the dimension of force per area.

The choice of $\Lch$ and $\Tch$ is arbitrary, but should relate
to the problem at hand. In capsule simulations, the
characteristic length $\Lch$ could for example be selected as the
radius of the sphere with the same volume as the (initial)
capsule \citep{Pozrikidis2005,Lac2004,Zhao2010}, also known as
the ``volume-equivalent cell radius''. Note that due to incompressibility,
equation \eqref{eq:stokes-2}, the volume enclosed by the capsule is
constant throughout the simulation, so the volume-equivalent cell
radius is the same at any point in time.

An alternative convention is to select $\Lch$ as the radius of
the sphere with the same \emph{surface area} as the initial
capsule \citep{Veerapaneni2011,Agarwal2022}, i.e., an
``area-equivalent cell radius''. This makes sense for
capsules with conserved surface area (inextensible capsules), but
in our formulation, the surface area may change during the
simulation. Therefore, we select $\Lch$ to be the
volume-equivalent cell radius as in \cite{Pozrikidis2005}.
Thus, the dimensionless capsule volume should always be $4\pi/3$.

In order to select the characteristic time $\Tch$, let us first
consider the different time scales involved in the formulation.
The time scale associated with elastic shear relaxation of the
capsule may be defined as
\begin{equation}
  \Ts = \frac{\mu \Lch}{\Es},
\end{equation}
and the time scale associated with elastic dilatation relaxation
as
\begin{equation}
  \Td = \frac{\mu \Lch}{\Ed}.
\end{equation}
We can also define a time scale $\Tbg$ associated with the
background flow $\bvec{u}_\infty(\bvec{x})$. For example, for an
unbounded shear flow
\begin{equation}
  \label{eq:background-flow-shear-flow}
  \bvec{u}_\infty^\text{shear}(\bvec{x}) = (\dot{\gamma} x_2, 0, 0),
  \qquad \dot{\gamma} \in \Reals,
\end{equation}
we define
\begin{equation}
  \label{eq:time-scale-shear-flow}
  \Tbg^\text{shear} = \dot{\gamma}^{-1}.
\end{equation}
For an unbounded Poiseuille flow
\begin{equation}
  \label{eq:background-flow-Poiseuille}
  \bvec{u}_\infty^\text{Pois}(\bvec{x}) = a (b^2 - x_2^2 - x_3^2, 0, 0),
  \qquad a \in \Reals, \quad b \geq 0,
\end{equation}
we define
\begin{equation}
  \label{eq:time-scale-Poiseuille}
  \Tbg^\text{Pois} = \frac{\Lch}{\Ubg},
\end{equation}
where $\Ubg = ab^2/2$ is the mean flow velocity over the circular
cross-section of radius $b$. For a general pipe-bounded flow, we
use equation \eqref{eq:time-scale-Poiseuille} with $\Ubg$ being the mean
flow velocity over the cross-section of the pipe (assuming this
is constant). See \cref{app:pipe-background-flow} for details on
how the background flow is selected in the pipe-bounded case.

Any one of the time scales above could be selected as the
characteristic time. We choose the time scale of the elastic
shear relaxation, i.e., $\Tch = \Ts$. This has the effect that
the dimensionless elastic shear modulus $\Estilde = \Es /
(\mu\Lch/\Tch)$ should always be one. This choice is similar to
the one made by \cite{Agarwal2024}, who did however select $T$ to
be twice as large, i.e., $2\Ts$.

With the above choices, the dimensionless governing equations
will have the same form as in
\cref{sec:formulation-unbounded,sec:formulation-pipe},
but with $\mu=1$ and $\Es=1$. For example, equation
\eqref{eq:stokeslet} for the stokeslet becomes
\begin{equation}
  \label{eq:stokeslet-nondim}
  \tilde{\bmat{G}}(\tilde{\bvec{r}}) = \frac{1}{8\pi} \left(
  \frac{\bmat{I}}{\lVert \tilde{\bvec{r}} \rVert} + \frac{\tilde{\bvec{r}}
  \otimes \tilde{\bvec{r}}}{\lVert \tilde{\bvec{r}} \rVert^3}
  \right),
\end{equation}
and equation \eqref{eq:stress-tensor} for the in-plane elastic
stress tensor becomes
\begin{equation}
  \label{eq:stress-tensor-nondim}
  \tilde{\bLambda}(\tilde{\bvec{x}})
  = \frac{1}{2J(\bvec{x})} \Big(I_1(\bvec{x}) + 1\Big) \bmat{V}^2(\bvec{x}) +
  \frac{J(\bvec{x})}{2} \Big(\Edtilde I_2(\bvec{x}) - 1\Big) \bmat{P}(\bvec{x}),
\end{equation}
where $\Edtilde = \Ed/\Es$.

In the rest of this paper, we will omit the tilde above the
dimensionless quantities, and it will be clear from the context
if a symbol refers to the dimensionless or physical quantity.

\subsubsection{Dimensionless parameters}

We define a number of dimensionless parameters that characterize
the flow setup and may be varied. These include:
\begin{itemize}
  \item
    The dimensionless elastic dilatation modulus $\Ed/\Es$, which
    is the ratio between dilatation and shear resistance of the
    capsule membrane. Throughout this paper, we fix $\Ed/\Es=10$.
  \item
    The confinement ratio $\Con$ for pipe-bounded (or unbounded
    Poiseuille) flow is defined as
    \begin{equation}
      \label{eq:confinement-ratio}
      \Con = \frac{\Lch}{b},
    \end{equation}
    where $\Lch$ is the characteristic capsule length (volume-equivalent cell radius), and $b$ in general can be understood as a characteristic
    radius of the pipe cross-section. This is the ratio between
    capsule size and pipe size.
  \item
    The capillary number $\Ca$ for the background flow is defined:
    \begin{itemize}
      \item
        For unbounded shear flow as
        \begin{equation}
          \Ca^\text{shear} = \frac{\Tch}{\Tbg^\text{shear}}
          = \dot{\gamma} T
          = \frac{\dot{\gamma} \mu \Lch}{\Es},
        \end{equation}
        which is the same as the dimensionless shear rate.
      \item
        For unbounded Poiseuille flow or general pipe-bounded
        flow, the capillary number can be defined either as
        \citep{Pozrikidis2005}
        \begin{equation}
          \label{eq:capillary-number-pois-1}
          \Ca_1^\text{Pois} = \frac{\Tch}{\Tbg^\text{Pois}}
          = \frac{\Tch\Ubg}{\Lch}
          = \frac{\mu\Ubg}{\Es},
        \end{equation}
        or as \citep{Agarwal2022}
        \begin{equation}
          \Ca_2^\text{Pois} = a\Lch\Tch
          = \frac{2 \Ubg \Lch \Tch}{b^2}
          = \frac{2 \mu \Ubg \Lch^2}{b^2 \Es}
          = 2 \Ca_1^\text{Pois} \Con^2.
        \end{equation}
        Note that $\Ca_1^\text{Pois}$ is fixed if $\Ubg$ is
        fixed, while $\Ca_2^\text{Pois}$ is fixed if $a$ is fixed,
        where $\Ubg = ab^2/2$ as in equations
        \eqref{eq:background-flow-Poiseuille}--\eqref{eq:time-scale-Poiseuille}.
        This corresponds to two different ways of adjusting the
        flow profile as the pipe radius $b$ varies, essentially
        fixing either the mean flow velocity or the pressure
        gradient.
    \end{itemize}
  \item
    The reduced volume $\rv$ of the capsule is defined as
    \begin{equation}
      \label{eq:reduced-volume}
      \rv = \frac{6\sqrt{\pi} V}{A^{3/2}},
    \end{equation}
    where $V$ is the volume enclosed by the capsule membrane, and
    $A$ is the surface area of the membrane. This is the ratio of
    the capsule volume $V$ and the volume of the sphere with the
    same surface area $A$ as the capsule. Note that in our
    formulation (for extensible capsules), the surface area $A$
    may change during the simulation, so the reduced volume $\rv$
    is not constant. However, the reduced volume $\rvr$
    of the relaxed reference configuration $\gamma_\text{r}$ as
    defined in \cref{sec:formulation-unbounded} can be viewed as
    a parameter which is constant in time.
\end{itemize}
Finally, we can define the Reynolds number $\Rey$ associated with the
fluid flow as
\begin{equation}
  \label{eq:reynolds-number}
  \Rey = \frac{\rho \Lch^2}{\mu \Tbg}
  = \begin{cases}
    \rho \Lch^2 \dot{\gamma} / \mu = K \, \Ca^\text{shear} & \text{(shear flow)}, \\
    \rho \Lch \Ubg / \mu = K \, \Ca_1^\text{Pois} & \text{(Poiseuille flow)},
  \end{cases}
\end{equation}
where $\rho$ is the density of the blood plasma. In typical flows, $K = \rho \Lch \Es / \mu^2 \approx \num{8.4e-3}$.
Evidently, $\Rey$ is not an independent parameter, but a function
of the capillary number (since the fluid properties are fixed).
For the Stokes equations \eqref{eq:stokes-1}--\eqref{eq:stokes-2}
to be a valid model, the Reynolds number must be significantly
less than 1; for instance, \cite{Pozrikidis2005} states that the
Reynolds number is typically less than $0.01$ in the
microcirculation. We will assume that $\Rey$ as computed by
equation \eqref{eq:reynolds-number} is always less than 0.1,
which means that the capillary number must be less than 11.

\section{Numerical method}
\label{sec:method}

This section describes the discretization of the geometry and the numerical methods used to perform the computations. In \cref{sec:method-capsule} we describe the methods needed for the capsule both in unbounded and pipe-bounded flows. This includes the partition-of-unity representation of the surface, differentiation, integration, upsampling and blending. Then, in \cref{sec:method-pipe}, we describe how the pipe is discretized, and methods needed specifically for pipe-bounded flows, including Ewald summation and the boundary integral equation solve. Finally, in \cref{sec:method-time-stepping}, we describe the time-stepping method for evolving the capsule surface.

\subsection{Discretization of the capsule}
\label{sec:method-capsule}

Our representation of the capsule surface $\gamma$ is based on \cite{Agarwal2024}, in which a partition-of-unity parametrization with $\numpatches=6$ overlapping patches is used. There is assumed to exist a smooth bijective map $\phi : \Sphere \to \gamma$ from the unit sphere $\Sphere$ to the capsule surface $\gamma$. Defining a set $\{\mathcal{P}_i^0\}_{i=1}^{\numpatches}$ of overlapping patches on $\Sphere$ and corresponding partition-of-unity functions $\psi_i^0$ such that $\supp \psi_i^0 \subset \mathcal{P}_i^0$ and $\sum_{i=1}^{\numpatches} \psi_i^0(\bvec{x}_0) = 1$ for all $\bvec{x}_0 \in \Sphere$, we obtain patches for $\gamma$ by setting $\mathcal{P}_i = \phi(\mathcal{P}_i^0)$, and partition-of-unity functions $\psi_i = \psi_i^0 \circ \phi^{-1}$. Naturally, it holds that $\sum_{i=1}^{\numpatches} \psi_i(\bvec{x})=1$ for all $\bvec{x} \in \gamma$.

\begin{figure}
  \centering
  \includegraphics{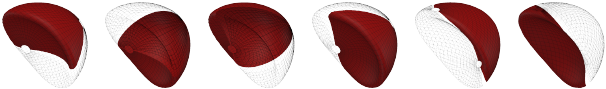}
  \caption{Illustration of the six hemispherical overlapping patches $\{\mathcal{P}_i\}$ of a deformed capsule.}
  \label{fig:capsule-patches}
\end{figure}

The patches $\{\mathcal{P}_i^0\}$ on the unit sphere are selected to be the six hemispheres that result from slicing the unit sphere by the Cartesian coordinate planes. Each patch is parametrized using local coordinates $(u,v) \in \mathcal{U} = (0,\pi) \times (0,\pi)$, which also provides a parametrization of each patch $\mathcal{P}_i$ on $\gamma$. The resulting patches on a deformed capsule surface $\gamma$ are shown in \cref{fig:capsule-patches}.
The partition-of-unity functions $\psi_i^0$ are constructed based on the bump function
\begin{equation}
    B(t) = \begin{cases}
        \exp\left( \frac{2 \E^{-1/t}}{t-1} \right), & t \in [0,1], \\
        0, & t > 1.
    \end{cases}
\end{equation}
Set $\beta_i(\bvec{x}_0) = B(d_i(\bvec{x}_0)/r_0)$ for any $\bvec{x}_0 \in \Sphere$, where $d_i(\bvec{x}_0)$ denotes the great circle distance between $\bvec{x}_0$ and the center point of $\mathcal{P}_i^0$ (i.e., the point on $\mathcal{P}_i^0$ corresponding to local coordinates $(u,v) = (\pi/2, \pi/2)$), and $r_0=5\pi/12$. Then,
\begin{equation}
    \psi_i^0(\bvec{x}_0) := \frac{\beta_i(\bvec{x}_0)}{\sum_{j=1}^{\numpatches} \beta_j(\bvec{x}_0)},
    \qquad \bvec{x}_0 \in \Sphere.
\end{equation}
For details we refer to \cite[section~3.1]{Agarwal2024}.

The capsule is discretized by introducing an equidistant tensor product grid on $\mathcal{U}$, with grid points $(u_k, v_l) = (\coordspace k, \coordspace l)$, $k, l \in \{1,\ldots,m-1\}$, $\coordspace=\pi/m$. Note that there are $(m-1)^2$ grid points per patch, i.e., $\numpatches (m-1)^2$ grid points in total on $\gamma$. The bijective map $\phi$ is needed at the first initialization of the capsule surface, but not afterwards, since the grid points $\{\bvec{x}_{kl}^i\} \subset \gamma$ are stored and updated in a Lagrangian fashion, and the values of $\{\psi_i\}$ at the grid points are precomputed.

\paragraph{Blending.}
Given a set of functions $\{f_i : \mathcal{P}_i \to \Reals \}_{i=1}^{\numpatches}$, we can define a blending operator $\mathcal{B}$ by
\begin{equation}
    \label{eq:blending}
    [\mathcal{B} \{f_i\}](\bvec{x}) = \sum_{i=1}^{\numpatches} \bar{f}_i(\bvec{x}) \psi_i(\bvec{x}), \qquad
    \bvec{x} \in \gamma,
\end{equation}
where $\bar{f}_i$ is the zero-extension of $f_i$, given by $\bar{f}_i(\bvec{x}) = f(\bvec{x})$ if $\bvec{x} \in \mathcal{P}_i$ and zero otherwise. The blending operator is merely a weighted average of $\{f_i\}$ based on the partition of unity $\{\psi_i\}$. It is clear that $\mathcal{B}$ is a linear operator, and also a projection (i.e., $\mathcal{B}^2 f = \mathcal{B}f$).

In the discrete setting, we instead have function values $\{f^i_{kl}\}_{i=1,k=1,l=1}^{\numpatches,m-1,m-1}$ at the grid points only. We then define a discrete blending operator $\mathcal{B}_{\gridspace}$ by
\begin{equation}
    \label{eq:discrete-blending}
  [\mathcal{B}_{\gridspace} \{f^i_{kl}\}]^j_{\kappa\lambda} =
  f^j_{\kappa\lambda}\psi_j(\bvec{x}^j_{\kappa\lambda})
  + \sum_{\substack{i=1\\i \neq j}}^{\numpatches}
  [\mathcal{I}_i\{f^i_{kl}\}_{k,l}](\bvec{x}^j_{\kappa\lambda})
  \psi_i(\bvec{x}^j_{\kappa\lambda}),
\end{equation}
where $\mathcal{I}_i$ is a cubic spline interpolation operator such that $\mathcal{I}_i \{g_{kl}\} : \gamma \to \Reals$. The operator $\mathcal{B}_{\gridspace}$ is also linear, but it is no longer a projection due to the interpolation.

Blending is used to ensure that quantities that have been computed patchwise have a well-defined global representation on $\gamma$. It is used mainly when computing derivatives, but we also blend the grid points themselves to stabilize the time-stepping, as will be pointed out below in \cref{sec:method-time-stepping}.

\paragraph{Upsampling.}
Certain computations, most prominently the evaluation of the layer potential, benefit from using a refined grid. Here, we use an upsampling factor of 4, i.e., we construct a refined grid for $\mathcal{U}$ using $(4m-3)^2$ points, which leads to $\numpatches (4m-3)^2$ upsampled grid points for $\gamma$. To upsample a set of function values $\{f^i_{kl}\}_{i=1,k=1,l=1}^{\numpatches,m-1,m-1}$ defined on the coarse grid, we first use patch extension as described in \cite[section~3.5]{Agarwal2024}, and then use cubic spline interpolation on each patch to interpolate onto the refined grid.

Unlike in \cite{Agarwal2024}, where derivatives were computed on the coarse grid and then upsampled, the only data that we upsample here are the grid points $\{\bvec{x}^i_{kl}\}$ themselves. We then perform differentiation and other computations needed to obtain surface area elements, interfacial forces, etc., directly on the refined grid. Computing directly on the refined grid is more expensive, since it has around 16 times as many points as the coarse grid, but it improves the accuracy, since the derivative information may have a higher bandwidth than the surface itself (and thus may require higher resolution).

(In the following, we will sometimes use $\{\bvec{x}^i_{kl}\}$ to refer also to the grid points on the refined grid, for convenience. Whether the coarse or refined grid is to be understood should be clear from context.)

\paragraph{Differentiation.}
Computing derivatives on $\gamma$ is needed for tangent and normal vectors (which are used to compute the surface deformation gradient $\bmat{F}_\text{S}$ and stress tensor $\bLambda$ from \cref{sec:formulation-unbounded}, and also surface area elements needed for integration), and for computing the surface divergence needed for interfacial forces, cf.\ \eqref{eq:interfacial-force}. As mentioned, these computations all take place on the refined grid. Apart from this difference, the computations are the same as in \cite[section~3.5]{Agarwal2024}, i.e., the function to be differentiated is first extended through patch extension, then standard central finite differences are applied to each patch separately, and finally the result is blended to ensure coherency between patches.

We have tried finite differences of second and fourth order (even higher orders lead to no improvement since the cubic spline interpolation limits the accuracy). While fourth-order differences give higher accuracy in a static setting and for simple observables (such as capsule surface area), the difference in accuracy between second and fourth order becomes negligible when considering the result from the singular integration (i.e., the Stokes layer potential), especially as the capsule deforms. We observed that second-order differences were slightly more stable, so we always use second-order finite differences for the capsule.

\paragraph{Smooth integration.}
We use the trapezoidal rule on each patch separately, which has spectral accuracy due to the compact support of the partition-of-unity functions $\{\psi_i\}$. More specifically, the integral of a smooth function $f : \gamma \to \Reals$ is computed as
\begin{equation}
    \label{eq:smooth-integration}
    \int_\gamma f(\bvec{x}) \, \mathrm{d}S
    = \sum_{i=1}^{\numpatches} \int_{\mathcal{P}_i} f(\bvec{x}) \psi_i(\bvec{x}) \, \mathrm{d}S
    \approx \sum_{i=1}^{\numpatches} \sum_{k,l} f(\bvec{x}^i_{kl}) \psi_i(\bvec{x}^i_{kl}) W(\bvec{x}^i_{kl}) \coordspace^2,
\end{equation}
where $\coordspace=\pi/m$ and $W$ is the surface area element of $\gamma$ (computed using finite differences). The right-hand side of \cref{eq:smooth-integration} can be understood either on the coarse or refined grid.

\paragraph{Singular integration.}
Computing the Stokes single-layer potential in \eqref{eq:single-layer} for $\bvec{x} \in \gamma$ requires the integration of a singular function \eqref{eq:stokeslet}. We follow \cite{Agarwal2024} and use the regularized stokeslet from \cite{Tlupova2019}, which replaces \eqref{eq:stokeslet} by
\begin{equation}
  \label{eq:regularized-stokeslet}
  \bmat{G}_\delta(\bvec{r}) = \frac{1}{8\pi\mu}
  \left(
  \frac{\bmat{I}}{\lVert \bvec{r} \rVert}
  s_1^{\#} \left( \frac{\lVert \bvec{r} \rVert}{\delta} \right)
  +
  \frac{\bvec{r} \otimes \bvec{r}}{\lVert \bvec{r} \rVert^3}
  s_2^{\#} \left( \frac{\lVert \bvec{r} \rVert}{\delta} \right)
  \right),
\end{equation}
where $\delta$ is a positive parameter known as the regularization length. The special functions that appear in \cref{eq:regularized-stokeslet} are given by
\begin{equation}
    s_1^{\#}(t) = \erf(t) -
    \frac{2}{3}t(2t^2-5)\frac{\text{e}^{-t^2}}{\sqrt{\pi}}, \qquad
    s_2^{\#}(t) = \erf(t) -
    \frac{2}{3}t(4t^4-14t^2+3)\frac{\text{e}^{-t^2}}{\sqrt{\pi}}.
\end{equation}
With \cref{eq:regularized-stokeslet}, the integrand of the layer potential \cref{eq:single-layer} becomes a smooth function, so it can be integrated using the trapezoidal rule described above. The choice of $\delta$ is important, as it must balance the regularization error, which behaves like $\bigO(\delta^5)$, and the quadrature error, which behaves like $\bigO(\gridspace \E^{-c_0(\delta/\gridspace)^2})$ \citep{Tlupova2019}, where $\gridspace$ is a measure of the local mesh spacing on $\gamma$. Here, we let $\delta$ depend on the source point, i.e., \cref{eq:single-layer} is regularized and discretized as ($\bvec{x} \in \gamma$)
\begin{equation}
    \label{eq:regularized-single-layer}
  \bvec{\mathcal{S}}_\delta[\bvec{f}](\bvec{x})
  = \int_\gamma \bmat{G}_{\delta(\bvec{y})}(\bvec{x}-\bvec{y}) \bvec{f}(\bvec{y})
  \, \mathrm{d}S(\bvec{y})
  \approx
  \sum_{i=1}^{\numpatches} \sum_{k,l} \bmat{G}_{\delta^i_{kl}}(\bvec{x} - \bvec{x}^i_{kl}) \bvec{f}(\bvec{x}^i_{kl}) \psi_i(\bvec{x}^i_{kl}) W(\bvec{x}^i_{kl}) \coordspace^2.
\end{equation}
In this context, we always discretize \cref{eq:regularized-single-layer} on the refined grid.
We choose $\delta^i_{kl}$ based on the local mesh spacing, more specifically,
\begin{equation}
    \label{eq:regularization-length-choice}
    \delta^i_{kl} = \max \{
    \lVert \bvec{x}^i_{kl} - \bvec{x}^i_{k+\Delta k,l+\Delta l} \rVert : \Delta k, \Delta l \in \{-1,0,1\}
    \}.
\end{equation}
This is a minor difference compared to \cite{Agarwal2024}, in which $\delta$ was selected as the maximum of the above quantity over each patch (i.e., $\delta$ was constant within a patch). We have observed that selecting $\delta$ per source point is beneficial in cases where $\gridspace$ has large variability over $\gamma$.

An important observation to make regarding the evaluation of \cref{eq:regularized-single-layer} is that the regularized layer potential can be evaluated at any $\bvec{x} \in \gamma$; it does not have to be at a grid point. Therefore, we can evaluate \cref{eq:regularized-single-layer} directly at the points of the coarse grid (rather than the refined grid), which is what we need to time-step \eqref{eq:time-evolution}. This is a difference compared to \cite{Agarwal2024}, where the potential was evaluated on the refined grid and then downsampled. Evaluating directly on the coarse grid is significantly faster, and no less accurate. 

\paragraph{Summary.}
At this stage, the final output is the regularized layer potential \eqref{eq:regularized-single-layer} at all grid points of the coarse grid (time-stepping the capsule comes later in \cref{sec:method-time-stepping}). In order to compute it, the coarse grid points are first upsampled to a refined grid, where tangent and normal vectors are computed using finite differences. From these, the surface area element $W$ and also the interfacial force $\bvec{f}$ can be computed. Both of these enter into \eqref{eq:regularized-single-layer}, which is evaluated on the coarse grid.

The overall expected order of accuracy is second order, from the finite differences. The computational complexity is $\bigO(\numpatches m^2)$ for blending, upsampling and finite differences. Computing the regularized Stokes layer potential at all points of the coarse grid is $\bigO(\numpatches m^4)$ using direct summation, which is also the overall complexity of the method. This can be improved using fast Ewald summation, as we shall see in \cref{sec:method-pipe}.

\subsection{Discretization of the pipe}
\label{sec:method-pipe}

The pipe segment $\Gamma$ within the primary simulation box $\mathcal{B}$ is assumed to be diffeomorphic to a cylinder segment, and can be parametrized using coordinates $(u,v) \in \mathcal{T} = [0,2\pi) \times [0,2\pi)$. For simplicity, we assume the parametrization to be of the form
\begin{equation}
    \label{eq:pipe-parametrization}
    \begin{cases}
        x_1 = \frac{\Lper}{2\pi} u, \\
        x_2 = f_2(u,v), \\
        x_3 = f_3(u,v), \\
    \end{cases}
\end{equation}
where $f_2$ and $f_3$ are $2\pi$-periodic in both $u$ and $v$.
We introduce an equidistant tensor product grid on $\mathcal{T}$, with points $(u_k,v_l) = (\coordspaceupipe k, \coordspacevpipe l)$, $k \in \{0, \ldots, \mupipe-1\}, l \in \{0, \ldots, \mvpipe-1\}$, $\coordspaceupipe = 2\pi / \mupipe$, $\coordspacevpipe = 2\pi / \mvpipe$. There are in total $\mupipe \mvpipe$ grid points $\{\bvec{x}_{kl}\}$ on $\Gamma$. The discussion below follows \cref{sec:method-capsule} closely, but note that the pipe consists of a single patch, and no partition of unity is needed.

\paragraph{Upsampling.}
The refined grid has $(4 \mupipe + 1)(4 \mvpipe + 1)$ grid points, and we use Fourier interpolation to interpolate values from the coarse grid to the refined grid, since functions on $\Gamma$ can be assumed to be periodic in both $u$ and $v$. As an exception, when upsampling the grid points $\{\bvec{x}_{kl}\}$, periodicity of the $x_1$ component cannot be assumed. Instead, we use the known form in \cref{eq:pipe-parametrization} to evaluate $x_1$ analytically.

\paragraph{Differentiation.}
We use standard fourth order central finite differences, taking periodicity in both $u$ and $v$ into account. For the pipe, only the normal vectors $\bvec{n}$ and surface area elements $W$ are needed.

\paragraph{Smooth integration.}
We use the trapezoidal rule, which is spectrally accurate since the integrand is periodic. The integral of a smooth function $f : \Gamma \to \Reals$ is computed as
\begin{equation}
    \int_\Gamma f(\bvec{x}) \, \mathrm{d}S
    \approx
    \sum_{k,l} f(\bvec{x}_{kl}) W(\bvec{x}_{kl}) \coordspaceupipe \coordspacevpipe.
\end{equation}

\paragraph{Singular integration.}
For a pipe, the Hebeker potential \cref{eq:hebeker-potential}, which includes both a single-layer and a double-layer potential, is needed. The single layer is regularized in the same way as for the capsule. For the double layer, the regularized stresslet from \cite{Tlupova2019} is used,
\begin{equation}
    \label{eq:regularized-stresslet}
    T_{jlm}^\delta(\bvec{r}) = -\frac{6}{8\pi} \frac{r_j r_l r_m}{\lVert \bvec{r} \rVert^5} s_3^{\#} \left( \frac{\lVert \bvec{r} \rVert}{\delta} \right), \qquad
    s_3^{\#}(t) = \erf(t) -
    \frac{2}{9}t(8t^6 - 36t^4 + 6t^2 + 9)\frac{\text{e}^{-t^2}}{\sqrt{\pi}},
\end{equation}
with the same $\delta$ used for both the stokeslet and stresslet (set in the same way as for the capsule).

\paragraph{Fast Ewald summation.}
The periodic summation in \cref{eq:periodic-flow-field,eq:boundary-integral-equation} is evaluated by a fast spectrally accurate Ewald summation method from \cite{Bagge2023}. This method can evaluate singly periodic sums of the form
\begin{equation}
    \bvec{\mathcal{K}}[\bvec{q}](\bvec{x}) = \sum_{n=-\infty}^\infty \int_\Gamma \bmat{K}(\bvec{x} - \bvec{y} - \bvec{p}_n) \bvec{q}(\bvec{y}) \, \mathrm{d}S(\bvec{y}),
\end{equation}
which encompasses the Stokes single layer and double layer, by splitting the kernel into two parts $\bmat{K} = \bmat{K}^\mathrm{N} + \bmat{K}^\mathrm{F}$, where the first part is singular and short-range, while the second part is smooth and long-range. The second part is treated in Fourier space, where it converges rapidly, using the fast Fourier transform (FFT). The algorithm is similar to the nonuniform FFT \citep{Dutt1993,Greengard2004}. The treatment of this second part is described completely in \cite{Bagge2023}, so we do not repeat the details here. The computational complexity is $\bigO(N \log N)$ for $N$ total grid points.

The first, singular, part requires some care as it must be combined with the special quadrature method -- in this case, the regularization of the kernel. Let us consider the stokeslet as an example. The plain stokeslet $\bmat{G}$ is decomposed using the Hasimoto split \citep{Hasimoto1959} such that
\begin{equation}
\label{eq:hasimoto-near-field-stokeslet}
  \bmat{G}^\mathrm{N}(\bvec{r})
  = \bmat{G}(\bvec{r})
  \left( \erfc(\xi \lVert \bvec{r} \rVert)
  + \frac{2\xi \lVert \bvec{r} \rVert \E^{-\xi^2 \lVert \bvec{r} \rVert^2}}{\sqrt{\pi}}
  \right)
  - \frac{1}{8\pi\mu} \bmat{I} \frac{4\xi \E^{-\xi^2 \lVert \bvec{r} \rVert^2}}{\sqrt{\pi}},
\end{equation}
where $\xi > 0$ is a parameter controlling the decay. Note that $\bmat{G}^\mathrm{N}(\bvec{r})$ is still singular at $\bvec{r} = \bvec{0}$. One option would be to try to regularize $\bmat{G}^\mathrm{N}(\bvec{r})$ directly, but doing so is not straightforward. Instead, we use an idea from \cite{afKlinteberg2016}: Since there is a cutoff radius $\rc$ such that $\lVert \bmat{K}^\mathrm{N}(\bvec{r}) \rVert \approx 0$ whenever $\lVert \bvec{r} \rVert > \rc$, we can write
\begin{multline}
    \label{eq:near-field-trick}
    \bvec{\mathcal{K}}^\mathrm{N}[\bvec{q}](\bvec{x}) \approx \sum_{n=-\infty}^\infty \int_{\Gamma^*(\bvec{x},n)} \bmat{K}^\mathrm{N}(\bvec{x} - \bvec{y} - \bvec{p}_n) \bvec{q}(\bvec{y}) \, \mathrm{d}S(\bvec{y})
    =
    \\
    =
    \sum_{n=-\infty}^\infty \int_{\Gamma^*(\bvec{x},n)} \bmat{K}(\bvec{x} - \bvec{y} - \bvec{p}_n) \bvec{q}(\bvec{y}) \, \mathrm{d}S(\bvec{y})
    -
    \sum_{n=-\infty}^\infty \int_{\Gamma^*(\bvec{x},n)} \bmat{K}^\mathrm{F}(\bvec{x} - \bvec{y} - \bvec{p}_n) \bvec{q}(\bvec{y}) \, \mathrm{d}S(\bvec{y}),
\end{multline}
where $\Gamma^*(\bvec{x},n) = \{ \bvec{y} \in \Gamma : \lVert \bvec{x} - \bvec{y} - \bvec{p}_n \rVert \leq \rc \}$. Now, the second integral (involving $\bmat{K}^\mathrm{F}$) in \cref{eq:near-field-trick} has a smooth integrand and requires no special treatment. The first integral (involving $\bmat{K}$) can be regularized by replacing $\bmat{K}$ by $\bmat{K}_\delta$, i.e., the standard regularizations already presented above in \cref{eq:regularized-stokeslet,eq:regularized-stresslet}. All integrals are local since $\Gamma^*(\bvec{x},n)$ takes the place of $\Gamma$. The complexity of evaluating $\bvec{\mathcal{K}}^\mathrm{N}$ at $N$ grid points is $\bigO(N)$, so the overall complexity is $\bigO(N \log N)$ from the Fourier space computations.

\paragraph{Integral equation solve.}
The integral equation \eqref{eq:boundary-integral-equation} is discretized in the $\mupipe \mvpipe$ grid points on the wall, which leads to a linear system of size $\mupipe \mvpipe \times \mupipe \mvpipe$. This system can be solved iteratively using e.g.\ GMRES, which requires a modest number of iterations (at most 15--20 iterations without preconditioner for relative tolerance $10^{-7}$ for typical pipe shapes, independent of pipe resolution). The solution process is not very sensitive to the value of the Hebeker parameter $\eta$, so from now on, we fix $\eta=0.1$.

For the problem sizes considered in this paper, however, it is typically faster to construct the whole coefficient matrix of the $\mupipe \mvpipe \times \mupipe \mvpipe$ linear system, make an LU decomposition of it, and solve the system directly using back-substitution. For example, for the highest pipe resolution in \cref{tab:pipe-data} (\cref{sec:results-pipe-flow}), the coefficient matrix is of size $4096 \times 4096$, and two back-substitutions take about \SI{21}{ms} on an Intel Ice Lake node (using MATLAB).

\paragraph{Summary.}
The computations from \cref{sec:method-capsule} for the capsule remain, with the difference that the Stokes single-layer potential \cref{eq:single-layer} from the capsule is now evaluated using the fast Ewald summation method, and on both the coarse capsule and pipe grids. Then, in order to solve the integral equation \eqref{eq:boundary-integral-equation}, we must first compute surface area elements $W$ and normal vectors $\bvec{n}$ for the pipe, which is done on its refined grid (using finite differences). We then solve the integral equation using either LU decomposition or GMRES. Finally, the Hebeker potential \cref{eq:hebeker-potential} from the pipe is evaluated on the coarse grid of the capsule.

The pipe resolution $(\mupipe, \mvpipe)$ is related to the capsule resolution $m$, so that both $\mupipe$ and $\mvpipe$ are $\bigO(m)$. In practice, we use $\mupipe = \frac{2}{3} m \Lper/5$ and $\mvpipe = a \mupipe$, where $a$ depends on the pipe shape (for the shapes in \cref{tab:pipe-data}, $a$ varies between 2 and 4).

For the pipe, the order of accuracy is limited by the selected order of the finite differences, which is fourth order. However, the overall accuracy is still second order, from the capsule. The computational cost is dominated by the Ewald potential evaluations, which have complexity $\bigO(m^2 \log m^2)$. Solving the integral equation using LU decomposition has complexity $\bigO(m^4)$, while using GMRES has complexity $\bigO(m^2 \log m^2)$. Switching from LU to GMRES for large enough $m$ makes the overall complexity of the method $\bigO(m^2 \log m^2)$.

\subsection{Time-stepping}
\label{sec:method-time-stepping}

The time evolution of the capsule membrane is governed by \cref{eq:time-evolution}, which after discretization in space becomes
\begin{equation}
    \label{eq:discrete-time-evolution}
    \dot{\bvec{x}}^i_{kl} = \bvec{u}(\bvec{x}^i_{kl}),
\end{equation}
for grid points $\{\bvec{x}^i_{kl}\}$ on the coarse capsule grid. Here, $\bvec{u}(\bvec{x})$ is the numerical evaluation of either \cref{eq:flow-field} or \cref{eq:periodic-flow-field}, in the unbounded or pipe-bounded setting, respectively, as described above. For time-stepping \eqref{eq:discrete-time-evolution}, we use an explicit adaptive Runge--Kutta method of order 3(2) by \cite{Bogacki1989}. We also tried higher-order time-stepping methods such as the Runge--Kutta--Fehlberg method of order 4(5) \citep{Fehlberg1969}, but typically, the third-order method will be more efficient due to CFL-like step-size constraints.

The time-stepping method has a relative error tolerance $\odetol$, which can be related to the capsule resolution $m$ as follows. We want to select $\odetol$ such that the temporal error is about the same size as, or slightly smaller than, the spatial error. This way, $m$ controls the overall accuracy, and the time-step size does not become unnecessarily small. Using numerical experiments, we have found that setting
\begin{equation}
    \label{eq:odetol-rule}
    \odetol = \tau_0 \left( \frac{12}{m} \right)^{\log_2(10)}
\end{equation}
leads to a good balance. This sets $\odetol=\tau_0$ for $m=12$ and reduces $\odetol$ by a factor of 10 as $m$ doubles. Of course, the behavior encoded by \cref{eq:odetol-rule} is not second-order in $m$ ($\log_2(10) \approx 3.3$), but it matches how the spatial error tends to behave in practice for moderate $m$. The value of $\tau_0$ is selected as $10^{-5}$ for experiments in unconfined flows (shear flow and relaxation), and as $10^{-4}$ for experiments in pipe-confined flows.

\paragraph{Parameters.}
With the parameter choices mentioned above, the method as described so far has a single parameter controlling the error, namely the capsule resolution $m$. All other numerical parameters are either fixed (such as the Hebeker parameter $\eta=0.1$) or determined through $m$ directly (such as the pipe resolution $(\mupipe, \mvpipe)$ and time-stepping tolerance $\odetol$) or indirectly (such as the regularization length $\delta$ given by \eqref{eq:regularization-length-choice}). The parameters for the fast Ewald summation method have not been mentioned here, but they are set from an error tolerance according to \cite{Bagge2023}; for convenience, we fix this Ewald tolerance to $10^{-6}$, which is small enough not to influence the overall error.

\paragraph{Accuracy and complexity.}
The method as a whole is second-order accurate in space and third-order accurate in time. The time-stepping method requires three evaluations of $\bvec{u}(\bcdot)$ per time step, each of which has computational complexity $\bigO(m^2 \log m^2)$ assuming the fast Ewald summation is used.

\paragraph{Blending.}
In order to stabilize the scheme, additional blending of grid point positions is needed. If this is not done, we observe that individual grid points may escape from the capsule surface and lag behind it, causing instabilities. To see why, note that some grid points in fact have the partition-of-unity value $\psi_i(\bvec{x}^i_{kl}) = 0$ (this is because the support of $\psi_i$ is actually not all of $\mathcal{P}_i$). Let $\bvec{x}^1_*$ be such a grid point on patch $\mathcal{P}_1$. Suppose $\gamma$ is initially in its relaxed configuration $\gamma_\mathrm{r}$, and then perturb only $\bvec{x}^1_*$ by moving it away from the surface $\gamma$ altogether. The interfacial force $\bvec{f}(\bvec{x}^1_*)$ may try to pull $\bvec{x}^1_*$ back to $\gamma$, but since $\psi_i(\bvec{x}^i_{kl}) = 0$, this term disappears from the Stokes potential \cref{eq:regularized-single-layer}. Even when evaluating $\bvec{\mathcal{S}}_\delta[\bvec{f}](\bvec{x}^1_*)$ at the point itself, the contribution from $\bvec{f}(\bvec{x}^1_*)$ is zero. None of the other terms in \cref{eq:regularized-single-layer} will provide a force pulling $\bvec{x}^1_*$ back to the capsule surface; on the contrary, the neighbors of $\bvec{x}^1_*$ on the grid may even be pulled in the opposite direction (towards $\bvec{x}^1_*$). Thus, the velocity $\bvec{u}(\bvec{x}^1_*)$ at $\bvec{x}^1_*$ will \emph{not} bring the grid point back towards $\gamma$.

Blending the velocities\footnote{When blending a vector quantity such as velocity or position, \cref{eq:discrete-blending} is applied to each component separately.} $\bvec{u}(\bvec{x}^i_{kl})$ after computing them on the coarse capsule grid will not help with this, as the problem remains that none of the velocities will try to bring $\bvec{x}^1_*$ back. Instead, we blend the grid point positions $\bvec{x}^i_{kl}$ themselves, which will bring $\bvec{x}^1_*$ back to $\gamma$. In practice, we blend both velocities and positions. Velocities are blended as soon as they have been computed on the coarse capsule grid (i.e., as part of the evaluation of the right-hand side of \cref{eq:discrete-time-evolution}), while positions are blended immediately after each completed time step. This correction improves the stability and does not seem to affect the time-stepper adversely in any way.

\paragraph{Spectral filtering.}
Even when blending velocities and positions as mentioned above, the simulation eventually becomes unstable, especially for challenging cases such as reduced volumes $\rvr < 1$, large capillary numbers, or flows that induce high curvature in the capsule shape. In order to stabilize the simulations further, we develop a spectral filtering method that is described separately in \cref{sec:spectral_filtering}.

\section{Stability analysis for a spherical capsule}
\label{sec:stability-analysis}

The motivation for the analysis in this section is to understand the stability and stiffness of the capsule evolution equation \eqref{eq:time-evolution} for the quiescent unbounded case ($\bvec{u}_\infty = \bvec{0}$), i.e., during pure relaxation. The evolution equation is then given by
\begin{equation}
\label{eq:relaxation-evolution}
    \dot{\bvec{x}} = \bvec{\mathcal{S}}[\bvec{f}] (\bvec{x}),
\end{equation}
which can be understood as an operator $\bvec{\mathcal{S}} \forceop$ acting on a perturbation $\bvec{\delta}$ from the relaxed reference configuration $\gammar$: $\dot{\bvec{\delta}} = \bvec{\mathcal{S}} \forceop \bvec{\delta}$. The spectra of the operators $\bvec{\mathcal{S}}$ and $\forceop$ determine the stability of this differential equation, and any time-step restrictions due to stiffness.

A similar analysis was made in \cite[section~5.1]{Veerapaneni2011}, for the operator $\bvec{\mathcal{S}}$, and for the inextensibility and bending operators that appear in their formulation. Here, we extend the analysis to the Skalak model for the interfacial force that appears in our formulation, cf.\ \cref{eq:interfacial-force,eq:stress-tensor}, and this is the main novel contribution of this section. The analysis is carried out in the case of a spherical relaxed capsule, i.e., $\gammar = \Sphere$.

We start by defining the scalar and vector spherical harmonics, which are convenient basis functions on the unit sphere, as in \cite{Veerapaneni2011}.

\begin{definition}
Let the \emph{(scalar) spherical harmonic} function of degree $\spha$ and order $\sphb$ ($\spha = 0, 1, 2, \ldots; \sphb = -\spha, \ldots, \spha$) be the function $\sph_\spha^\sphb : [0,\pi] \times [0,2\pi) \to \Complex$ given by
\begin{equation}
    \sph_\spha^\sphb(\theta, \varphi) = \frac{1}{\sqrt{2\pi}}
    \sqrt{
      \left( \spha + \frac{1}{2} \right) \frac{(\spha-\sphb)!}{(\spha+\sphb)!}
    } P_\spha^\sphb(\cos \theta) \E^{\I \sphb \varphi},
\end{equation}
where $P_\spha^\sphb$ is the associated Legendre function of degree $\spha$ and order $\sphb$.
\end{definition}

\begin{definition}
  \label{def:vsph}
  Let the \emph{vector spherical harmonics} of degree $\spha$ and order $\sphb$ be three sets of functions, all $[0,\pi] \times [0, 2\pi) \to \Complex^3$, given by
  \begin{align}
    \label{eq:def-vsphv}
    \vsphv_\spha^\sphb(\theta,\varphi) &=
    \gradsphere \sph_\spha^\sphb(\theta,\varphi)
    - (\spha+1)\sph_\spha^\sphb(\theta,\varphi) \nr(\theta,\varphi), \\
    \label{eq:def-vsphw}
    \vsphw_\spha^\sphb(\theta,\varphi) &=
    \gradsphere \sph_\spha^\sphb(\theta,\varphi)
    + \spha\sph_\spha^\sphb(\theta,\varphi) \nr(\theta,\varphi), \\
    \label{eq:def-vsphx}
    \vsphx_\spha^\sphb(\theta,\varphi) &=
    \nr(\theta,\varphi) \times
    \gradsphere \sph_\spha^\sphb(\theta,\varphi),
  \end{align}
  where $\Sphere$ is the unit sphere embedded in $\Reals^3$ and $\nr$ is its outward-pointing unit normal.
\end{definition}

The vector spherical harmonics form an orthogonal basis for the set of square-integrable vector-valued functions defined on $\Sphere$, i.e., $L^2(\Sphere; \Reals^3)$, with inner product $\langle \bvec{u}, \bvec{v} \rangle = \int_{\Sphere} \bvec{u}^\CT \bvec{v} \, \mathrm{d}S$, where $(\bcdot)^\CT$ denotes the conjugate transpose.

The following theorem due to \cite[Theorem~5.1]{Veerapaneni2011} give the eigenfunctions of the Stokes operator on the unit sphere.

\begin{theorem}[Veerapaneni et al., 2011]
  \label{thm:stokes-eigenfunctions}
    When $\gamma = \Sphere$, the vector spherical harmonics $\vsphv_\spha^\sphb$ and $\vsphw_\spha^\sphb$ are eigenfunctions of the Stokes single-layer operator $\bvec{\mathcal{S}}$, defined by \cref{eq:single-layer}, and
    \begin{align}
      \bvec{\mathcal{S}}[\vsphv_\spha^\sphb] = \frac{\spha}{(2\spha+1)(2\spha+3)} \vsphv_\spha^\sphb,
      \qquad \spha \geq 0, \\[5pt]
      \bvec{\mathcal{S}}[\vsphw_\spha^\sphb] = \frac{\spha+1}{(2\spha-1)(2\spha+1)} \vsphw_\spha^\sphb,
      \qquad \spha \geq 1.
    \end{align}
\end{theorem}

Furthermore, $\bvec{\mathcal{S}}[\vsphx_\spha^\sphb] = (2\spha+1)^{-1} \vsphx_\spha^\sphb$ for $\spha \geq 1$ \citep{CoronaVeerapaneni2018}.

Based on the formulas for the interfacial force, \cref{eq:interfacial-force,eq:stress-tensor}, we introduce the following two operators for the shear and dilatation parts. The full force is $\forceop = \Es \forceops + \Ed \forceopd$.

\begin{definition}
    For fixed reference configuration $\gammar = \Sphere$, let the operators $\forceops$ and $\forceopd$, both mapping functions $\gammar \to \Reals^3$ to $\gamma \to \Reals^3$, be given by
  \begin{align}
    \label{eq:force-op-s-def}
    \forceops[\bvec{\delta}](\bvec{x})
    &=
    \gradgamma \bcdot \left(
      \frac{1}{2 J(\bvec{x})} \Big( I_1(\bvec{x}) + 1 \Big)
      \bmat{V}^2(\bvec{x})
      - \frac{J(\bvec{x})}{2} \bmat{P}(\bvec{x})
    \right),
    \\
    \label{eq:force-op-d-def}
    \forceopd[\bvec{\delta}](\bvec{x})
    &=
    \gradgamma \bcdot \left(
      \frac{J(\bvec{x})}{2} I_2(\bvec{x}) \bmat{P}(\bvec{x})
    \right),
  \end{align}
  where $\bvec{\delta}$ is the perturbation from $\gammar$ to $\gamma$, i.e., $\gamma = \{ \bvec{x}_\mathrm{r} + \bvec{\delta}(\bvec{x_\mathrm{r}}) :
  \bvec{x}_\mathrm{r} \in \gammar \}$, and the definitions of all quantities above are as in \cref{sec:formulation-unbounded}.
\end{definition}

We now state the two main results of this section, namely the linearizations of the operators just defined and their spectra.

\begin{proposition}[Linearization]
    \label{prop:force-linearization}
    For small $\lVert \bvec{\delta} \rVert_{L^2}$, it holds that $\forceops[\bvec{\delta}] = \forceopslin[\bvec{\delta}] + \bigO(\lVert \bvec{\delta} \rVert_{L^2}^2)$ and $\forceopd[\bvec{\delta}] = \forceopdlin[\bvec{\delta}] + \bigO(\lVert \bvec{\delta} \rVert_{L^2}^2)$, where
    \begin{align}
        \label{eq:prop-linear-force-fs}
        \forceopslin[\bvec{\delta}] &= \frac{1}{2} (\gradsphere - 2\nr) (\gradsphere \bcdot \bvec{\delta})
        + \frac{1}{2} \bmat{P}_\mathrm{r} (\Deltasphere \bvec{\delta})
        - \gradsphere (\nr \bcdot \bvec{\delta})
        + \bmat{P}_\mathrm{r} \bvec{\delta},
        \\
        \label{eq:prop-linear-force-fd}
        \forceopdlin[\bvec{\delta}] &= (\gradsphere - 2 \nr) (\gradsphere \bcdot \bvec{\delta}).
    \end{align}
    Here, $\nr$ is the outward unit normal of the sphere, and $\bmat{P}_\mathrm{r} = \bmat{I} - \nr \otimes \nr$.
\end{proposition}

\begin{theorem}[Eigenfunctions]
    \label{thm:force-eigenfunctions}
    Let $\bvec{A}_0^0 = \vsphv_0^0$ and
    \begin{align}
        \bvec{A}_\spha^\sphb &= (\spha+2) \vsphv_\spha^\sphb
        + (\spha-1) \vsphw_\spha^\sphb, \\
        \bvec{B}_\spha^\sphb &= \spha (\spha-1) \vsphv_\spha^\sphb
        - (\spha+1)(\spha+2) \vsphw_\spha^\sphb,
    \end{align}
    $\spha \geq 1$, $\lvert \sphb \rvert \leq \spha$. Then $\bvec{A}_\spha^\sphb$, $\bvec{B}_\spha^\sphb$ and $\vsphx_\spha^\sphb$ are eigenfunctions of $\forceopdlin$, and
    \begin{align}
        \label{eq:eig-fd-a}
        \forceopdlin[\bvec{A}_\spha^\sphb] &= -(\spha^2 + \spha + 4) \bvec{A}_\spha^\sphb,
        \qquad \spha \geq 0, \\
        \label{eq:eig-fd-b}
        \forceopdlin[\bvec{B}_\spha^\sphb] &= \bvec{0}, \qquad \spha \geq 1, \\
        \label{eq:eig-fd-x}
        \forceopdlin[\vsphx_\spha^\sphb] &= \bvec{0}, \qquad \spha \geq 1.
    \end{align}

    Furthermore, let $\bvec{C}_0^0 = \vsphv_0^0$, $\bvec{C}_1^\sphb = \vsphv_1^\sphb$, $\bvec{D}_1^\sphb = \vsphw_1^\sphb$ and
    \begin{align}
        \bvec{C}_\spha^\sphb &= 2 \spha (\spha-1)(\spha+2) \vsphv_\spha^\sphb
        + K_{+} \vsphw_\spha^\sphb, \\
        \bvec{D}_\spha^\sphb &= 2 \spha (\spha-1)(\spha+2) \vsphv_\spha^\sphb
        + K_{-} \vsphw_\spha^\sphb,
    \end{align}
    $\spha \geq 2$, $\lvert \sphb \rvert \leq \spha$, where $K_{\pm} = -3(\spha^2 + \spha + 1)
    + (2\spha+1)R_{\pm}$ and $R_{\pm} = \pm \sqrt{\spha^4 + 2 \spha^3 - \spha^2 - 2\spha + 9}$.
    Then $\bvec{C}_\spha^\sphb$, $\bvec{D}_\spha^\sphb$ and $\vsphx_\spha^\sphb$ are eigenfunctions of $\forceopslin$, and
    \begin{align}
        \label{eq:eig-fs-c}
        \forceopslin[\bvec{C}_\spha^\sphb] &= -(\spha-1)\frac{\spha^2 + 3\spha + 5 + R_{+}}{\spha^2 + 3\spha - 1 - R_{+}} \bvec{C}_\spha^\sphb, \qquad \spha \geq 0, \\[5pt]
        \label{eq:eig-fs-d}
        \forceopslin[\bvec{D}_\spha^\sphb] &= -(\spha-1)\frac{\spha^2 + 3\spha + 5 + R_{-}}{\spha^2 + 3\spha - 1 - R_{-}} \bvec{D}_\spha^\sphb, \qquad \spha \geq 1, \\[5pt]
        \label{eq:eig-fs-x}
        \forceopslin[\vsphx_\spha^\sphb] &= -\frac{(\spha-1)(\spha+2)}{2} \vsphx_\spha^\sphb, \qquad \spha \geq 1.
    \end{align}
\end{theorem}

The proofs of \cref{prop:force-linearization,thm:force-eigenfunctions} are found in \cref{app:stability-analysis}.

\Cref{thm:force-eigenfunctions} shows that the dilatation force operator $\forceopdlin$ has a large nullspace, while the nullspace of the shear force operator $\forceopslin$ is only six-dimensional and spanned by $\{\vsphw_1^\sphb, \vsphx_1^\sphb : \sphb \in \{ -1, 0, 1 \} \}$, which corresponds exactly to rigid-body translations and rotations.%
\footnote{The first few eigenvalues of $\forceopslin$ associated with $\bvec{C}_\spha^\sphb$ are $-2$, $-3$, $-(\sqrt{33}+15)/(9-\sqrt{33}) \approx \num{-6.372281}$ for $r=0,1,2$, i.e., none of them are zero.}
Using \cref{thm:force-eigenfunctions}, we can show that (i) the continuous differential equation \eqref{eq:relaxation-evolution} is Lyapunov-stable for small deformations of the sphere, and (ii) the only steady-state deformations are, as long as $\Es \neq 0$, rigid-body transformations and the normal vector field $\vsphv_0^0$ (which is in the nullspace of $\bvec{\mathcal{S}}$ and does not satisfy the incompressibility condition). The argument is as follows.

(i) Both of the operators $\bvec{\mathcal{S}}$ and $\forceoplin = \Es \forceopslin + \Ed \forceopdlin$ are block-diagonal in the sense that they preserve the space $\mathcal{V}_\spha^\sphb$ spanned by $\{ \vsphv_\spha^\sphb, \vsphw_\spha^\sphb, \vsphx_\spha^\sphb \}$ for fixed $\spha$ and $\sphb$. (This space is one-dimensional for $r=0$ and three-dimensional for $r>0$.) Thus, the composite operator $\mathcal{L} = \bvec{\mathcal{S}} \forceoplin$ is also block-diagonal, with blocks given by $\mathcal{L}_j = \bvec{\mathcal{S}}_j \forceoplin_j$ (no sum over $j$), where $j=0, 1, 2, \ldots$ is a block index. Each block $\bvec{\mathcal{S}}_j$ is Hermitian positive semidefinite, while $\forceoplin_j$ is Hermitian negative semidefinite. Since $\bvec{\mathcal{S}}_j$ and $\forceoplin_j$ have different eigenbases, $\mathcal{L}_j$ is not Hermitian, but it is similar to a Hermitian negative semidefinite operator $\mathcal{B}_j = \mathcal{R}_j^{-1} \mathcal{L}_j \mathcal{R}_j$, where $\mathcal{R}_j = \mathcal{S}_j^{1/2}$, which shows that $\mathcal{L}_j$ has non-positive real eigenvalues. Using this similarity relation, one can show that the solution to $\dot{\bvec{\delta}}_j = \mathcal{L}_j \bvec{\delta}_j$ satisfies
\[
\lVert \bvec{\delta}_j(t) \rVert_{L^2} \leq C_j \lVert \bvec{\delta}_j(0) \rVert_{L^2}, \qquad\text{for all $t \geq 0$},
\]
where $C_j$ is the condition number of $\mathcal{R}_j$. From the eigenvalues of $\mathcal{S}_j$ (\cref{thm:stokes-eigenfunctions}), we see that $C_j$ is uniformly bounded by $\sqrt{10}$. Thus, the solution to the full system $\dot{\bvec{\delta}} = \mathcal{L} \bvec{\delta}$ satisfies the same bound $\lVert \bvec{\delta}(t) \rVert_{L^2} \leq \sqrt{10} \lVert \bvec{\delta}(0) \rVert_{L^2}$, which shows Lyapunov-stability of the stationary solution $\bvec{\delta}=\bvec{0}$.

(ii) For the scalar block $j=0$, corresponding to the space spanned by $\vsphv_0^0$, we get $\mathcal{L}_0=0$, i.e., $\vsphv_0^0$ belongs to the nullspace of $\mathcal{L}$. For blocks with $j > 0$, the $\mathcal{S}_j$ operator is invertible, so the nullspace of $\mathcal{L}_j$ is that of $\forceoplin_j$. Since $(\forceopslin)_j$ and $(\forceopdlin)_j$ are both Hermitian negative semidefinite, we have $\ker \forceoplin_j = \ker (\forceopslin)_j \cap \ker (\forceopdlin)_j$, and in this case, the intersection is the same as $\ker (\forceopslin)_j$.

The stiffness of the operators is governed by their condition numbers, which grow as $\bigO(\spha^2)$ for both $\forceopslin$ and $\forceopdlin$. Thus, an explicit time-stepper would have the step-size restriction $\Delta t < \bigO(\spha^{-2})$ for solving $\dot{\bvec{\delta}} = \widetilde{\forceop}\bvec{\delta}$. This is consistent with observations in \cite[section~7]{Zhao2010}. Including the Stokes operator $\bvec{\mathcal{S}}$ can be expected to improve the behavior, perhaps to $\Delta t < \bigO(r^{-1 - \alpha})$ for some $\alpha \in [0,1]$. (\cite{Zhao2010} observe $\bigO(\spha^{-1.2})$.) In practice, the largest $\spha$ to consider here depends on what degree $\spha$ of spherical harmonics the grid can represent, which is related to the discussion in the next section. In most cases, one can expect $\spha_\text{max} = \bigO(m)$, so that $\Delta t < \bigO(m^{-1-\alpha})$.

\section{Spectral filtering}\label{sec:spectral_filtering}

Numerical instability in explicit simulations of elastic capsules generally arises from two sources: stiffness and aliasing. The high dilatation introduces severe stiffness that imposes a CFL-like restriction on the explicit time step. While this temporal instability is readily managed by using a relatively small $\Ed/\Es$ and employing sufficiently small time steps within our explicit time-integrator, refining the time step does not resolve the spatial instabilities. Specifically, the strong nonlinearities in the Skalak model generate aliasing errors. If left unaddressed, these errors cause a spurious amplification of high-frequency modes that eventually destabilizes the simulation entirely.

To stabilize the capsule simulations, we develop a spectral filtering method as an anti-aliasing method based on spherical harmonic expansions.  The surface is represented using its spherical harmonic coefficients, high-frequency modes are truncated, and the filtered spectrum is used to reconstruct a smoothed surface. While spectral filtering is a standard approach, our key contribution is a detailed analysis of how to choose the truncation index $p_{\max}$. Specifically, $p_{\max}$ is selected based on the interpolation error on the POU grid, ensuring that we retain only the spherical harmonic modes that can be accurately resolved.

Algorithm \ref{alg:spectral_filter} outlines the filtering procedure. 
Starting with a partition-of-unity (POU) discretized surface $\gamma$, we denote the surface positions on the POU grid by $\bmat{x}_m\in \mathbb{R}^{6(m - 1)^2\times 3}$. A corresponding spherical harmonic grid is defined on $\gamma$, using the same parameterization as the POU grid and the discretization scheme from \cite{Veerapaneni2011}. Surface positions on the spherical harmonic grid, $\bmat{x}_{q}\in \mathbb{R}^{2q(q + 1)\times 3}$, are obtained via interpolation, with $q = 4m$. 
We then compute the discrete spherical harmonic transform (DSHT) to obtain coefficients up to degree $p = 2m$, denoted by $\bmat{\hat{x}}_p$. (See \cref{appendix:DSHT} for the definition and properties of the DSHT and its inverse.) These coefficients are truncated at $p_{\max} \leq p$, and an inverse transform yields the filtered positions $\bmat{x}_{q}^{p_{\max}}$ on the spherical harmonic grid. Finally, we interpolate back to the POU grid to recover the filtered surface positions $\bmat{x}^{p_{\max}}_m$.

\begin{algorithm}[H]
\caption{Spectral Filtering Method}
\begin{algorithmic}[1]
\State Interpolate to spherical harmonic grid ($q = 4m$): $\bmat{x}_q = I^{q}_m\bmat{x}_m$
\State Compute coefficients up to $p = 2m$ using DSHT: $\bmat{\hat{x}}_p = \bmat{Q}^*_{qp}\bmat{x}_q$
\State Truncate coefficients at $p_{\max}$: $\bmat{\hat{x}}_{p_{\max}} = {\mathds{1}}_{[0,\, p_{\max}]} \bmat{\hat{x}}_p$
\State If $\epsilon_{p_{\max}}^\text{tail} = \sqrt{1 - \frac{\Vert\bmat{\hat{x}}_{p_{\max}}\Vert_2^2}
{\Vert\bmat{\hat{x}}_p\Vert_2^2}} > \text{tol}_{\min}$:
\begin{enumerate}
  \item Compute $\bmat{x}_q^{p_{\max}} = \bmat{Q}_{qp} \bmat{\hat{x}}_{p_{\max}}$
  \item Interpolate back: $\bmat{x}_m^\text{filt} = \bmat{x}^{p_{\max}}_m = I^{m}_q\bmat{x}_q^{p_{\max}}$
\end{enumerate}
\State Else: $\bmat{x}_m^\text{filt} = \bmat{x}_m$
\end{algorithmic}\label{alg:spectral_filter}
\end{algorithm}

The complexity is dominated by the DSHT, which can be reduced to $\mathcal{O}(p^2 (\log p)^2 + p^2 \log p)$ using FFTs in longitude and fast Legendre transforms \citep{mohlenkamp1999fast} in latitude. In our current implementation, we use FFTs only, resulting in $\mathcal{O}(p^3)$ complexity.

\subsection{Parameter selection}

The key parameters in the spectral filter are $p_{\max}$ and $\text{tol}_{\min}$. By default, we choose
\begin{equation}
p_{\max} = m.
\end{equation}
This choice is justified by analysis of the interpolation error of the surface function on the POU grid (see \cref{appendix:parameter_filtering} for details). In short, the error can be estimated as
\begin{equation}
\epsilon_m^\text{interp} \leq \sqrt{\sum_{n = 0}^{\infty} \left\Vert  \Delta \bmat{Y}_n^m \right\Vert_2^2  c_n^2},
\end{equation}
where $c_n$ is the normalized magnitude of the $n$th mode of the spherical harmonic coefficients of the surface function, and $\left\Vert \Delta \bmat{Y}_n^m \right\Vert_2$ is a term depending on the interpolation error of the $n$th mode of the spherical harmonic basis $Y_n$ on the POU grid with resolution $m$. 
For cubic spline interpolation, we find 
\begin{equation}
\begin{aligned}
\left\Vert  \Delta \bmat{Y}_n^m \right\Vert_2 & \approx m^{-4.3}n^{4.3}, \text{ for } 0\leq n \leq m, \\
\left\Vert \Delta \bmat{Y}_n^m \right\Vert_2 & > 1, \text{ for } n > m.
\end{aligned}
\end{equation}
Therefore, we truncate at $p_{\max} = m$ (when $m^{-4.3}n^{4.3} = 1$) to avoid amplifying the high-frequency components of the spherical harmonic spectrum. Since $\left\Vert  \Delta \bmat{Y}_n^m \right\Vert_2$ reflects the interpolation error of the $n$th mode of the spherical harmonic basis $Y_n$, this truncation can also be interpreted as removing modes that cannot be reliably captured on a POU grid with resolution $m$, i.e., modes whose interpolation error is $\mathcal{O}(1)$.

For the tolerance $\text{tol}_{\min}$, one can either enforce truncation at every step by setting $\text{tol}_{\min} = 0$, or use a theoretically motivated threshold to avoid unnecessary filtering when the high-frequency content is already negligible. Specifically, we set
\begin{equation}
\text{tol}_{\min} = \frac{1}{\sqrt{2k_0 - 1}}{p_{\max}}^{-\frac{2k_0 - 1}{2}},\ k_0 = 4.3,
\end{equation}
as justified by the analysis in \cref{appendix:parameter_filtering-parameter_selection}.

\subsection{Aggressive spectral filtering}
\label{sec:aggressive-spectral-filtering}

In cases where accurate surface derivatives are required, especially in the presence of high-curvature regions during capsule deformation, we apply a more aggressive truncation
\begin{equation}
p_{\max} = m^{0.68}.
\end{equation}
This value is derived by estimating the error of the second derivatives of the interpolated surface function (See \cref{appendix:parameter_filtering-aggressive_filtering} for details)
\begin{equation}
\epsilon_m^{\text{interp}, (2)} \leq \sqrt{\sum_{n = 0}^{\infty}  \left(m^{-4.3}n^{6.3}\right)^2  c_n^2 }.
\end{equation}
Truncating when $m^{-4.3}n^{6.3} = 1$ yields $p_{\max} = m^{0.68}$.
We focus on second-order derivatives because they are the highest-order derivatives in the surface force formulation used in our model.

\subsection{Summary of the method}
\label{sec:method-summary}

To advance the simulation from $t$ to $t+\Delta t$, the capsule positions are first upsampled to a refined partition-of-unity (POU) grid, where surface derivatives are computed using second-order finite differences and blended to obtain consistent values across overlapping patches. Next, the interfacial force is computed. The regularized Stokes layer potentials are then evaluated to find the fluid velocity at the capsule grid points. For pipe-confined flows, the boundary integral equation on the pipe wall is solved to enforce the no-slip condition. The explicit, adaptive RK3(2) time-integrator uses these velocities to update the capsule positions. Finally, to maintain stability, spectral filtering is applied to the updated surface geometry to mitigate aliasing.

\paragraph{Convergence rate.} 
The overall numerical scheme is second-order accurate in space and third-order accurate in time. The spatial error is dominated by the second-order finite differences used to compute the surface derivatives and the interfacial force on the capsule surface. The temporal discretization uses a third-order Runge--Kutta scheme. We dynamically couple the ODE time-stepping tolerance $\odetol$ to the spatial resolution $m$ (as described in \cref{eq:odetol-rule}).

\paragraph{Computational complexity.}
The computational cost per time step is governed by three main components: flow field evaluation, integral equation solving (for confined flows), and spectral filtering. For $N \sim \bigO(m^2)$ total grid points, we distinguish between the theoretical optimum and our current practical implementation:
\begin{itemize}
    \item \textit{Flow evaluation and pipe boundary solve:} Theoretically, evaluating the Stokes layer potentials with fast Ewald summation should yield a complexity of $\bigO(m^2 \log m)$, and solving the pipe boundary integral equation iteratively (e.g., via GMRES) scales as $\bigO(m^2 \log m)$. However, in our current implementation, we use exact LU decomposition (with back-substitution) for the integral equation, resulting in an actual working complexity of $\bigO(m^4)$ per time step.
    \item \textit{Filtering:} Spectral filtering relies on the discrete spherical harmonic transform (DSHT) up to degree $p=m$. Theoretically, by combining FFTs in longitude with fast Legendre transforms in latitude, this can be executed in $\bigO(m^2 (\log m)^2)$ operations. In our current implementation, however, we only utilize FFTs and omit the fast Legendre transform, leading to a practical complexity of $\bigO(m^3)$ per invocation. 
\end{itemize}
Consequently, the fully optimized theoretical complexity of the method is $\bigO(m^2 (\log m)^2)$ per time step, and the overall complexity of our current working implementation is $\bigO(m^4)$.

\section{Results}
\label{sec:results}

In this section, we assess the numerical effectiveness of our spectral filtering scheme through stability and convergence analyses, and we use the stabilized simulations to investigate the dynamic behaviors of a biconcave capsule in various flow settings. Specifically, we consider unconfined shear flow, relaxation flow following shear, and pressure-driven flow through a pipe. The first part of this section focuses on demonstrating the robustness and accuracy of the method, while the second part highlights selected simulation results and quantities of interest (QoIs) that characterize the capsule’s physical behavior in these flow settings.

Let us define some quantities of interest that will be used in this section. The \emph{Taylor asphericity} (in the $x_1x_2$-plane) of the capsule membrane is defined as
\begin{equation}
    D = \frac{\sqrt{\lambda_1} - \sqrt{\lambda_2}}{\sqrt{\lambda_1} + \sqrt{\lambda_2}},
\end{equation}
where $\lambda_1$ and $\lambda_2$ are the two eigenvalues of the $2 \times 2$-block of the moment of inertia tensor $\bmat{I}$ corresponding to the first two coordinates. In other words, $\lambda_1 = A+B$ and $\lambda_2=A-B$, where
\begin{align}
    A &= I_{11} + I_{22}, \\
    B &= \sqrt{(I_{11}-I_{22})^2 + 4 I_{12}^2}.
\end{align}
We define the moment of inertia tensor of the capsule as
\begin{equation}
    I_{ij} = \int_\Omega \left( \lVert \bvec{x} \rVert^2 \delta_{ij} - x_i x_j \right) \, \mathrm{d}V,
\end{equation}
where $\Omega$ is the volume enclosed by the capsule membrane.
The Taylor asphericity is zero for a sphere and one for a line segment.
It is, together with the instantaneous reduced volume $\rv$, defined in \cref{eq:reduced-volume}, a good indicator of how deformed the capsule shape is. As a function of time, it shows whether the capsule is in a stable, transient, or oscillating configuration.

The \emph{strain energy} \citep{Skalak1973} is defined as
\begin{equation}
    \strain = \int_\gamma \frac{1}{J(\bvec{x})} \left(
    \frac{\Es}{4} \left[
    \frac{1}{2}I_1(\bvec{x})^2 + I_1(\bvec{x}) - I_2(\bvec{x})
    \right]
    + \frac{\Ed}{8} I_2(\bvec{x})^2
    \right) \, \mathrm{d} S,
\end{equation}
in which the capsule membrane is also of interest, with all quantities as defined in connection to \cref{eq:stress-tensor}. This is the energy that the interfacial force tries to minimize (see \ref{appendix:strain_energy_functional-force}\textbf{} for proof). The strain energy has a minimum at the relaxed configuration (in fact, it is zero there, see \ref{appendix:strain_energy_functional-zero_condition}\textbf{} for details). It is therefore a measurement of how far the capsule is from the relaxed configuration.

We summarize the numerical experiments conducted in this section:
\begin{itemize}
    \item We evaluate the stability and accuracy of the spectral filtering scheme in both unconfined and confined flows (Tables \ref{tab:shear_stability}--\ref{tab:shear_vloss} and \cref{appendix:stable_conv_tbls}). To demonstrate accuracy, we report the convergence of long-time quantities of interest (QoIs) with increasing capsule resolution, alongside long-time volume conservation. 
    
    \item We investigate the dynamics and quantities of interest for a capsule in unconfined flows (\Cref{fig:shear_qois_selected,fig:relax_qois_selected}). For shear flow, we track the Taylor asphericity, inclination angle, and strain energy; for relaxation flow, we measure the strain energy and the Hausdorff error relative to the initial configuration.
    
    \item We explore the dynamics of a capsule in a standard Poiseuille flow through a circular pipe (Figures \ref{fig:circle-evolution-y0}--\ref{fig:circle-lateral-conclusion} and \Cref{tab:circle-arrival-time}). We report the transient dynamics, terminal shapes, Taylor asphericity, and strain energy. We characterize lateral migration and quantify its dependence on the capillary number.
    
    \item We examine the dynamics of capsules in pressure-driven flows through pipes with arbitrary, non-circular cross-sections (Figures \ref{fig:pipe-shapes}--\ref{fig:arbitrary-lateral-drift} and Tables \ref{tab:pipe-data}--\ref{tab:arbitrary-lateral-drift}). We report the transient behavior, terminal shapes, Taylor asphericity, and lateral drift.
\end{itemize}

\subsection{Stability and convergence}
\label{sec:results-stability}

We evaluate the stability and resolution convergence of the spectral filtering scheme in both unconfined (shear and relaxation) and confined (Poiseuille) flows. For each flow, we assess whether simulations reach the intended final time, whether the results converge with increasing capsule resolution, and whether volume is conserved. We compare the stability of our method against two baselines: an unfiltered version of our method, and the reparametrization approach of \cite{Veerapaneni2011} modified with the deformation gradient correction for capsules from \cite{Agarwal2022}.

For unconfined shear flow, Tables \ref{tab:shear_stability}--\ref{tab:shear_vloss} summarize the results. \Cref{tab:shear_stability} reports the maximum time the simulation remains stable under different methods, with varying reduced volume $\rvr$, capillary number $\Ca$, and capsule resolution $m$. \Cref{tab:shear_conv_short} shows relative position errors at short-time ($T=1$), while \Cref{tab:shear_conv_long} reports relative errors in position (Pos), Taylor asphericity (Asp), and inclination angle (Inc) at long-time ($T=50$). \Cref{tab:shear_vloss} shows relative volume loss at $T = 50$. The relative errors of the three QoIs are defined as
\begin{equation}
\begin{aligned}
\epsilon^\text{Pos}(t) & = \frac{\max_i\Vert \boldsymbol{x}_i(t) - \boldsymbol{x}_i^*(t)\Vert_2}{\max_i\Vert \boldsymbol{x}_i^*(t)\Vert_2}, \\
\epsilon^\text{Asp}(t) & = \Vert D(t) - D^*(t)\Vert_2, \\
\epsilon^\text{Inc}(t) & = \frac{1}{\pi}\Vert \theta(t) - \theta^*(t)\Vert_2,
\end{aligned}
\end{equation}
where $\boldsymbol{x}_i$ represents the $i_\mathrm{th}$ points in the discretization mesh. $\left\{\boldsymbol{x}_i^*\right\}_i$, $D^*$, and $\theta^*$ represents the reference values, which are obtained by evolving the system with capsule resolution $m = 96$.

\begin{table}[tbp]
\centering
\setlength{\aboverulesep}{1pt}
\setlength{\belowrulesep}{1pt}
\setlength{\tabcolsep}{4pt}
\small
\begin{tabular}{rrrrrrrrrrrrrr}
\toprule
\multirow{2}{*}{$\rvr$} & \multirow{2}{*}{method} & \multicolumn{4}{c}{$m = 24$} & \multicolumn{4}{c}{$m = 48$} & \multicolumn{4}{c}{$m = 96$} \\
\cmidrule(lr){3-6} \cmidrule(lr){7-10} \cmidrule(lr){11-14}
 & & $\Ca = 0.125$ & $0.25$ & $0.5$ & $1$ & $\Ca = 0.125$ & $0.25$ & $0.5$ & $1$ & $\Ca = 0.125$ & $0.25$ & $0.5$ & $1$ \\
\midrule
$1$&none&50&50&9&1&50&50&8&1&50&50&3&1\\
    &filter&50&50&50&50&50&50&50&50&50&50&50&50\\
    &reparam&50&50&9&1&50&50&8&1&50&50&2&1\\
\midrule
$0.89$&none&50&15&2&1&30&3&1&1&27&2&1&0\\
    &filter&50&50&50&50&50&50&50&50&50&50&50&50\\
    &reparam&50&16&2&1&29&3&1&1&7&2&1&0\\
\midrule
$0.78$&none&33&15&2&1&28&3&1&1&6&2&1&0\\
    &filter&50&50&50&50&50&50&50&50&50&50&50&50\\
    &reparam&31&5&2&1&8&3&1&1&5&2&1&0\\
\midrule
$0.64$&none&35&15&2&1&31&4&2&1&28&3&1&0\\
    &filter&50&50&50&50&50&50&50&50&50&50&50&50\\
    &reparam&33&7&2&1&10&4&1&0&5&2&1&0\\
\bottomrule
\end{tabular}
\caption{Maximum time the simulation remains stable (up to final time $T = 50$) for single biconcave capsule in shear flow with varying reduced volume $\rvr$, capillary number $\Ca$, capsule resolution $m$, and simulation method. Three methods are compared: unfiltered (none), spectral filtering (filter), and reparametrization (reparam). A value of 50 indicates full stability throughout the entire simulation window.}\label{tab:shear_stability}
\end{table}

\begin{table}[tbp]
\centering
\setlength{\aboverulesep}{1pt}
\setlength{\belowrulesep}{1pt}
\setlength{\tabcolsep}{3pt}
\small
\begin{tabular}{c>{\columncolor[gray]{0.9}}c>{\columncolor[gray]{0.9}}ccc>{\columncolor[gray]{0.9}}c>{\columncolor[gray]{0.9}}ccc}
\toprule
\multirow{2}{*}{$\rvr$} & \multicolumn{2}{c}{$\Ca = 0.125$} & \multicolumn{2}{c}{$\Ca = 0.25$} & \multicolumn{2}{c}{$\Ca = 0.5$} & \multicolumn{2}{c}{$\Ca = 1$} \\
\cmidrule(lr){2-3} \cmidrule(lr){4-5} \cmidrule(lr){6-7} \cmidrule(lr){8-9}
 & $m = 24$ & $m = 48$ & $m = 24$ & $m = 48$ & $m = 24$ & $m = 48$ & $m = 24$ & $m = 48$ \\
\midrule
$1$&$5.5\times 10^{-5}$&$2.4\times 10^{-6}$&$7.7\times 10^{-5}$&$4.6\times 10^{-6}$&$6.8\times 10^{-5}$&$7.3\times 10^{-6}$&$1.1\times 10^{-4}$&$1.9\times 10^{-5}$\\
\midrule
$0.89$&$5.7\times 10^{-5}$&$8.5\times 10^{-6}$&$6.4\times 10^{-5}$&$1.0\times 10^{-5}$&$8.7\times 10^{-5}$&$1.7\times 10^{-5}$&$2.4\times 10^{-4}$&$2.8\times 10^{-5}$\\
\midrule
$0.78$&$1.0\times 10^{-4}$&$1.5\times 10^{-5}$&$1.2\times 10^{-4}$&$1.6\times 10^{-5}$&$1.1\times 10^{-4}$&$1.7\times 10^{-5}$&$4.0\times 10^{-4}$&$4.1\times 10^{-5}$\\
\midrule
$0.64$&$1.8\times 10^{-4}$&$2.5\times 10^{-5}$&$3.2\times 10^{-4}$&$2.8\times 10^{-5}$&$6.5\times 10^{-4}$&$2.9\times 10^{-5}$&$1.1\times 10^{-3}$&$9.0\times 10^{-5}$\\
\bottomrule
\end{tabular}
\caption{Relative position error at $T = 1$ for single biconcave capsule in shear flow with spectral filtering, for different reduced volumes $\rvr$ and capillary numbers $\Ca$. The reference solution is computed with $m = 96$.}\label{tab:shear_conv_short}
\end{table}

\begin{table}[tbp]
\centering
\setlength{\aboverulesep}{1pt}
\setlength{\belowrulesep}{1pt}
\setlength{\tabcolsep}{2.5pt}
\small
\begin{tabular}{cc>{\columncolor[gray]{0.9}}c>{\columncolor[gray]{0.9}}ccc>{\columncolor[gray]{0.9}}c>{\columncolor[gray]{0.9}}ccc}
\toprule
\multirow{2}{*}{$\rvr$} & \multirow{2}{*}{QoI} & \multicolumn{2}{c}{$\Ca = 0.125$} & \multicolumn{2}{c}{$\Ca = 0.25$} & \multicolumn{2}{c}{$\Ca = 0.5$} & \multicolumn{2}{c}{$\Ca = 1$} \\
\cmidrule(lr){3-4} \cmidrule(lr){5-6} \cmidrule(lr){7-8} \cmidrule(lr){9-10}
&& $m = 24$ & $m = 48$ & $m = 24$ & $m = 48$ & $m = 24$ & $m = 48$ & $m = 24$ & $m = 48$ \\
\midrule
$1$&Pos&$1.3\times 10^{-3}$&$5.4\times 10^{-4}$&$3.3\times 10^{-3}$&$1.1\times 10^{-3}$&$1.0\times 10^{-2}$&$2.0\times 10^{-3}$&$7.3\times 10^{-2}$&$8.0\times 10^{-3}$\\
    &Asp&$2.8\times 10^{-4}$&$1.3\times 10^{-4}$&$3.5\times 10^{-4}$&$1.2\times 10^{-4}$&$6.2\times 10^{-4}$&$1.5\times 10^{-4}$&$3.4\times 10^{-3}$&$3.5\times 10^{-4}$\\
    &Inc&$1.7\times 10^{-4}$&$7.8\times 10^{-5}$&$1.9\times 10^{-4}$&$6.9\times 10^{-5}$&$2.8\times 10^{-4}$&$6.4\times 10^{-5}$&$1.3\times 10^{-3}$&$1.2\times 10^{-4}$\\
\midrule
$0.89$&Pos&$2.1\times 10^{-2}$&$3.4\times 10^{-2}$&$7.2\times 10^{-3}$&$5.5\times 10^{-4}$&$3.1\times 10^{-2}$&$3.3\times 10^{-3}$&$8.8\times 10^{-2}$&$9.9\times 10^{-3}$\\
    &Asp&$3.7\times 10^{-4}$&$9.1\times 10^{-5}$&$9.9\times 10^{-4}$&$6.9\times 10^{-5}$&$1.8\times 10^{-3}$&$2.0\times 10^{-4}$&$8.8\times 10^{-4}$&$9.3\times 10^{-5}$\\
    &Inc&$3.4\times 10^{-3}$&$2.2\times 10^{-3}$&$4.8\times 10^{-4}$&$2.8\times 10^{-5}$&$1.4\times 10^{-3}$&$1.4\times 10^{-4}$&$1.3\times 10^{-3}$&$1.3\times 10^{-4}$\\
\midrule
$0.78$&Pos&$3.3\times 10^{-2}$&$4.7\times 10^{-2}$&$2.8\times 10^{-2}$&$1.5\times 10^{-2}$&$5.3\times 10^{-2}$&$5.8\times 10^{-3}$&$8.8\times 10^{-1}$&$4.8\times 10^{-2}$\\
    &Asp&$1.4\times 10^{-3}$&$1.1\times 10^{-3}$&$3.1\times 10^{-4}$&$2.7\times 10^{-5}$&$1.8\times 10^{-3}$&$2.5\times 10^{-4}$&$2.7\times 10^{-2}$&$4.2\times 10^{-4}$\\
    &Inc&$1.1\times 10^{-2}$&$4.8\times 10^{-3}$&$1.6\times 10^{-3}$&$3.4\times 10^{-4}$&$1.4\times 10^{-3}$&$1.3\times 10^{-4}$&$4.3\times 10^{-3}$&$2.6\times 10^{-4}$\\
\midrule
$0.64$&Pos&$8.2\times 10^{-2}$&$5.7\times 10^{-2}$&$2.8\times 10^{-1}$&$1.3\times 10^{-1}$&$2.0\times 10^{-1}$&$2.0\times 10^{-2}$&$1.4\times 10^{0}$&$1.4\times 10^{-1}$\\
    &Asp&$7.0\times 10^{-3}$&$3.0\times 10^{-3}$&$2.7\times 10^{-3}$&$2.7\times 10^{-4}$&$1.3\times 10^{-2}$&$1.1\times 10^{-3}$&$1.0\times 10^{-1}$&$6.4\times 10^{-3}$\\
    &Inc&$2.8\times 10^{-2}$&$1.0\times 10^{-2}$&$1.8\times 10^{-2}$&$4.1\times 10^{-3}$&$1.1\times 10^{-2}$&$9.7\times 10^{-4}$&$4.1\times 10^{-2}$&$3.2\times 10^{-3}$\\
\bottomrule
\end{tabular}
\caption{Relative errors at final time $T = 50$ for single biconcave capsule in shear flow with spectral filtering. Relative errors are reported for position (Pos), Taylor asphericity (Asp), and inclination angle (Inc), for different reduced volumes $\rvr$ and capillary numbers $\Ca$. The reference solution is computed with $m = 96$.}\label{tab:shear_conv_long}
\end{table}

\begin{table}[tbp]
\centering
\setlength{\aboverulesep}{1pt}
\setlength{\belowrulesep}{1pt}
\setlength{\tabcolsep}{4pt}
\small
\begin{tabular}{c>{\columncolor[gray]{0.9}}c>{\columncolor[gray]{0.9}}c>{\columncolor[gray]{0.9}}cccc}
\toprule
\multirow{2}{*}{$\rvr$} & \multicolumn{3}{c}{$\Ca = 0.125$} & \multicolumn{3}{c}{$\Ca = 0.25$} \\
\cmidrule(lr){2-4} \cmidrule(lr){5-7} 
 & $m = 24$ & $m = 48$ & $m = 96$ & $m = 24$ & $m = 48$ & $m = 96$ \\
\midrule
$1$&$4.6\times 10^{-4}$&$2.6\times 10^{-4}$&$7.8\times 10^{-6}$&$6.9\times 10^{-4}$&$2.7\times 10^{-4}$&$1.1\times 10^{-5}$\\
\midrule
$0.89$&$1.5\times 10^{-3}$&$4.2\times 10^{-4}$&$2.8\times 10^{-4}$&$3.0\times 10^{-3}$&$1.4\times 10^{-4}$&$2.1\times 10^{-5}$\\
\midrule
$0.78$&$8.4\times 10^{-3}$&$1.6\times 10^{-3}$&$6.5\times 10^{-4}$&$1.4\times 10^{-2}$&$3.0\times 10^{-3}$&$1.3\times 10^{-3}$\\
\midrule
$0.64$&$1.3\times 10^{-2}$&$1.8\times 10^{-3}$&$3.4\times 10^{-4}$&$6.9\times 10^{-2}$&$1.5\times 10^{-2}$&$1.1\times 10^{-3}$\\
\bottomrule
\end{tabular}
\begin{tabular}{c>{\columncolor[gray]{0.9}}c>{\columncolor[gray]{0.9}}c>{\columncolor[gray]{0.9}}cccc}
\toprule
\multirow{2}{*}{$\rvr$}  & \multicolumn{3}{c}{$\Ca = 0.5$} & \multicolumn{3}{c}{$\Ca = 1$} \\
\cmidrule(lr){2-4} \cmidrule(lr){5-7} 
 & $m = 24$ & $m = 48$ & $m = 96$ & $m = 24$ & $m = 48$ & $m = 96$ \\
\midrule
$1$&$1.5\times 10^{-3}$&$3.7\times 10^{-4}$&$2.2\times 10^{-5}$&$1.0\times 10^{-2}$&$1.0\times 10^{-3}$&$7.2\times 10^{-5}$\\
\midrule
$0.89$&$7.2\times 10^{-3}$&$7.4\times 10^{-4}$&$5.9\times 10^{-5}$&$1.3\times 10^{-2}$&$1.5\times 10^{-3}$&$2.0\times 10^{-4}$\\
\midrule
$0.78$&$1.8\times 10^{-2}$&$1.8\times 10^{-3}$&$9.6\times 10^{-5}$&$1.6\times 10^{-1}$&$8.2\times 10^{-3}$&$8.4\times 10^{-4}$\\
\midrule
$0.64$&$1.0\times 10^{-1}$&$8.9\times 10^{-3}$&$9.5\times 10^{-4}$&$4.3\times 10^{-1}$&$3.2\times 10^{-2}$&$2.6\times 10^{-3}$\\
\bottomrule
\end{tabular}
\caption{Relative volume loss at final time $T = 50$ for single biconcave capsule in shear flow with spectral filtering, for different reduced volumes $\rvr$ and capillary numbers $\Ca$.}\label{tab:shear_vloss}
\end{table}

A simulation is considered stable if it reaches the final time $T = 50$ without a nonsmooth increase in the gradient norm. Spectral filtering is the only method that achieves stability across all tested combinations of $\rvr$ and $\Ca$. At short time ($T = 1$), fourth-order convergence is observed in position error for the spherical case ($\rvr=1$) at $\Ca=0.125$ and $0.25$, and error decay is seen in all other cases. At long time ($T = 50$), position error is the most stringent metric but remains below $15\%$ at $m=48$ for all cases. Taylor asphericity and inclination angle also converge. Volume loss decreases with increasing $m$, and the worst-case value at $m = 96$ is $0.26\%$, further demonstrating the reliability of our scheme. 

Interestingly, for $\rvr < 1$, long-time errors do not necessarily increase with $\Ca$. As shown in \cref{fig:shear_qois_selected}, all such cases exhibit sustained shape oscillations, but lower $\Ca$ leads to longer periods, and the capsule may not complete a full cycle by $T = 50$. As a result, the errors at that time are more likely to reflect transient dynamics, which can lead to larger errors than in cases with higher $\Ca$.

For unconfined relaxation flow, the biconcave capsule evolves under shear flow until $T = 1$ and then relaxes in quiescent fluid over $T \in [1, 51]$. Stability and convergence results are shown in \cref{appendix:stable_conv_tbls} (Tables \ref{tab:relax_stability}--\ref{tab:relax_vloss}). All filtered simulations remain stable. All QoIs converge as resolution increases, and the volume is conserved. 

For Poiseuille flow through a pipe, results are presented in \cref{appendix:stable_conv_tbls} (Tables \ref{tab:pois_stability}--\ref{tab:pois_vloss}). This case is more difficult to stabilize because of the high-curvature regions. To maintain stability, we employ aggressive spectral filtering with a truncation degree of $p_{\max} = m^{0.68}$. This approach successfully stabilizes all cases at sufficiently high resolutions, including those with small $\rvr$. 
Convergence studies show decay in both position and asphericity errors as resolution increases. At $T = 100$, the worst-case position error is $11\%$ at $m = 48$, and the maximum volume loss is $2\%$ at $m = 96$.
To further assess the accuracy, we compute the residual $\Vert\dot{\boldsymbol{x}} - \boldsymbol{u}(\boldsymbol{x})\Vert$ on an upsampled mesh (\cref{tab:pois_res}). Even with aggressive filtering, the residuals at $m=48$ remain on the order of $10^{-5}$ or smaller across all cases. 
These results demonstrate that, with suitable truncation, the spectral filtering scheme enables robust and accurate simulations even in demanding confined geometries.

To further characterize the behavior of the spectral filtering scheme, we monitor its activation frequency and its interaction with the time-integrator. In shear flow simulations, the filtering is typically invoked during $70\%$ to $100\%$ of the Runge--Kutta (RK) time steps. The notable exceptions are the spherical capsules ($\rvr = 1$) and certain low-to-moderate capillary number cases for $\rvr = 0.89$, where the filtering is rarely triggered. For the Poiseuille flow simulations, we set the filtering tolerance to zero, meaning the filtering is unconditionally applied at every step. Importantly, comparing the filtered simulations against stable unfiltered baselines reveals that the filtering step does not degrade the RK acceptance rate. The adaptive solver rejects no additional steps when filtering is active and, in some instances, accepts slightly more steps, demonstrating that the spectral filtering successfully stabilizes the system without disrupting the temporal continuity of the solver.

Furthermore, we quantify the perturbation introduced by the filtering by measuring the relative magnitude of the truncated high-frequency tail and the resulting change in strain energy. Across all shear flow simulations, the relative magnitude of the truncated tail oscillates around or remains below $10^{-5}$, $10^{-6}$, and $10^{-7}$ for resolutions $m = 24$, $48$, and $96$, respectively. The relative difference in strain energy before and after filtering follows a similar trend, bounded by $10^{-3}$, $10^{-4}$, and $10^{-5}$ for the same respective resolutions. In the Poiseuille flow cases, the aggressive filtering strategy naturally results in larger perturbations: the truncated tail magnitudes are bounded by approximately $10^{-2}$, $10^{-3}$, and $10^{-4}$ for $m = 24$, $48$, and $96$, while the relative strain energy differences are bounded by $10^{-1}$, $10^{-2}$, and $10^{-3}$. For both flow types, the numerical interventions introduced by the filter decay rapidly with increasing resolution.

\subsection{Simulation results and quantities of interest}

We now examine the physical behavior of the biconcave capsule in each flow setting, using selected QoIs to illustrate the dynamics. 

\subsubsection{Unconfined shear flow and relaxation}
\label{sec:results-unconfined}

\begin{figure}[tbp]
    \centering
    \begin{subfigure}{\textwidth}
        \centering
        \includegraphics[width=\textwidth]{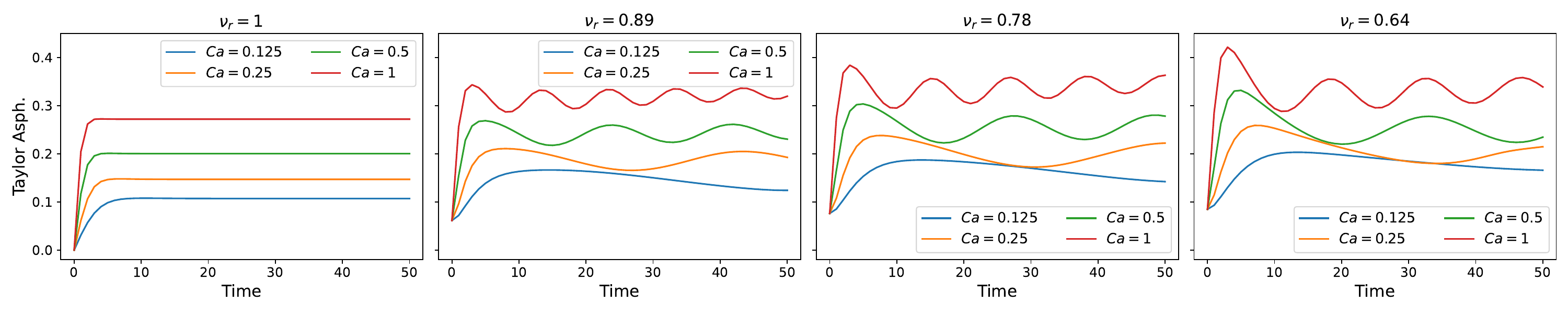}
        \caption{Taylor asphericity}
    \end{subfigure}
    \begin{subfigure}{\textwidth}
        \centering
        \includegraphics[width=\textwidth]{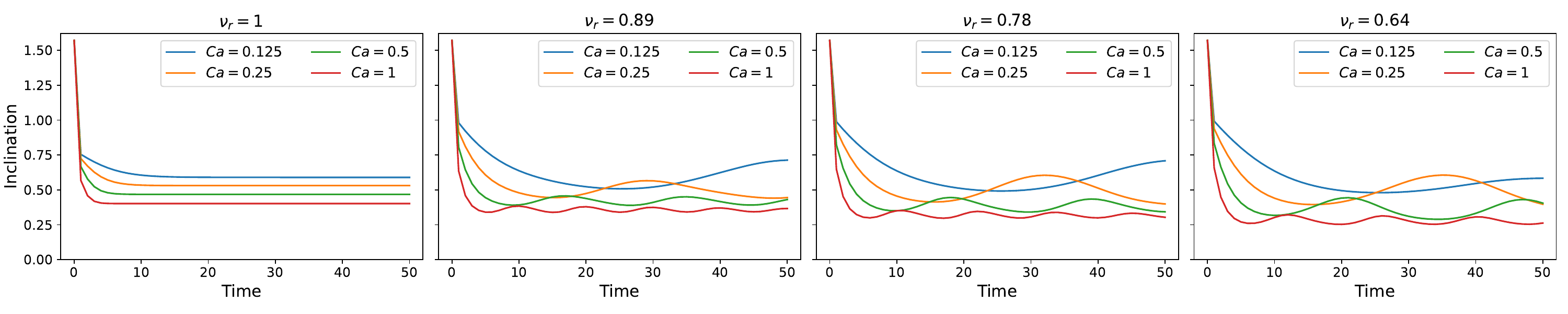}
        \caption{Inclination angle}
    \end{subfigure}
    \begin{subfigure}{\textwidth}
        \centering
        \includegraphics[width=\textwidth]{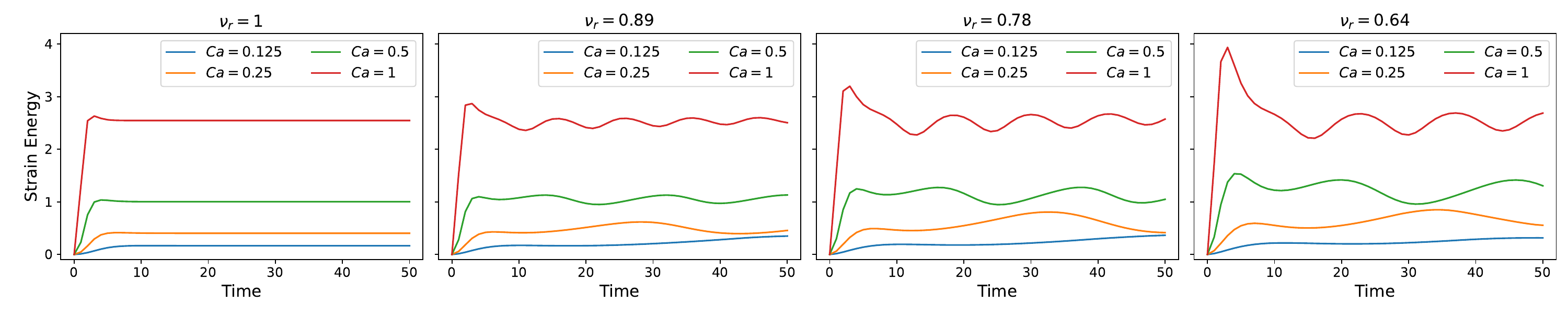}
        \caption{Strain energy}
    \end{subfigure}
    \caption{Selected quantities of interest (QoIs) for single biconcave capsule in shear flow under different reduced volumes $\rvr$ and capillary numbers $\Ca$. Each row corresponds to a distinct QoI.}
    \label{fig:shear_qois_selected}
\end{figure}

Unless otherwise stated, the simulations in this section use spectral filtering with capsule resolution $m = 96$.
In unconfined shear flow, \cref{fig:shear_qois_selected} shows the evolution of Taylor asphericity, inclination angle, and strain energy over time for different $\rvr$ and $\Ca$. For all $\rvr < 1$, the capsule undergoes sustained shape oscillations and does not reach a steady state within the simulation window. Increasing $\Ca$ shortens the oscillation period and increases peak deformation, while decreasing $\rvr$ lengthens the oscillation period. These trends highlight the capsule’s sensitivity to both membrane properties and flow strength.

\begin{figure}[tbp]
    \centering
    \begin{subfigure}{\textwidth}
        \centering        \includegraphics[width=\textwidth]{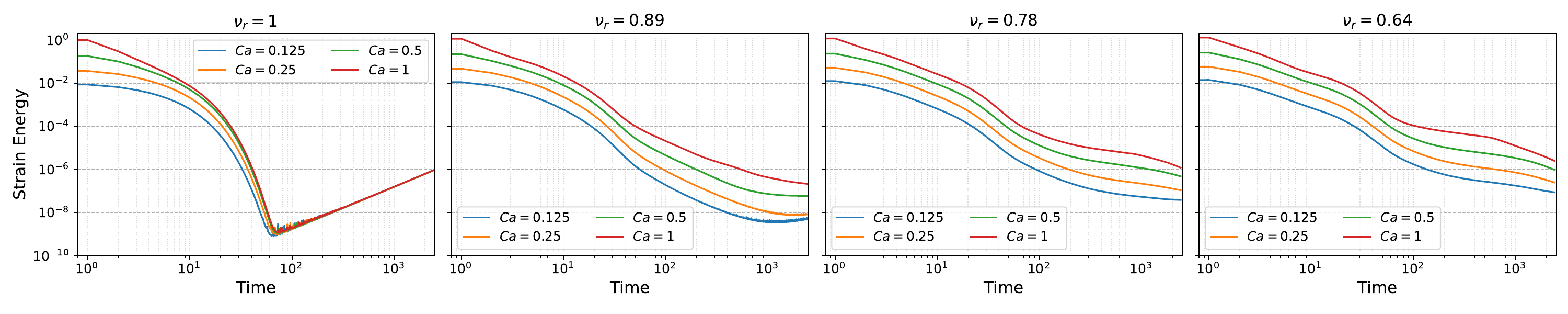}
        \caption{Strain energy}
    \end{subfigure}
    \begin{subfigure}{\textwidth}
        \centering
        \includegraphics[width=\textwidth]{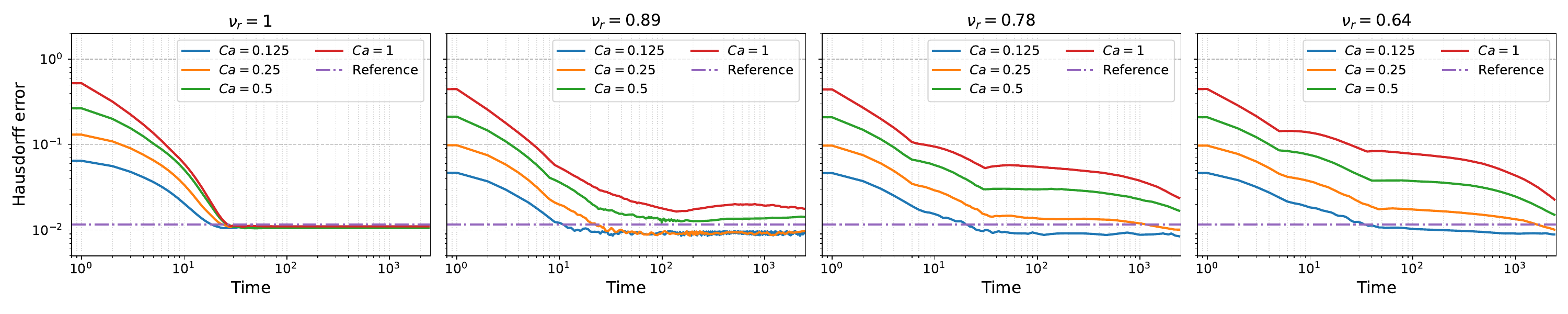}
        \caption{Hausdorff error}
    \end{subfigure}
    \caption{Selected quantities of interest (QoIs) for single biconcave capsule in relaxation flow under different reduced volumes $\rvr$ and capillary numbers $\Ca$. Each row corresponds to a distinct QoI.}
    \label{fig:relax_qois_selected}
\end{figure}

In unconfined relaxation flow, the capsule is first deformed in shear flow for $T = 1$ and then allowed to relax. \Cref{fig:relax_qois_selected} shows the evolution of strain energy and the Hausdorff error with respect to the stress-free reference configuration (the initial configuration). To quantify shape recovery, we define the normalized Hausdorff error after translational and rotational alignment
\begin{equation}
\epsilon^\text{Haus}(t) = \frac{\max_i\min_j\Vert \boldsymbol{x}_i(t) - \boldsymbol{x}_j^\text{ref} \Vert_2}{\max_i\Vert \boldsymbol{x}_i^\text{ref}\Vert_2},
\end{equation}
where $\left\{\boldsymbol{x}_i^\text{ref}\right\}_i$ represents the stress-free reference configuration (aligned with $\left\{\boldsymbol{x}_i(t)\right\}_i$ by translation and rotation). The resolution of the reference configuration is $m^\text{ref} = 192$. A reference tolerance $\epsilon^*$ is computed by
\begin{equation}
\epsilon^* = \frac{\max_i\min_j\Vert \boldsymbol{x}_i^{\mathrm{ref}, \text{up}} - \boldsymbol{x}_j^\mathrm{ref}\Vert_2}{\max_i\Vert \boldsymbol{x}_i^\mathrm{ref}\Vert_2} \approx \frac{\sqrt{2}}{2} \frac{\pi}{m^{\text{ref}}},
\end{equation}
where $\boldsymbol{x}_i^{\mathrm{ref}, \text{up}}$ is the reference configuration at resolution $2m^\text{ref}$. When $\epsilon^{\text{Haus}}(t) < \epsilon^*$, the shape is considered recovered.

During relaxation, the strain energy decreases monotonically at first and then either decays more slowly or shows a slight rebound. This behavior is consistent with the model structure: the interfacial force acts to minimize the strain energy (see \cref{appendix:strain_energy_functional-force} for proof). 
However, for small reduced volume $\rvr$ and large capillary number $\Ca$, the Hausdorff error decays much more slowly: after an initial period of relatively rapid relaxation, it remains nearly stagnant over long times before decreasing again very gradually, or in some cases, appears to stagnate altogether, remaining above $\epsilon^*$. This indicates that the capsule does not fully return to its reference shape. It remains an open question whether the shape would eventually reach the reference configuration without prohibitively long simulation times.
This outcome can be explained in two ways:
the strain energy becomes trapped at a local minimum, reflecting the fact that the functional is not convex (see \cref{appendix:strain_energy_functional-non_convexity} for the non-convexity proof); or
the strain energy continues to converge toward zero, but the system settles into a configuration that is not the reference shape, yet still has zero strain energy and cannot be related to the reference by a rigid transform. Such configurations do exist (see \cref{appendix:strain_energy_functional-non_convexity} for details and further discussion). Note that this behavior is unlikely to be an artifact of the filtering process. We run unfiltered relaxation simulations up to $T = 50$ (within which stagnation is already evident); all stable cases behave identically to the corresponding filtered simulations.

\subsubsection{Pressure-driven flow through a pipe}
\label{sec:results-pipe-flow}

We now consider a spherical capsule ($\rvr=1$) confined in a pipe. First, we consider a pipe with circular cross-section, and then, a pipe with arbitrary constant cross-section. The goal of this section is to show that our method is stable over a long time and can be used to study relevant aspects of the capsule's long-term dynamics, such as lateral drift. Throughout this section, the relaxed reference configuration of the capsule is a unit sphere, and this is also the initial configuration. The background flow through the pipe is induced by a constant pressure gradient as described in \cref{app:pipe-background-flow}.
The simulations in this section use capsule resolution $m=48$ unless otherwise stated. The periodic length is set to $\Lper=5$ (cf.\ \cref{sec:formulation-pipe}).

\begin{figure}[p]
  \centering
  Pipe cross-sections and background flows, with initial capsule outlines\\
  {\footnotesize%
  \begin{minipage}[t]{0.5\textwidth}\vskip0pt\centering
    (a) Case~A\hspace*{18pt}~\\
    \includegraphics[width=\textwidth,trim=0pt 0pt 0pt 28pt,clip]{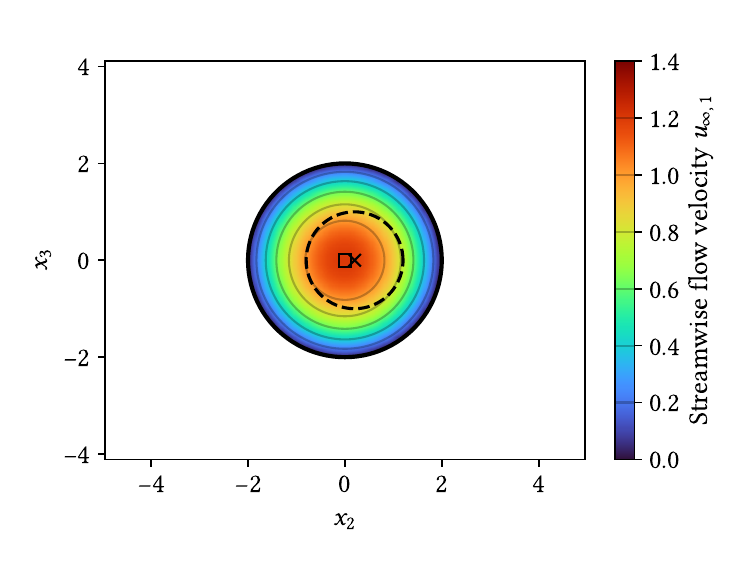}%
  \end{minipage}%
  \begin{minipage}[t]{0.5\textwidth}\vskip0pt\centering
    (b) Case~B\hspace*{18pt}~\\
    \includegraphics[width=\textwidth,trim=0pt 0pt 0pt 28pt,clip]{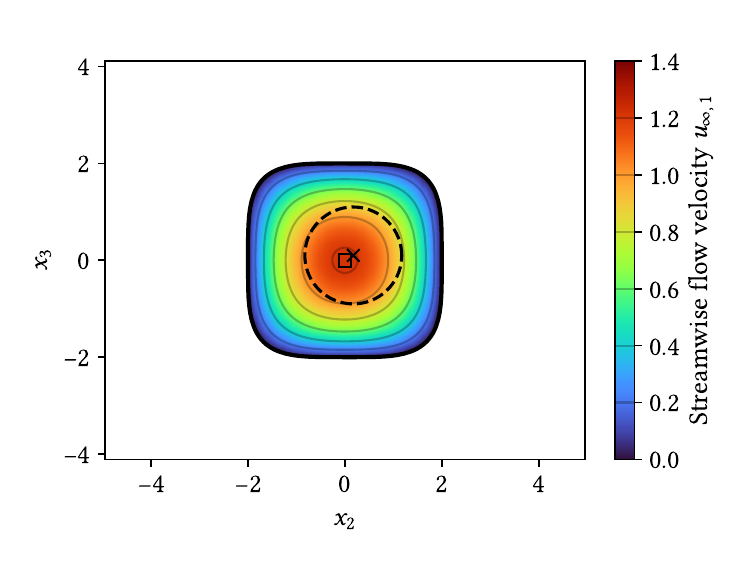}%
  \end{minipage}\\
  \begin{minipage}[t]{0.5\textwidth}\vskip0pt\centering
    (c) Case~C\hspace*{18pt}~\\
    \includegraphics[width=\textwidth,trim=0pt 0pt 0pt 28pt,clip]{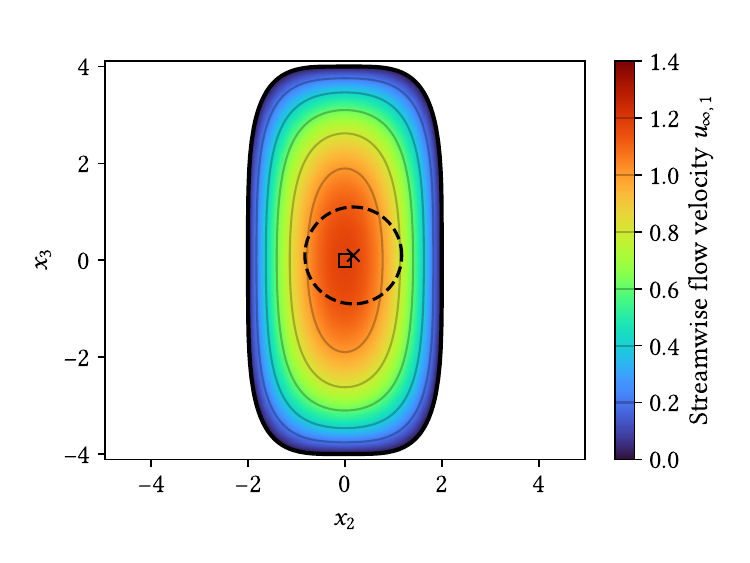}%
  \end{minipage}%
  \begin{minipage}[t]{0.5\textwidth}\vskip0pt\centering
    (d) Case~D\hspace*{18pt}~\\
    \includegraphics[width=\textwidth,trim=0pt 0pt 0pt 28pt,clip]{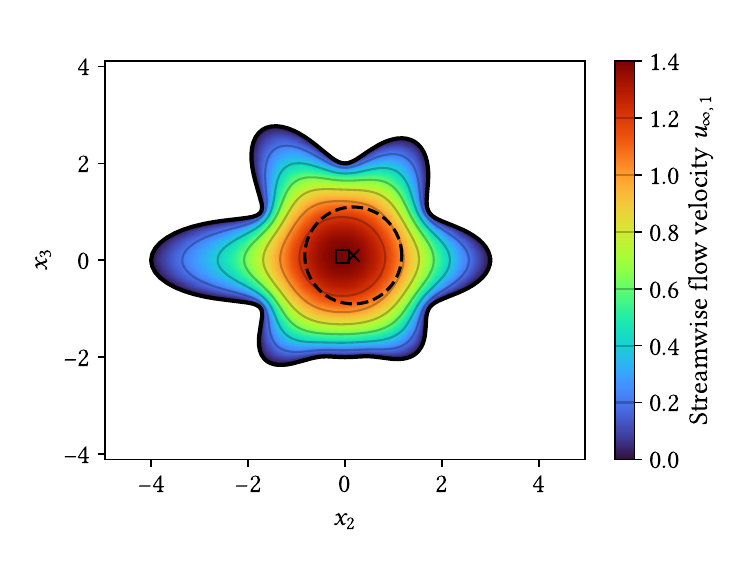}%
  \end{minipage}\\
  \begin{minipage}[t]{0.5\textwidth}\vskip0pt\centering
    (e) Case~E\hspace*{18pt}~\\
    \includegraphics[width=\textwidth,trim=0pt 0pt 0pt 28pt,clip]{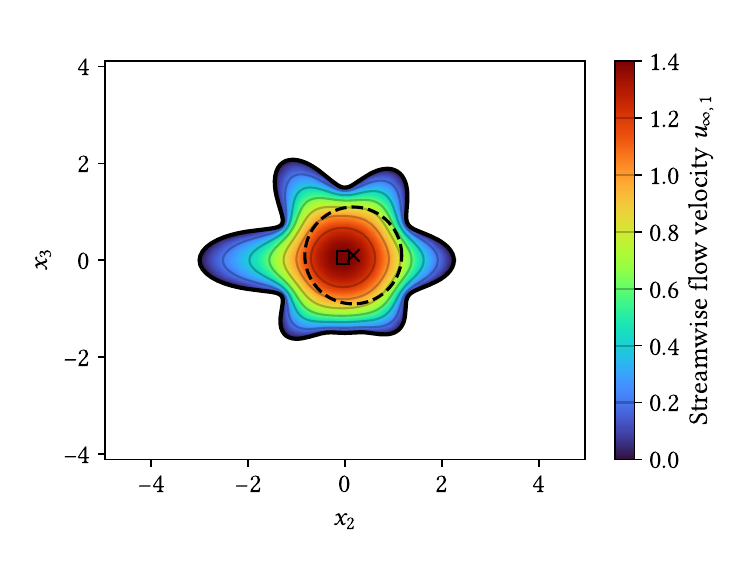}%
  \end{minipage}%
  }
  \caption{Pipe cross-sections and background
  flows $u_{\infty,1}$ for average capillary number $\Ca = 0.6$. The location of the maximum background flow is marked with a square
  ($\Box$). The outline of the initial capsule shape at $t=0$ is
  shown as a dashed curve, and the capsule center is marked by a
  cross ($\times$).}
  \label{fig:pipe-shapes}
\end{figure}

\paragraph{Circular cross-section.}
Let the pipe be a cylinder with radius 2, so that the confinement ratio $\Con$ (cf.\ \cref{eq:confinement-ratio}) is $1/2$. For this pipe shape, the pressure-induced flow is known analytically, and is given by \cref{eq:background-flow-Poiseuille} with $b=2$ and $a=2 \,\Ca/b^2$, where the capillary number $\Ca:=\Ca_1^\text{Pois}$ is given by \cref{eq:capillary-number-pois-1} and is the dimensionless mean flow velocity over the pipe cross-section. Here, we consider capillary numbers ranging from $\Ca=0.15$ to $\Ca=0.6$. The pipe and its flow are shown in \cref{fig:pipe-shapes}~(a) for $\Ca=0.6$. Note that the coordinate system $(x_1, x_2, x_3)$ is selected such that the $x_1$-axis is the centerline of the pipe cylinder. The pipe resolution is $\frac{2}{3}m$ in the periodic (axial) direction and $\frac{4}{3}m$ in the circumferential direction ($m$ being the capsule resolution).

Let the center of the capsule be defined by
\begin{equation}
    \label{eq:center-of-capsule}
    \centerv = \frac{\int_\gamma \bvec{x} \, \mathrm{d}S}{\int_\gamma \mathrm{d}S},
\end{equation}
and denote its time-derivative by $\dot{\centerv}$. The initial location of the capsule center is
\begin{equation}
    \centervinit = (0, \centercyinit, 0),
\end{equation}
where $|\centercyinit| < b-1$. Here, we consider $\centercyinit = 0, 0.05, 0.2$. The simulations for this pipe do not use spectral filtering since it is not needed for stability.

\begin{figure}[tbp]
  \centering
  Capsule shape evolution for $\centercyinit=0$\\
  {\footnotesize%
  \begin{minipage}[t]{0.5\textwidth}\vskip0pt\centering
    \hspace*{30pt}(a) $\Ca = 0.6$\\
    \includegraphics[width=\textwidth]{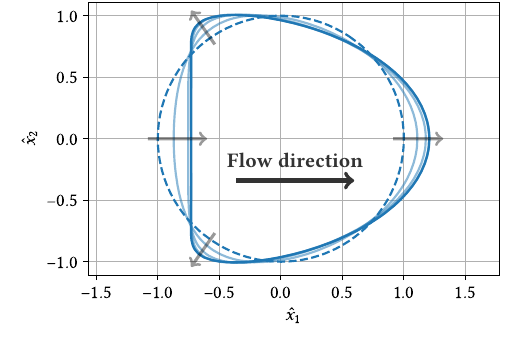}%
  \end{minipage}%
  \begin{minipage}[t]{0.5\textwidth}\vskip0pt\centering
    \hspace*{30pt}(b) $\Ca = 0.15$\\
    \includegraphics[width=\textwidth]{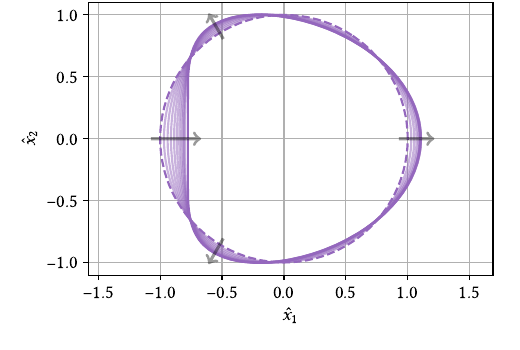}%
  \end{minipage}%
  }
  \caption{Capsule outline over time for $\centercyinit=0$. (a) $\Ca=0.6$, shown
  at times $t=0, 1, 2, \ldots, 5$. (b) $\Ca=0.15$, shown at times
  $t=0, 1, 2, \ldots, 10$ and $20$. The initial contour is dashed, the final
  is solid, and intermediate contours are semitransparent. The arrows indicate
  directions of motion. Note that the origin of the $(\hat{x}_1,\hat{x}_2)$ coordinate system follows
  the capsule center over time, $\hat{x}_i = x_i - c_i$.}
  \label{fig:circle-evolution-y0}
\end{figure}

\begin{figure}[tbp]
  \centering
  Capsule shape evolution for $\centercyinit=0.2$, $\Ca=0.6$\\
  {\footnotesize%
  \begin{minipage}[t]{0.5\textwidth}\vskip0pt\centering
    \hspace*{30pt}(a) Times $0 \leq t \leq 5$\\
    \includegraphics[width=\textwidth]{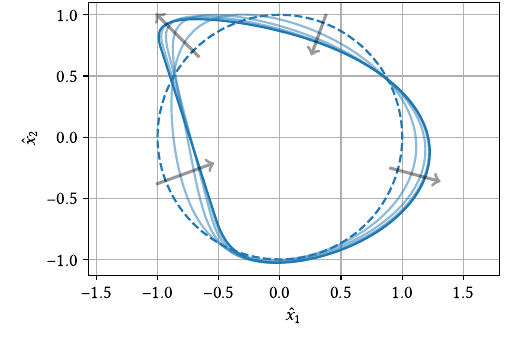}%
  \end{minipage}%
  \begin{minipage}[t]{0.5\textwidth}\vskip0pt\centering
    \hspace*{30pt}(b) Times $5 \leq t \leq 100$\\
    \includegraphics[width=\textwidth]{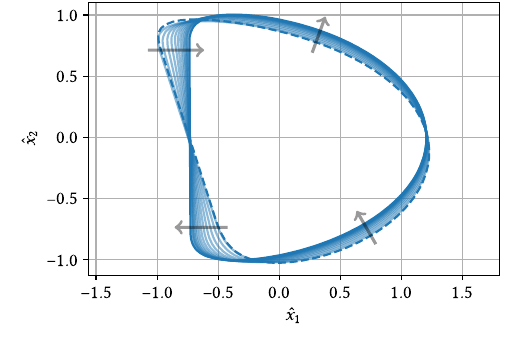}%
  \end{minipage}%
  }
  \caption{Capsule outline over time for $\centercyinit=0.2$. (a) $\Ca=0.6$, shown
  at times $t=0, 1, 2, \ldots, 5$. (b) same case, shown at times $t=5, 10, 15, \ldots,
  70$ and $100$. Same notes as in \cref{fig:circle-evolution-y0} apply.}
  \label{fig:circle-evolution-y02}
\end{figure}

A capsule initialized at the pipe centerline ($\centercyinit=0$) will deform into a steady rotationally symmetric shape which depends on the capillary number $\Ca$ (more deformed for higher $\Ca$), cf.\ \cref{fig:circle-evolution-y0}. A capsule initialized off-center ($\centercyinit \neq 0$) instead deforms into an asymmetric shape which induces a lateral drift towards the pipe centerline via the interfacial force model \cref{eq:interfacial-force,eq:stress-tensor}. During the migration towards the centerline, the capsule continuously deforms and eventually ends up at the same steady symmetric shape as when $\centercyinit=0$, cf.\ \cref{fig:circle-evolution-y02}.

\begin{figure}[tbp]
  \centering\footnotesize
  \begin{minipage}[t]{0.5\textwidth}\vskip0pt\centering
    \includegraphics[width=\textwidth,trim=0cm 3cm 0cm 0.5cm,clip]{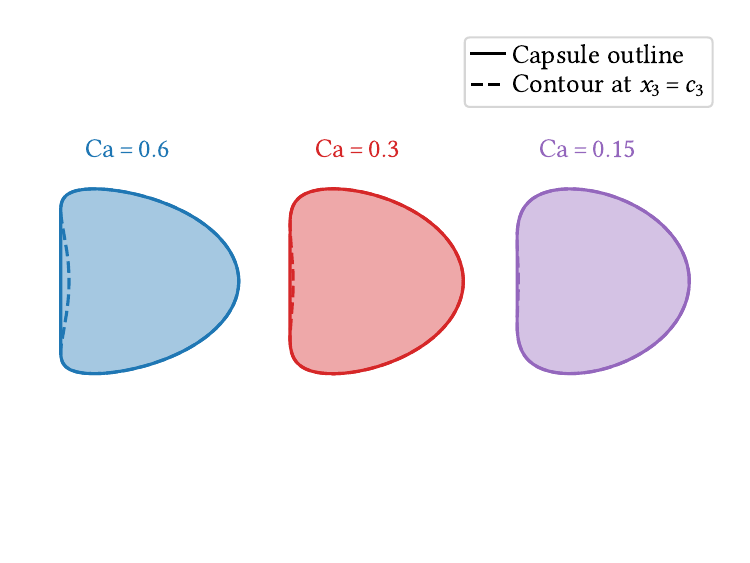}%
  \end{minipage}%
  \caption{Steady capsule shapes after long time for different capillary numbers.
  See text for explanation.}
  \label{fig:circle-final-1}
\end{figure}

\begin{figure}[tbp]
  \centering\footnotesize
  \begin{minipage}[t]{0.5\textwidth}\vskip0pt\centering
    \hspace*{30pt}(a) Steady capsule outlines\\
    \includegraphics[width=\textwidth,trim=0pt 0pt 0pt 28pt,clip]{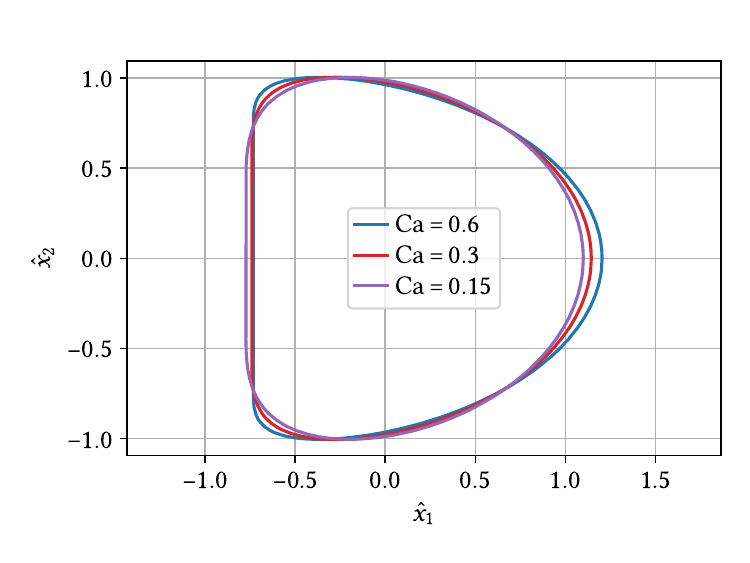}%
  \end{minipage}%
  \begin{minipage}[t]{0.5\textwidth}\vskip0pt\centering
    \hspace*{30pt}(b) Steady capsule contours at $x_3 = \centerc_3$\\
    \includegraphics[width=\textwidth,trim=0pt 0pt 0pt 28pt,clip]{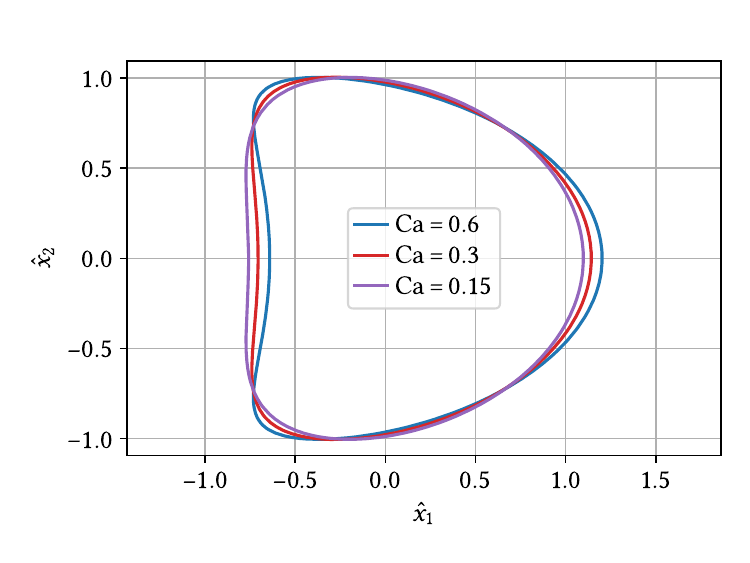}%
  \end{minipage}%
  \caption{Steady capsule shapes after long time for different capillary numbers, superimposed.
  (a) Capsule outlines. (b) Capsule centerline contours (slice at $x_3 = \centerc_3$).}
  \label{fig:circle-final-2}
\end{figure}

The final steady capsule shapes are shown in \cref{fig:circle-final-1,fig:circle-final-2}. They are visualized in two different ways: The \emph{outline} is generated by projecting the whole capsule surface onto the $x_1x_2$-plane and taking the boundary of the resulting planar region (by constructing a so-called ``alpha shape''). The \emph{contour} is generated by intersecting the capsule surface with a plane, in this case the plane $x_3 = \centerc_3$ which is parallel to the $x_1x_2$-plane and passes through the capsule center. Thus, the contour shows a single slice of the capsule. Note the slightly concave area at the back of the capsule, like the bottom of a bottle.

\begin{figure}[ptb]
  \centering\footnotesize
  \begin{minipage}[t]{0.5\textwidth}\vskip0pt\centering
    \hspace*{30pt}(a) Capsule reduced volume\\
    \includegraphics[width=\textwidth,trim=0pt 0pt 0pt 28pt,clip]{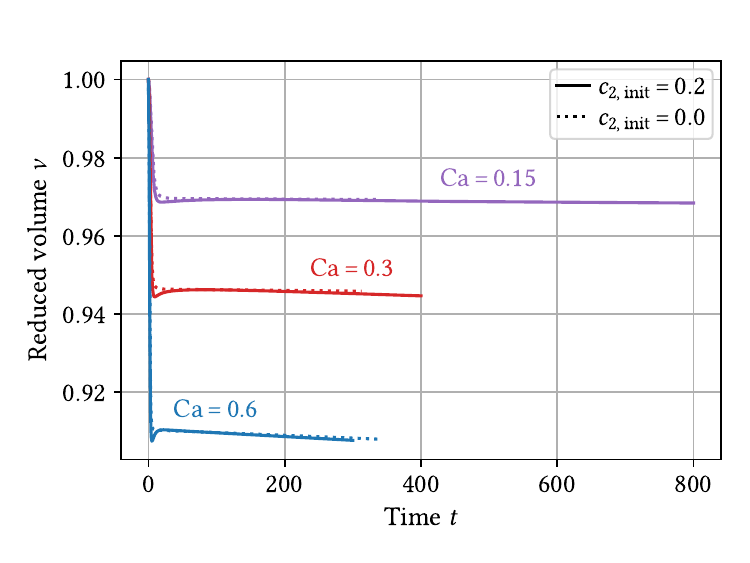}%
  \end{minipage}%
  \begin{minipage}[t]{0.5\textwidth}\vskip0pt\centering
    \hspace*{30pt}(b) Capsule Taylor asphericity\\
    \includegraphics[width=\textwidth,trim=0pt 0pt 0pt 28pt,clip]{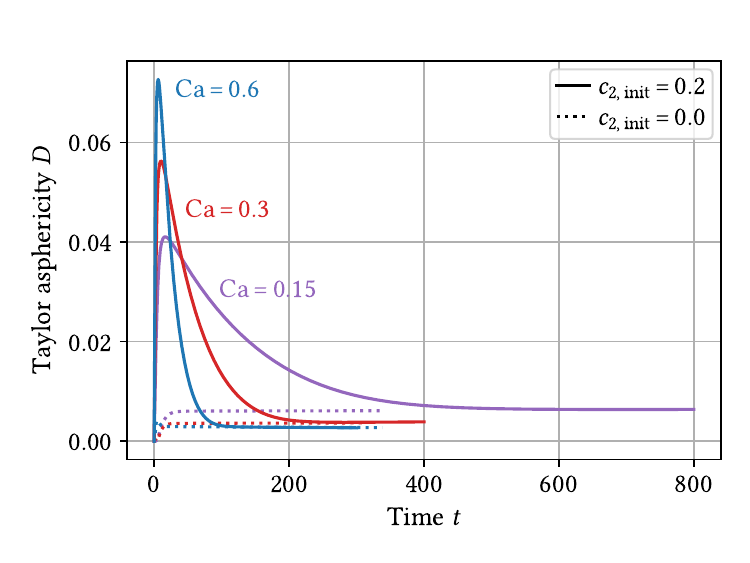}%
  \end{minipage}%
  \caption{Capsule reduced volume $\rv$ and Taylor asphericity $D$ as functions
  of time for different capillary numbers $\Ca$.}
  \label{fig:circle-reduced-volume}
\end{figure}

\begin{figure}[ptb]
  \centering\footnotesize
  \begin{minipage}[t]{0.5\textwidth}\vskip0pt\centering
    \hspace*{30pt}Capsule strain energy\\
    \includegraphics[width=\textwidth,trim=0pt 0pt 0pt 28pt,clip]{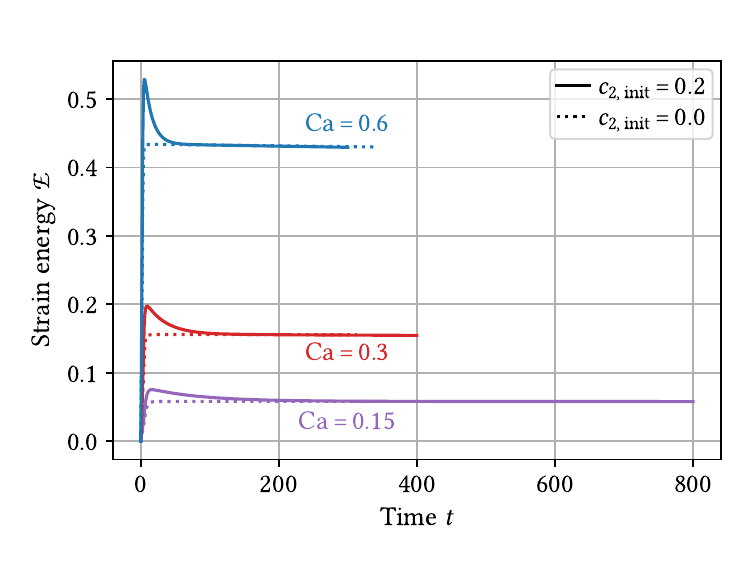}%
  \end{minipage}%
  \caption{Strain energy $\strain$ of capsule as a function of time
  for different capillary numbers $\Ca$.}
  \label{fig:circle-strain-energy}
\end{figure}

The instantaneous reduced volume $\rv$, Taylor asphericity $D$ and strain energy $\strain$ of the capsule membrane are plotted in \cref{fig:circle-reduced-volume,fig:circle-strain-energy}. These show how the capsule initially deforms due to the flow, and eventually finds a new equilibrium shape. It takes longer for the off-center capsule ($\centercyinit=0.2$) to reach equilibrium, since it has to migrate to the pipe centerline.

\begin{figure}[ptb]
  \centering\footnotesize
  \begin{minipage}[t]{0.5\textwidth}\vskip0pt\centering
    \hspace*{20pt}(a) Streamwise velocity of capsule center\\
    \includegraphics[width=\textwidth,trim=0pt 0pt 0pt 28pt,clip]{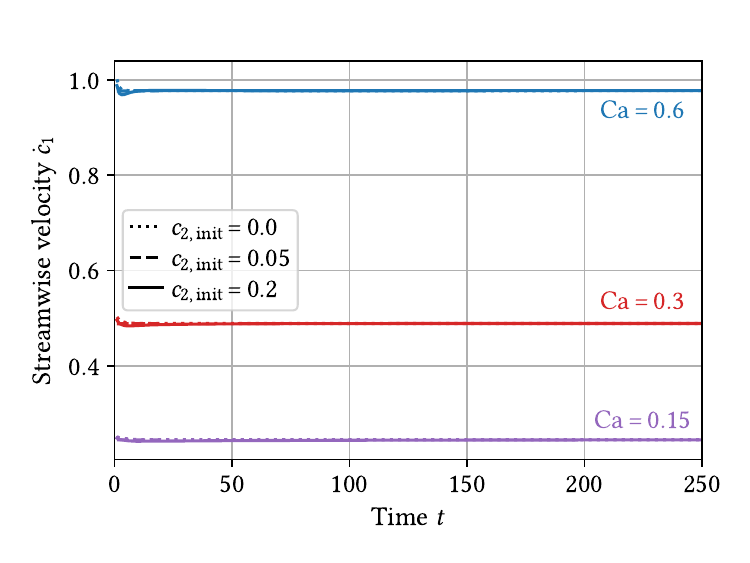}%
  \end{minipage}%
  \begin{minipage}[t]{0.5\textwidth}\vskip0pt\centering
    \hspace*{25pt}(b) Steady streamwise velocity\\
    \includegraphics[width=\textwidth,trim=0pt 0pt 0pt 28pt,clip]{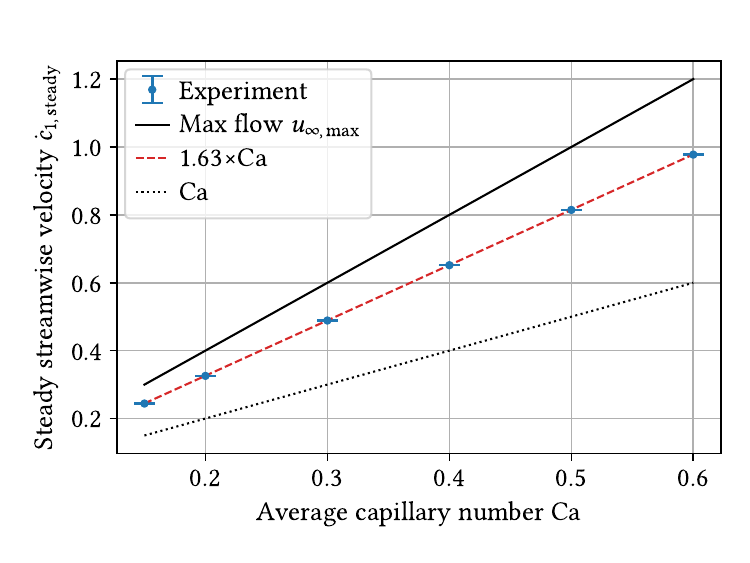}%
  \end{minipage}%
  \caption{(a) Streamwise velocity $\dcenterc_1$ of capsule center as a
  function of time. (b) Steady streamwise velocity $\dcentercxsteady$ of
  capsule center, measured at $t=90$, as a function of average
  capillary number $\Ca$ of the background flow profile.
  The steady velocity is well described by
  the relation $\dcentercxsteady = 1.63 \, \Ca$.
  The error bars are estimated by comparing $m=48$ with $m=24$.}
  \label{fig:circle-streamwise}
\end{figure}

The streamwise velocity of the capsule center, i.e., $\dcenterc_1$, quickly reaches a constant value $\dcentercxsteady$ that depends on the capillary number $\Ca$, as shown in \cref{fig:circle-streamwise}. The dependence is well modeled by the relation
\begin{equation}
    \dcentercxsteady = 1.63 \, \Ca.
\end{equation}
Note that the average flow velocity in the pipe is $\Ca$ and the maximum flow velocity (at the center) is $\ubgmax = 2 \, \Ca$; the capsule velocity is larger than the former but smaller than the latter. Numerically, we find that $\dcentercxsteady$ is equal to the average of the total flow field over the steady capsule surface, i.e.,
\begin{equation}
    \label{eq:average-total-flow}
    \dcentercxsteady = \frac{1}{A} \int_\gamma u_1(\bvec{x}) \, \mathrm{d}S,
\end{equation}
where $A$ is the surface area of the membrane. Interestingly, replacing the total field $u_1$ by the background flow $u_{\infty,1}$ in \cref{eq:average-total-flow} leads to a slightly larger value, around $1.66 \, \Ca$. This means that the interfacial force on the capsule membrane is responsible for slowing down the capsule by about 1.8\% in the streamwise direction. Furthermore, the average of the background flow can be well approximated by the simplified formula
\begin{equation}
    \label{eq:average-bg-flow}
    \ubgavg := \frac{1}{A} \int_\gamma u_{\infty,1}(\bvec{x}) \, \mathrm{d}S \approx
    \ubgmax \left( 1 - \frac{2}{3} \areafrac \right),
\end{equation}
where $\ubgmax$ is the maximum background flow and $\areafrac$ is the ratio between the areas enclosed by the (steady) capsule outline and the pipe outline in the $x_2x_3$-plane. In practice, the capsule outline is very close to a unit circle, so $\areafrac$ can be approximated by $1/b^2$ for a cylindrical pipe of radius $b$. Thus, $\ubgavg$ can be approximated using only information about the pipe and its background flow. The formula in \cref{eq:average-bg-flow} is derived in \cref{app:average-bg-flow-derivation}.

\begin{figure}[htbp]
  \centering\footnotesize
  \begin{minipage}[t]{0.5\textwidth}\vskip0pt\centering
    \hspace*{25pt}(a) Lateral position of capsule center\\
    \includegraphics[width=\textwidth,trim=0pt 0pt 0pt 28pt,clip]{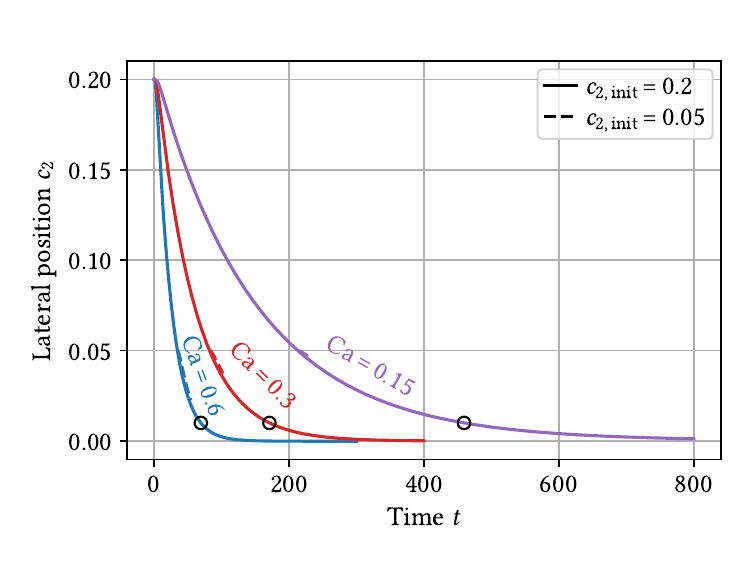}%
  \end{minipage}%
  \begin{minipage}[t]{0.5\textwidth}\vskip0pt\centering
    \hspace*{35pt}(b) Lateral velocity of capsule center\\
    \includegraphics[width=\textwidth]{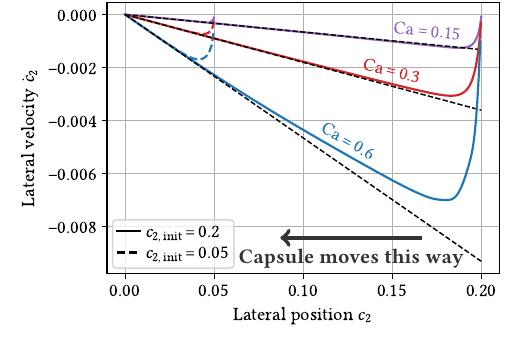}%
  \end{minipage}%
  \caption{Lateral position $\centerc_2$ and velocity $\dcenterc_2$ of capsule center (note: $\centerc_2=0$ is the centerline of the pipe). In
  (a), black circles ($\circ$) mark the observed times when
  the lateral center position $\centerc_2$ has been reduced to 5\% of its initial
  value, i.e., $\centerc_2=0.01$. In (b), the black dashed lines
  show curve fits of the form $\dcenterc_2 = \driftfactor \centerc_2$ to data close to $\centerc_2=0$.
  The obtained values of $\driftfactor$ are shown in \cref{fig:circle-lateral-conclusion}.}
  \label{fig:circle-lateral-data}
\end{figure}

\begin{figure}[htbp]
  \centering\footnotesize
  \begin{minipage}[t]{0.5\textwidth}\vskip0pt\centering
    \hspace*{35pt}Lateral drift factor\vphantom{p}\\
    \includegraphics[width=\textwidth,trim=0pt 0pt 0pt 28pt,clip]{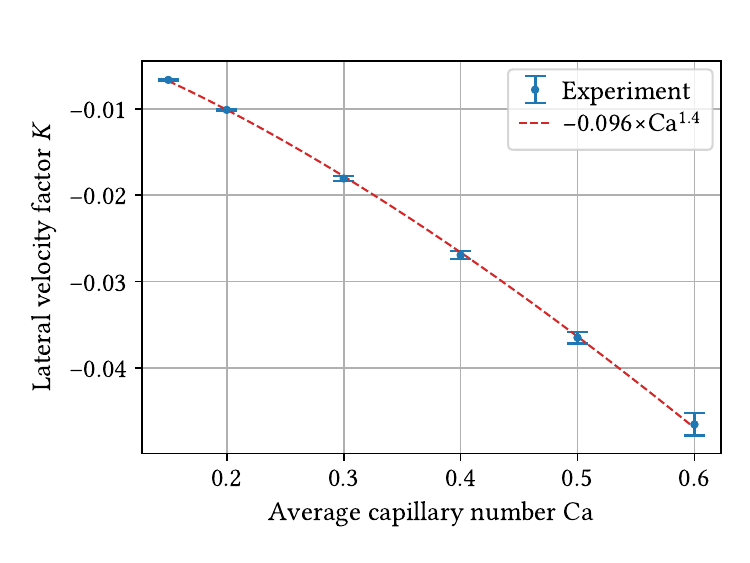}%
  \end{minipage}%
  \caption{Observed value of the factor $\driftfactor$ in the asymptotic formula
  $\dcenterc_2 \simeq \driftfactor \centerc_2$ for the lateral drift velocity, as a function of
  average capillary number $\Ca$ of the background flow profile.
  The factor is well approximated by the
  expression $\driftfactor = -0.096 \, \Ca^{1.4}$. The error is estimated by
  comparing $m=48$ with $m=24$.}
  \label{fig:circle-lateral-conclusion}
\end{figure}

As mentioned, when the capsule starts off-center ($\centercyinit \neq 0$), it will drift towards the pipe centerline. As shown in \cref{fig:circle-lateral-data}, the drift velocity is well approximated by the model
\begin{equation}
    \label{eq:lateral-drift-model}
    \dcenterc_2 = \driftfactor \centerc_2
\end{equation}
for $\centerc_2$ close to zero. Fitted values of $\driftfactor$ are shown in \cref{fig:circle-lateral-conclusion}; it turns out the factor is well described by
\begin{equation}
    \label{eq:circle-lateral-drift-factor}
    \driftfactor = -0.096 \, \Ca^{1.4}.
\end{equation}

Note that \cref{eq:lateral-drift-model} has an analytical solution, given by
\begin{equation}
    \label{eq:lateral-drift-solution}
    \centerc_2(t) = \centercyinit \text{e}^{\driftfactor t}.
\end{equation}
The time $\tau_r$ required for the solution to be reduced to $r \centercyinit$ is then
\begin{equation}
    \label{eq:drift-theoretical-time}
    \tau_r = \frac{\log(r)}{\driftfactor}.
\end{equation}
This is compared with observations in \cref{tab:circle-arrival-time}. Since the simulation has an initial transient (seen e.g.\ in the right part of \cref{fig:circle-lateral-data}~(b) and corresponding to the time interval in \cref{fig:circle-evolution-y02}~(a)), which is not taken into account by \cref{eq:drift-theoretical-time}, we measure $\Delta \tau := \tau_{0.05} - \tau_{0.5}$ to avoid the transient. The fitted model given by eqs.\ \eqref{eq:circle-lateral-drift-factor} through \eqref{eq:drift-theoretical-time} is then in good agreement with observed data.

\begin{table}[tbp]
  \centering
  \sisetup{table-format=3.2, table-auto-round=true, tight-spacing=true}
  \begin{tabular}{S[table-format=1.2, table-auto-round=false]SS[table-format=3.2+-2.2, table-auto-round=false]}
\toprule
$\Ca$ & {$\Delta\tau_\text{fit}$} & {$\Delta\tau_\text{obs}$} \\
\midrule
    0.15 & 341.52 & 351+-11 \\
    0.2  & 228.30 & 228.58+-0.11 \\
    0.3  & 129.41 & 127.5+-2.9 \\
    0.4  &  86.51 & 85.7+-1.6 \\
    0.5  &  63.30 & 63.65+-0.97 \\
    0.6  &  49.04 & 50.25+-0.76 \\
\bottomrule
  \end{tabular}
  \caption{Time $\Delta \tau = \tau_{0.05} - \tau_{0.5}$ for the
  lateral position $\centerc_2$ of the capsule center to go from 50\% to
  5\% of its initial value $\centercyinit$.
  Here, $\Delta\tau_\text{fit}$ is the theoretical value computed
  from \cref{eq:drift-theoretical-time,eq:circle-lateral-drift-factor};
  $\Delta\tau_\text{obs}$ is the value observed in the numerical experiments.
  The error in $\Delta\tau_\text{obs}$ is
  estimated by comparing $m=48$ with $m=24$. The estimated error for $\Ca=0.2$ is
  very small because the error switches signs around there.}
  \label{tab:circle-arrival-time}
\end{table}

\paragraph{Arbitrary cross-section.}
Next, consider a pipe with a cross-section in the $x_2x_3$-plane given by an arbitrary simple closed curve; the cross-section is assumed to be independent of $x_1$. Here, we consider the pipe shapes in \cref{fig:pipe-shapes}, with parametrizations given in \cref{app:pipe-shapes}.
Data for each pipe shape is given in \cref{tab:pipe-data}. The pipe resolution refers to the number of points in the axial ($x_1$) and circumferential directions. The cross-sections of the pipes are:
\begin{itemize}
    \item Case A (\iconAsmall): a circle of radius 2, i.e., the same as above.
    \item Case B (\iconBsmall): a superellipse with exponent 4 and semiaxes $(2,2)$.
    \item Case C (\iconCsmall): a superellipse with exponent 4 and semiaxes $(2,4)$.
    \item Case D (\iconDsmall): a ``starfish-like'' shape with inner radius 2.
    \item Case E (\iconEsmall): a ``starfish-like'' shape with inner radius 1.5.
\end{itemize}
The pressure-induced flow of each pipe is determined as described in \cref{app:pipe-background-flow}
with the capillary number (average background flow over pipe cross-section) fixed to $\Ca=0.6$. The capsule center is initialized a distance $\Delta c_\text{init} = 0.2$ from the origin in the $x_2x_3$-plane, at an angle $\varphi_\text{init}$ from the positive $x_2$-axis. For case~A, $\varphi_\text{init}=0^\circ$, and for the other cases, $\varphi_\text{init}=30^\circ$. For cases C and E, aggressive spectral filtering is used as described in \cref{sec:aggressive-spectral-filtering}; for the other cases, no spectral filtering is necessary.

\begin{table}[p]
  \centering\small
  \sisetup{table-format=-1.5, tight-spacing=true}
  \begin{tabular}{cS[table-format=2.3]cScc}
    \toprule
    Case & {Pipe area} & {Pipe resolution} & {Pressure gradient} & {Flow max.\ location} & {Max.\ value $\ubgmax$} \\
    \midrule
    A \iconAsmall & 12.566 & $32 \times 64$  & -1.2     & $(0, 0)$              & 1.200 \\
    B \iconBsmall & 14.833 & $32 \times 102$ & -1.0622  & $(0, 0)$              & 1.218 \\
    C \iconCsmall & 29.665 & $32 \times 154$ & -0.65832 & $(0, 0)$              & 1.175 \\
    D \iconDsmall & 20.154 & $32 \times 128$ & -1.1099  & $(-0.0509, 0.0755)$ & 1.399 \\
    E \iconEsmall & 11.336 & $32 \times 128$ & -1.9731  & $(-0.0382, 0.0566)$ & 1.399 \\
    \bottomrule
  \end{tabular}
  \caption{Data for each pipe shape and background flow, cases A--E.}
  \label{tab:pipe-data}
\end{table}

\begin{figure}[h]
  \centering
  Capsule shape evolution for case C (\iconCsmall)\\
  {\footnotesize%
  \begin{minipage}[t]{0.5\textwidth}\vskip0pt\centering
    \hspace*{30pt}(a) Times $0 \leq t \leq 5$\\
    \includegraphics[width=\textwidth]{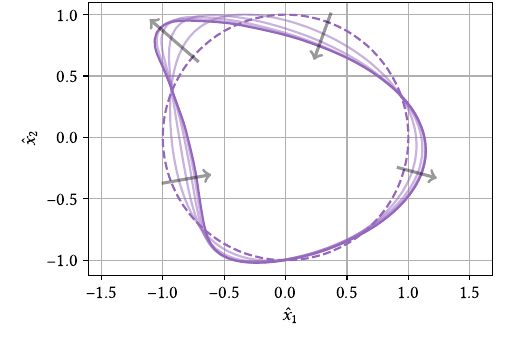}%
  \end{minipage}%
  \begin{minipage}[t]{0.5\textwidth}\vskip0pt\centering
    \hspace*{30pt}(b) Times $5 \leq t \leq 100$\\
    \includegraphics[width=\textwidth]{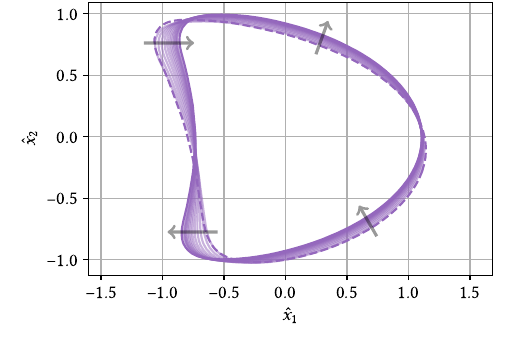}%
  \end{minipage}%
  }
  \caption{Capsule outline over time for case~C. (a) $t=0, 1, 2, \ldots, 5$.
  (b) $t=5, 10, 15, \ldots, 70$ and $100$.}
  \label{fig:arbitrary-evolution-C}
\end{figure}

\begin{figure}[h]
  \centering
  Capsule shape evolution for case D (\iconDsmall)\\
  {\footnotesize%
  \begin{minipage}[t]{0.5\textwidth}\vskip0pt\centering
    \hspace*{30pt}(a) Times $0 \leq t \leq 5$\\
    \includegraphics[width=\textwidth]{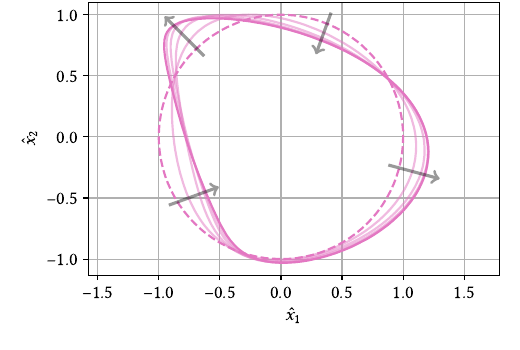}%
  \end{minipage}%
  \begin{minipage}[t]{0.5\textwidth}\vskip0pt\centering
    \hspace*{30pt}(b) Times $5 \leq t \leq 100$\\
    \includegraphics[width=\textwidth]{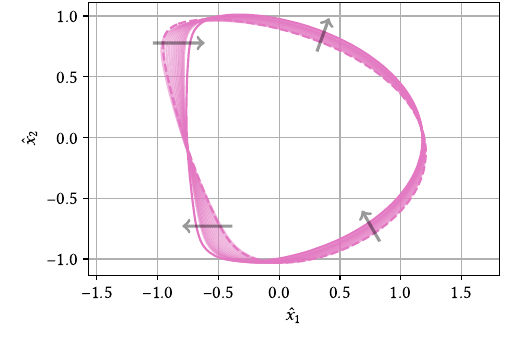}%
  \end{minipage}%
  }
  \caption{Capsule outline over time for case~D. (a) $t=0, 1, 2, \ldots, 5$.
  (b) $t=5, 10, 15, \ldots, 70$ and $100$.}
  \label{fig:arbitrary-evolution-D}
\end{figure}

\begin{figure}[h]
  \centering\footnotesize
  \begin{minipage}[t]{0.5\textwidth}\vskip0pt\centering
    \includegraphics[width=\textwidth,trim=0cm 3cm 0cm 0.5cm,clip]{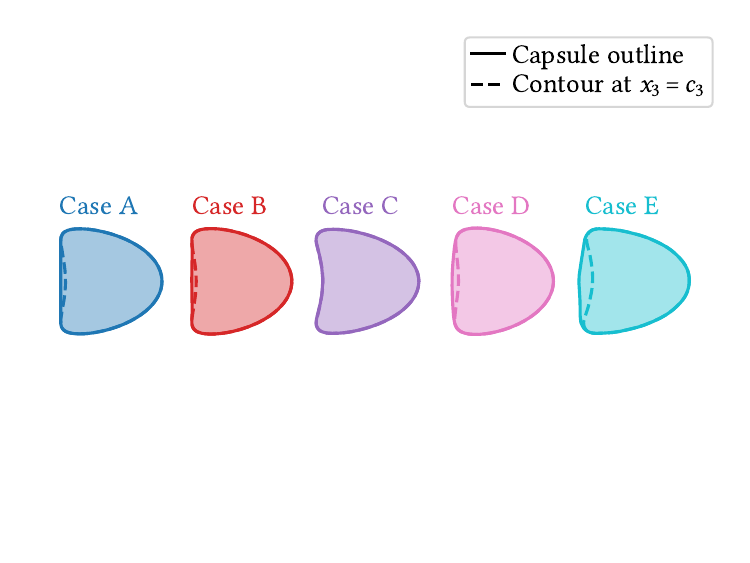}%
  \end{minipage}%
  \caption{Steady capsule shapes after long time for the different
  pipe cases. See text for explanation.}
  \label{fig:arbitrary-final-1}
\end{figure}

\begin{figure}[h]
  \centering\footnotesize
  \begin{minipage}[t]{0.5\textwidth}\vskip0pt\centering
    \hspace*{20pt}(a) Steady capsule outlines\\
    \includegraphics[width=\textwidth,trim=0pt 0pt 0pt 28pt,clip]{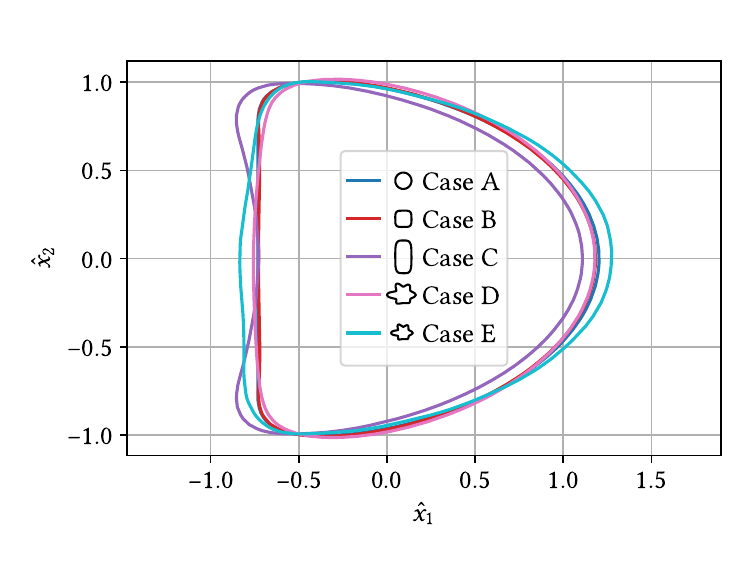}%
  \end{minipage}%
  \begin{minipage}[t]{0.5\textwidth}\vskip0pt\centering
    \hspace*{20pt}(b) Steady capsule contours at $x_3=\centerc_3$\\
    \includegraphics[width=\textwidth,trim=0pt 0pt 0pt 28pt,clip]{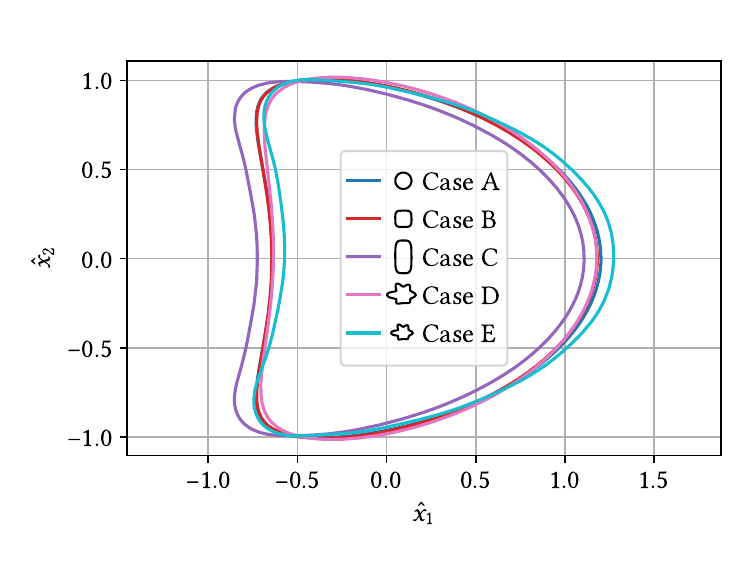}%
  \end{minipage}%
  \caption{Steady capsule shapes after long time for the different
  pipe cases, superimposed. (a) Capsule outlines. (b) Capsule centerline contours
  (slice at $x_3 = \centerc_3$).}
  \label{fig:arbitrary-final-2}
\end{figure}

The overall capsule behavior is similar to the case with a circular pipe (case~A), but with some differences. The capsule evolution for cases~C and D is shown in \cref{fig:arbitrary-evolution-C,fig:arbitrary-evolution-D}; compare with \cref{fig:circle-evolution-y02} for case~A, and additional figures in \cref{app:capsule-evolution}.
The final steady capsule shapes are shown in \cref{fig:arbitrary-final-1,fig:arbitrary-final-2}. Some differences are clear; for example, the final shape for case~C clearly lacks rotational symmetry.

\begin{figure}[h]
  \centering\footnotesize
  \begin{minipage}[t]{0.5\textwidth}\vskip0pt\centering
    \hspace*{20pt}Capsule Taylor asphericity\\
    \includegraphics[width=\textwidth,trim=0pt 0pt 0pt 28pt,clip]{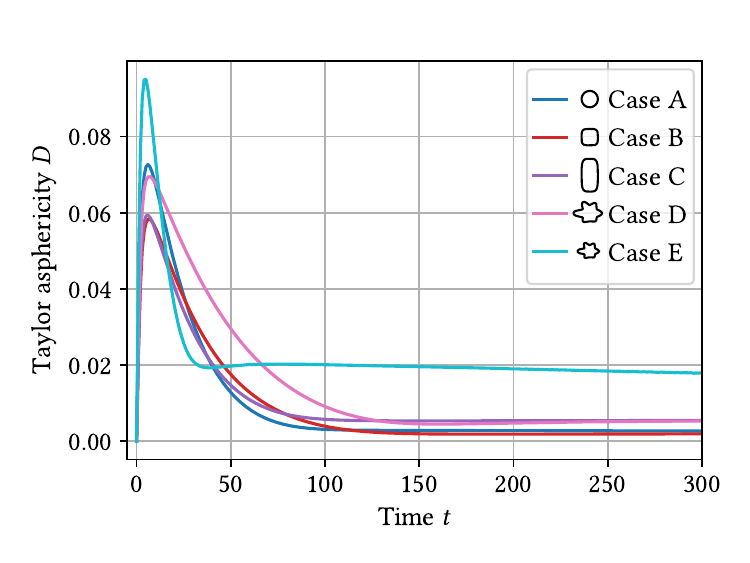}%
  \end{minipage}%
  \caption{Taylor asphericity $D$ of capsule for the different pipe cases.}
  \label{fig:arbitrary-taylor}
\end{figure}

The Taylor asphericity of the capsule is plotted in \cref{fig:arbitrary-taylor} for each case. This shows that the initial transient (from a spherical capsule to an asymmetrically deformed capsule) takes about as long in all cases (because it mainly depends on parameters such as $\Ed/\Es$ and $\Ca$ which are fixed). The subsequent lateral drift, on the other hand, varies more between the different pipe shapes (presumably because it depends on the local curvature of the flow profile).

\begin{figure}[h]
  \CenterFloatBoxes
  \begin{floatrow}
    \ffigbox{%
      \footnotesize
      \hspace*{35pt}Streamwise velocity of capsule center\\
      \includegraphics[width=0.5\textwidth,trim=0pt 0pt 0pt 28pt,clip]{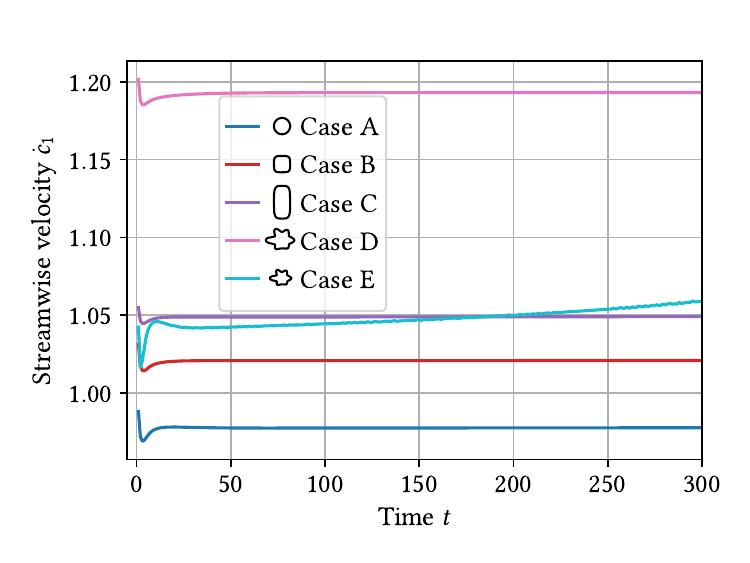}%
    }{%
      \caption{Streamwise velocity of capsule.}
      \label{fig:arbitrary-streamwise}
    }%
    \killfloatstyle\ftabbox[0.5\textwidth]{%
      \sisetup{table-format=1.3, table-auto-round=true, tight-spacing=true}
      \begin{tabular}{cSS}
        \toprule
        Case & {Velocity $\dcentercxsteady$} & {$\areafrac$} \\
        \midrule
        A \iconAsmall & 0.978 & 0.252 \\
        B \iconBsmall & 1.021 & 0.215 \\
        C \iconCsmall & 1.049 & 0.110 \\
        D \iconDsmall & 1.193 & 0.160 \\
        E \iconEsmall & 1.044 & 0.271 \\
        \bottomrule
      \end{tabular}
    }{%
      \caption{Steady streamwise velocity (at $t=100$).}
      \label{tab:arbitrary-streamwise}
    }%
  \end{floatrow}
\end{figure}

The streamwise velocity is shown in \cref{fig:arbitrary-streamwise,tab:arbitrary-streamwise}. It quickly reaches a steady value for all cases except case~E (probably due to large numerical errors). Numerically, \cref{eq:average-total-flow} holds for all pipe shapes. We can also consider the approximation on the right-hand side of \cref{eq:average-bg-flow}, with $\ubgmax$ taken from \cref{tab:pipe-data} and $\areafrac$ from \cref{tab:arbitrary-streamwise}. The difference between this approximation and the observed $\dcentercxsteady$ is around 2\% for cases~A and B, 4\% for case~C, 5\% for case~D, and 10\% for case~E. As explained in \cref{app:average-bg-flow-derivation},
the approximation is derived for case~A, so it is expected that the difference becomes larger for more complicated pipe shapes.

\begin{figure}[h]
  \centering\footnotesize
  \begin{minipage}[t]{0.5\textwidth}\vskip0pt\centering
    \hspace*{25pt}(a) Lateral position ($\centerch_2$) of capsule center\\
    \includegraphics[width=\textwidth,trim=0pt 0pt 0pt 28pt,clip]{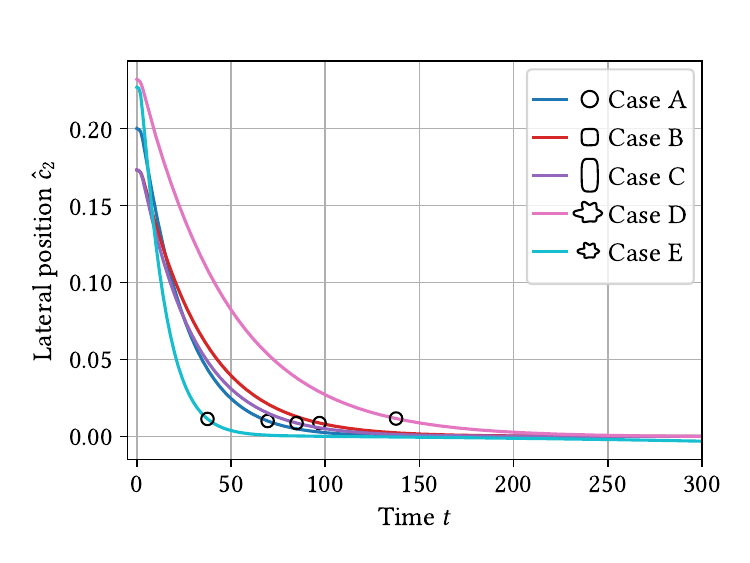}%
  \end{minipage}%
  \begin{minipage}[t]{0.5\textwidth}\vskip0pt\centering
    \hspace*{25pt}(b) Lateral position ($\centerch_3$) of capsule center\\
    \includegraphics[width=\textwidth,trim=0pt 0pt 0pt 28pt,clip]{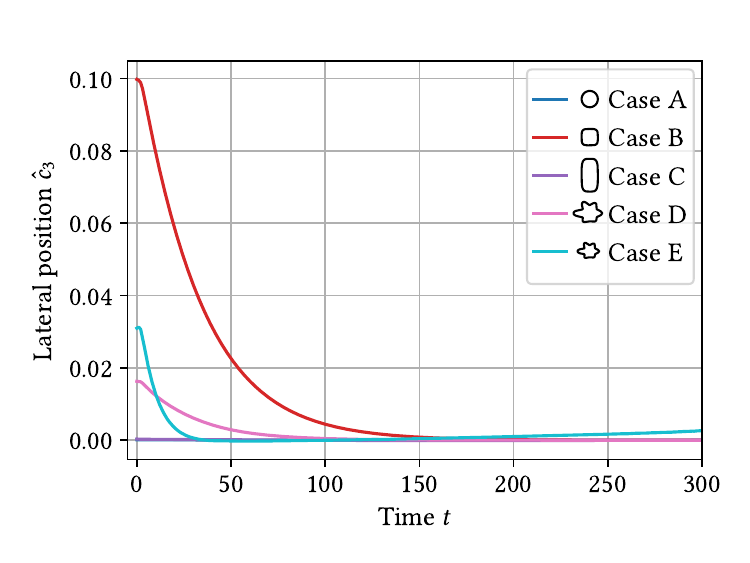}%
  \end{minipage}%
  \\
  \begin{minipage}[t]{0.5\textwidth}\vskip0pt\centering
    \hspace*{45pt}(c) Lateral velocity ($\dcenterch_2$) of capsule center\\
    \includegraphics[width=\textwidth]{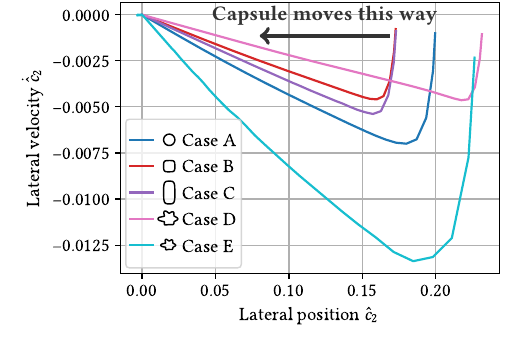}%
  \end{minipage}%
  \begin{minipage}[t]{0.5\textwidth}\vskip0pt\centering
    \hspace*{45pt}(d) Lateral velocity ($\dcenterch_3$) of capsule center\\
    \includegraphics[width=\textwidth]{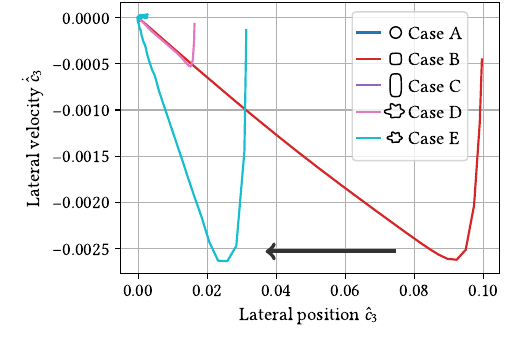}%
  \end{minipage}%
  \caption{Lateral position and velocity of capsule center. Here, $\centerch_i = \centerc_i - \centerc_{i,\text{steady}}$.}
  \label{fig:arbitrary-lateral-drift}
\end{figure}

Results for the lateral drift are shown in \cref{tab:arbitrary-lateral-drift,fig:arbitrary-lateral-drift}.
\begin{table}
  \centering
  \sisetup{table-format=-1.3, table-auto-round=true, tight-spacing=true}
  \begin{tabular}{ccSSS[table-format=2.2]S[table-format=2.2]}
    \toprule
    Case & {Final lateral position $\centerlatsteady$} & {$\driftfactor_2$} & {$\driftfactor_3$} & {$\Delta\tau_{2,\text{fit}}$} & {$\Delta\tau_{2,\text{obs}}$} \\
    \midrule
    A \iconAsmall & $(0, 0)$          & -0.047 & -0.047 &  49.04 &  50.46 \\
    B \iconBsmall & $(0, 0)$          & -0.032 & -0.032 &  71.18 &  71.76 \\
    C \iconCsmall & $(0, 0.100)$      & -0.036 & 0      &  63.10 &  63.07 \\
    D \iconDsmall & $(-0.059, 0.084)$ & -0.022 & -0.037 & 104.94 & 103.26 \\
    E \iconEsmall & $(-0.054, 0.069)$ & -0.089 & -0.120 &  25.74 &  26.21 \\
    \bottomrule
  \end{tabular}
  \caption{Lateral drift results. (i) Final lateral capsule center position $\centerlatsteady$. (ii) Lateral drift factors $\driftfactor_2$ and $\driftfactor_3$. (iii) Time $\Delta\tau_2 = \tau_{2,0.05} - \tau_{2,0.5}$ for the lateral position $\centerch_2$ of the capsule center to go from 50\% to 5\% of its initial value $\centerchyinit$; here, $\Delta\tau_{2,\text{fit}}$ is the
  theoretical value computed from \cref{eq:general-drift-theoretical-time};
  $\Delta\tau_{2,\text{obs}}$ is the value observed in the numerical experiments.}
  \label{tab:arbitrary-lateral-drift}
\end{table}
The capsule drifts towards a lateral position $\centerlatsteady = (\centerc_{2,\text{steady}}, \centerc_{3,\text{steady}})$ given in \cref{tab:arbitrary-lateral-drift}. For cases A and B, this final position coincides with the maximum background flow location, while for cases D and E it is close but not exactly the same. In case~C, the capsule drifts only in the $x_2$-direction but not in $x_3$. The reason is probably that the pipe diameter in the $x_3$ direction is too large to give a significant asymmetric deformation of the capsule in this direction. We introduce
\begin{equation}
    \centerch_i = \centerc_i - \centerc_{i,\text{steady}}, \qquad i=2,3,
\end{equation}
for the lateral capsule position adjusted by the final position, and generalize the model \cref{eq:lateral-drift-model} to
\begin{equation}
    \label{eq:general-drift-model}
    \dcenterlat = \driftfactormat \centerlat,
\end{equation}
where $\centerlat = (\centerc_2, \centerc_3)$ and $\driftfactormat$ is a tensor which we for simplicity here assume to be diagonal (although in general it may not be), i.e.,
\begin{equation}
    \driftfactormat = \begin{bmatrix}
        \driftfactor_2 & 0 \\
        0 & \driftfactor_3
    \end{bmatrix}.
\end{equation}
The values $\driftfactor_2$ and $\driftfactor_3$ are fitted to the data shown in \cref{fig:arbitrary-lateral-drift}~(c)--(d) close to $\centerlat = \bvec{0}$; the results are presented in \cref{tab:arbitrary-lateral-drift}. Both components of the solution to \eqref{eq:general-drift-model} are of the form of \eqref{eq:lateral-drift-solution}, and the time $\tau_{i,r}$ for the $i$th component to be reduced by a factor $r$ is
\begin{equation}
    \label{eq:general-drift-theoretical-time}
    \tau_{i,r} = \frac{\log(r)}{K_i}, \qquad i=2,3.
\end{equation}
This agrees well with observations, as shown in \cref{tab:arbitrary-lateral-drift}. The exact values of $K_i$ found in \cref{tab:arbitrary-lateral-drift} depend on the steady shape of the capsule and the background flow, as well as on the interfacial force model. We do not attempt to give any deeper analysis of these values here, but note that, in general, a narrower channel tends to lead to a larger magnitude of $K_i$ and therefore a larger drift velocity. This is probably because the flow has a higher local curvature and deforms the capsule more unevenly.

\section{Conclusions}
\label{sec:conclusions}

In this work, we developed a fast and stable method for simulating deformable capsules in both unbounded and pipe-confined Stokes flow. Our stability analysis shows that the relaxation of a spherical capsule is stable for small deformations even without bending stiffness, but it is known that physical instabilities can be introduced by the background flow \citep{Lac2004}, and numerical instabilities by aliasing errors \citep{Zhao2010}. Furthermore, biconcave capsules are observed to be more unstable than spheres. We devised a spectral filtering strategy to stabilize the scheme, with the main contribution being a detailed analysis for selecting the truncation index. Convergence tests for biconcave capsules with varying reduced volumes, in both unbounded and confined flows, confirm the accuracy and robustness of the method.

The stabilized scheme enables long-time simulations of capsules in pipes with a wide range of cross-sectional geometries. In circular pipes, we characterized the lateral migration of capsules and quantified its dependence on the capillary number. In particular, we identified the drift factor $\driftfactor$ in the asymptotic relation $\dcenterc_2 \simeq \driftfactor \centerc_2$ which governs the lateral drift velocity in this flow regime. We generalized the formulation to handle arbitrary constant cross-sections, allowing us to explore capsule dynamics beyond the circular case and opening the door to applications in more complex vessel geometries.

Future work includes refining the spectral filtering strategy by assigning different truncation indices to different force and velocity components, thereby balancing stability against the preservation of physical information. Another direction is to investigate alternative partition-of-unity functions that are easier to resolve on coarse meshes, which could improve computational efficiency. Additional extensions include developing semi-implicit time-stepping to reduce stiffness, as well as incorporating bending resistance and viscosity contrast to enhance the physical fidelity of the model.

\section*{Acknowledgements}

The authors acknowledge the Texas Advanced Computing Center (TACC)\footnote{URL: \url{http://www.tacc.utexas.edu}} at The University of Texas at Austin for providing HPC resources that have contributed to the research results reported within this paper. JB was supported in part by the Peter O'Donnell Jr.\ Postdoctoral Fellowship at the Oden Institute.

\bibliographystyle{jfm}
\bibliography{references}

\appendix
\section{Parametrization of biconcave capsules of arbitrary reduced volume}
\label{app:biconcave}

In this appendix, we describe how the biconcave shape of the relaxed capsule is constructed in the experiments where $\rvr < 1$. The shape is based on the biconcave disk model from \cite{Pozrikidis2005}, given by
\begin{align}
  \label{eq:biconcave-x}
  x_1 &= R \sin(\theta)\cos(\phi), \\
  \label{eq:biconcave-y}
  x_2 &= R \sin(\theta)\sin(\phi), \\
  \label{eq:biconcave-z}
  x_3 &= R k_0 \Big(1 + k_2 \sin^2(\theta) + k_4 \sin^4(\theta)\Big) \cos(\theta),
\end{align}
where $\theta \in [0,\pi]$, $\phi \in [0,2\pi)$, and $R$, $k_0$,
$k_2$, $k_4$ are shape parameters. The reduced volume of this shape can be computed as follows. Using that eqs.\ \eqref{eq:biconcave-x}--\eqref{eq:biconcave-z} define a surface of revolution, its enclosed volume can be expressed as
\begin{equation}
  \label{eq:biconcave-volume}
  V = \frac{4\pi R^3 k_0}{3} \Big(
  1 + \frac{2}{5} k_2 + \frac{8}{35} k_4
  \Big).
\end{equation}
The surface area has no closed-form expression, but can be
computed numerically to arbitrary precision as $A = 4\pi R^2 I_A(k_0,k_2,k_4)$,
where $I_A$ is the one-dimensional integral
\begin{equation}\small
  \label{eq:biconcave-area-integral}
  I_A(k_0,k_2,k_4) =
  \int_0^{\pi/2} \sin(\theta)
  \sqrt{
    \cos^2(\theta)
    +
    k_0^2 \sin^2(\theta)
    \Big[
      k_2 (3 \cos^2(\theta) - 1)
      +
      k_4 \sin^2(\theta) (4 - 5 \sin^2(\theta))
      - 1
    \Big]^2
  } \, \mathrm{d}\theta
  .
\end{equation}
Combining these, the expression for the reduced volume $\rvr$ of the
biconcave disk is
\begin{equation}
  \label{eq:biconcave-reduced-volume}
  \rvr = \frac{6\sqrt{\pi}V}{A^{3/2}}
  =
  k_0 \Big(
    1 + \frac{2}{5} k_2 + \frac{8}{35} k_4
  \Big)
  [I_A(k_0,k_2,k_4)]^{-3/2}.
\end{equation}
Note that the reduced volume is independent of $R$.

The task is now to select the shape parameters $k_0$, $k_2$, $k_4$ in \cref{eq:biconcave-reduced-volume} to produce a given reduced volume $\rvr$. However, specifying a reduced volume is not enough to uniquely determine the shape parameters. Furthermore, we would like shapes with higher reduced volume (closer to a sphere) to have lower curvature, which is not automatically  guaranteed for any choice of shape parameters.

To find shapes with the desired properties, we initialize a biconcave capsule with given shape parameters and let it relax to a sphere, recording the intermediate shapes at regular intervals.%
\footnote{More specifically, the initial biconcave capsule has parameters $R=1.38579052$,
$k_0=0.1035$, $k_2=9.67632850$, $k_4=-5.42512077$, which produce a reduced volume $\rvr=\nu_* \approx 0.644381503670110$. The relaxed configuration $\gammar$ is a sphere (shape parameters $R=1$, $k_0=1$, $k_2=0$, $k_4=0$), there is no background flow, the stiffness ratio is $\Ed/\Es=10$ and the capsule resolution is $m=12$. The capsule is evolved until time $t=30$.}
For every shape, the model eqs.\ \eqref{eq:biconcave-x}--\eqref{eq:biconcave-z} is fitted to the grid points in the $x_1x_3$-plane, where the model can be written as
\begin{equation}
  x_3 = \pm R k_0 \Big(1 + k_2 \frac{x_1^2}{R^2} + k_4 \frac{x_1^4}{R^4}\Big)
  \sqrt{1 - \frac{x_1^2}{R^2}}, \qquad \lvert x_1 \rvert \leq R.
\end{equation}
Determining $R$ from the grid point
with the largest $x_1$ coordinate, the fit becomes a linear
least-squares problem for determining $k_0$, $k_2$, $k_4$.

\begin{figure}[tb]
  \centering
  \includegraphics[width=\textwidth,trim=2.5cm 11cm 2.5cm 11cm,clip]{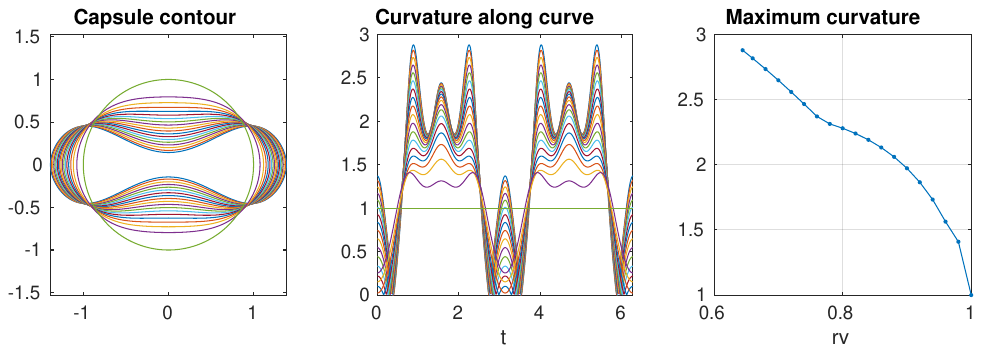}
  \caption{Capsule shape in $x_1x_3$-plane (left), measured curvature along
  the capsule curve (center), and maximum curvature as a function of
  reduced volume (right). Data for 19 different reduced volumes $\rvr =
  \nu_*$, 0.66, 0.68, $\ldots$, 0.98, 1.}
  \label{fig:biconcave-reduced-volume}
\end{figure}

Based on these fits, the algorithm for selecting the shape parameters given a reduced volume $\rvr$ is as follows:
\begin{enumerate}
  \item
    First select $k_0$ by inverting the 4th-degree polynomial equation (e.g.\ using Newton's method)
    \begin{align}
      \notag \rvr &= 1 - 0.164559954415538 \times (k_0-1)^2 + 0.570157143478058 \times (k_0-1)^3 \\
      \label{eq:rv-k0-relation}
      &\quad+ 0.290198279097787 \times (k_0-1)^4.
    \end{align}
  \item
    Fix the relation
    \begin{equation}
      \label{eq:k2-k4-relation}
      k_4 = -0.560659011318188 \times k_2,
    \end{equation}
    and determine $k_2$ by numerically solving
    \eqref{eq:biconcave-reduced-volume} to ensure the correct
    reduced volume $\rvr$.
  \item
    Finally, $R$ can be determined analytically from \cref{eq:biconcave-volume} given a desired volume $V$.
\end{enumerate}
This algorithm has been tested for reduced volumes in the range $\nu \in [\nu_*, 1]$, where $\nu_*\approx 0.64438$ is the reduced volume of the initial shape of the relaxation. As shown by \cref{fig:biconcave-reduced-volume}, the maximum curvature of the shapes decreases  monotonically as $\rvr$ increases.

\section{Pipes with arbitrary constant cross-section}
\label{app:pipes}

In this appendix, we describe how to construct a pressure-driven background flow for a pipe with arbitrary constant cross-section (\cref{app:pipe-background-flow}). More specifically, the cross-section of the pipe is an arbitrary simple closed curve in the $x_2x_3$-plane, independent of the axial coordinate $x_1$. We also give parametrizations (\cref{app:pipe-shapes}) of the test pipes used in \cref{sec:results-pipe-flow} (cf.\ \cref{fig:pipe-shapes}), and provide the derivation (\cref{app:average-bg-flow-derivation}) of the simplified formula \eqref{eq:average-bg-flow} for predicting the steady streamwise velocity of the capsule center.

\subsection{Computing pressure-driven background flow}
\label{app:pipe-background-flow}

\begin{figure}[h]
  \centering
  \begin{minipage}[t]{0.5\textwidth}\vskip0pt\centering
    \includegraphics[width=\textwidth]{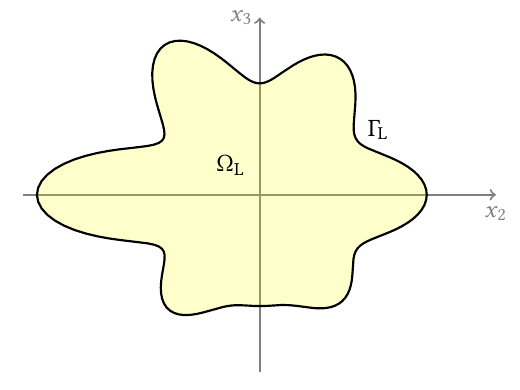}%
  \end{minipage}%
  \caption{An arbitrary pipe cross-section in the $x_2x_3$-plane, defined by a simple closed curve $\GammaL$ enclosing a region $\OmegaL$.}
  \label{fig:pipe-cross-section-sketch}
\end{figure}

Let the pipe cross-section be defined by the simple closed curve $\GammaL$ in the $x_2 x_3$-plane, which encloses a region $\OmegaL$ as shown in \cref{fig:pipe-cross-section-sketch}.
The full pipe surface is given by $\Gamma = \Reals \times \GammaL$ in $\Reals^3$ and encloses the fluid domain $\Omega = \Reals \times \OmegaL$. We seek a solution to the Stokes equations \eqref{eq:stokes-1}--\eqref{eq:stokes-2} for $\bvec{x} \in \Omega$. Due to symmetry, we assume that the flow velocity $\bvec{u}(\bvec{x})$ does not depend on $x_1$ (i.e., it is a function only of $x_2$ and $x_3$) and that $u_2=u_3=0$ (i.e., the flow points only in the $x_1$-direction), so that
\begin{equation}
    \bvec{u}(\bvec{x}) = (u_1(x_2, x_3), 0, 0).
\end{equation}
Under these assumptions, the Stokes equations reduce to a two-dimensional PDE
\begin{equation}
    \label{eq:poisson-problem}
    -\mu \DeltaL u_1 + \frac{\text{d}p}{\text{d}x_1} = 0,
\end{equation}
where $\DeltaL = \frac{\partial^2}{\partial x_2^2} + \frac{\partial^2}{\partial x_3^2}$ and the pressure $p$ is a function of $x_1$ only. Since $u_1$ is not allowed to depend on $x_1$, $\frac{\text{d}p}{\text{d}x_1}$ must be constant, and so the pressure has the form $p(x_1) = C x_1 + D$. Here, $C$ is the constant pressure gradient, and $D$ is an arbitrary constant. \Cref{eq:poisson-problem} is a two-dimensional Poisson problem and comes with no-slip boundary conditions $u_1 = 0$ on $\GammaL$.

To solve \eqref{eq:poisson-problem}, we write $u_1 = u_\text{part} + u_\text{hom}$, where $u_\text{part}$ is a particular solution satisfying \eqref{eq:poisson-problem} but not the boundary condition, and $u_\text{hom}$ satisfies the homogeneous version of \eqref{eq:poisson-problem}, i.e., $\DeltaL u_\text{hom} = 0$. A suitable particular solution is
\begin{equation}
    \label{eq:particular-solution}
    u_\text{part}(x_2, x_3) = a (b^2 - x_2^2 - x_3^2),
\end{equation}
cf.\ \cref{eq:background-flow-Poiseuille}, where $a$ and $b$ are arbitrary constants. Inserting this into \eqref{eq:poisson-problem}, we find that $C = -4\mu a$. The equation for $u_\text{hom}$ is then
\begin{equation}
    \label{eq:homogeneous-solution-pde}
    \begin{cases}
        \DeltaL u_\text{hom} = 0 & \text{in $\OmegaL$}, \\
        u_\text{hom} = -u_\text{part} & \text{on $\GammaL$}.
    \end{cases}
\end{equation}
This two-dimensional Laplace equation can easily be solved e.g.\ using a boundary integral method.

The procedure to compute the pressure-driven background flow is thus as follows: Given a desired pressure gradient $C$, compute $a=-C/(4\mu)$ which sets $u_\text{part}$ through \cref{eq:particular-solution}. Here, $b$ is arbitrary and does not influence the result; one can e.g.\ set $b=0$. Then, solve \cref{eq:homogeneous-solution-pde} for $u_\text{hom}$, and form $u_1 = u_\text{part} + u_\text{hom}$. The total pressure gradient of $u_1$ will be $C$.

In fact, from \cref{eq:poisson-problem} it is clear that $u_1$ is a linear function of $C$, so it is enough to compute $u_1$ for one value of $C$, e.g., $C=1$. The solution for any other value of $C$ can then be computed as
\begin{equation}
    \label{eq:pressure-gradient-scaling}
    u_1(x_2, x_3) = C \, u_1^{C=1}(x_2, x_3).
\end{equation}

\paragraph{Determining the pressure gradient from a capillary number.}
If the pressure gradient is not given, but the capillary number $\Ca$ is, an extra step is needed to determine $C$. Note that, using \cref{eq:capillary-number-pois-1} as the definition of the capillary number, it is simply the dimensionless mean flow velocity over the pipe cross-section. Thus, treating $C$ as unknown, we can integrate \cref{eq:pressure-gradient-scaling} and enforce
\begin{equation}
    \Ca \frac{\Es}{\mu} | \OmegaL | =
    \int_{\OmegaL} u_1 \, \text{d}S = C \int_{\OmegaL} u_1^{C=1} \, \text{d}S,
\end{equation}
where $|\OmegaL|$ denotes the area of $\OmegaL$. (If we work with dimensionless quantities, the factor $\Es/\mu$ disappears.) This equation can be directly solved for $C$, at the cost of one extra surface integration of $u_1^{C=1}$. This is what we do in \cref{sec:results-pipe-flow} to compute the pressure gradients given in \cref{tab:pipe-data} (with given $\Ca = 0.6$).

\subsection{Parametrizations of test pipes}
\label{app:pipe-shapes}

This section contains the parametrizations of the curves $\GammaL$ in the $x_2x_3$-plane that generate the pipe shapes used in \cref{sec:results-pipe-flow} shown in \cref{fig:pipe-shapes}.
\begin{itemize}
    \item
    Case A: The circle of radius 2 is defined by $x_2^2 + x_3^2 = 2^2$ and parametrized by $(x_2, x_3) = (2 \cos t, 2 \sin t)$, with $t \in [0, 2\pi)$.
    \item
    Case B: This is a superellipse defined by
    \begin{equation}
        \label{eq:superellipse}
        \left\lvert \frac{x_2}{a} \right\rvert^k + \left\lvert \frac{x_3}{b} \right\rvert^k = 1,
    \end{equation}
    here with exponent $k=4$ and semiaxes $(a,b) = (2,2)$. It is parametrized by
    \begin{equation}
        \begin{cases}
            x_2 = a R(t) \cos t, \\
            x_3 = b R(t) \sin t,
        \end{cases}
    \end{equation}
    where $R(t) = (\lvert\cos t\rvert^k + \lvert\sin t\rvert^k)^{-1/k}$ and $t \in [0, 2\pi)$.
    \item 
    Case C: Same as case~B but with $(a,b) = (2,4)$.
    \item 
    Case D: This starfish-like curve is parametrized by
    \begin{equation}
        \begin{cases}
            x_2 = c \rho(t) \cos t, \\
            x_3 = c \rho(t) \sin t,
        \end{cases}
    \end{equation}
    where $\rho(t) = 1 + \frac{1}{4} (1+\sin^2(\frac{t}{2})) (1+\cos(6t)) f(t)$, with $f(t) = 0.65 + 0.225 \sin(t) + 0.35 \cos(2t) - 0.075 \sin(3t)$, and the inner radius is $c=2$.
    \item 
    Case E: Same as case~D but with $c=1.5$.
\end{itemize}

\subsection{Derivation of \cref*{eq:average-bg-flow}}
\label{app:average-bg-flow-derivation}

We here derive \cref{eq:average-bg-flow} under the assumptions that the pipe cross-section is a circle of radius $b$ and the capsule is a sphere of radius $L$ centered at the middle of the pipe.

\begin{proposition}
Let $\GammaL$ be given by $x_2^2 + x_3^2 = b^2$, and let $u_{\infty,1}$ be the pressure-driven flow computed as in \cref{app:pipe-background-flow}. Furthermore, let $\gamma$ be the sphere $x_1^2 + x_2^2 + x_3^2 = L^2$. Then,
\begin{equation}
    \label{eq:average-bg-flow-proposition}
    \frac{1}{A} \int_\gamma u_{\infty,1}(\bvec{x}) \, \mathrm{d}S
    = \ubgmax \left( 1 - \frac{2}{3} \areafrac \right),
\end{equation}
where $A = \lvert \gamma \rvert$, $\ubgmax = \max_{\bvec{x} \in \OmegaL} u_{\infty,1}(\bvec{x})$ and $\areafrac = \lvert \gamma_\mathrm{L} \rvert / \lvert \OmegaL \rvert$. Here, $\gamma_\mathrm{L}$ is the projection of $\gamma$ onto the $x_2x_3$-plane.
\end{proposition}

The notation $\lvert S \rvert$ denotes the area of the surface $S$, and $\OmegaL$ is the region enclosed by $\GammaL$ as in \cref{app:pipe-background-flow}.

\begin{proof}
When $\GammaL$ is a circle, constructing the pressure-driven flow as described in \cref{app:pipe-background-flow} yields the radial solution
\begin{equation}
    u_{\infty,1}(\rho) = - \frac{C}{4\mu} (b^2 - \rho^2),
\end{equation}
where $\rho = (x_2^2 + x_3^2)^{1/2}$. Assuming that the pressure gradient $C$ is negative, we have
\begin{equation}
    \ubgmax = u_{\infty,1}(0) = -\frac{C b^2}{4\mu}.
\end{equation}
Furthermore, we have $A = 4\pi L^2$, $\lvert \gamma_\text{L} \rvert = \pi L^2$ and $\lvert \OmegaL \rvert = \pi b^2$, so that $\areafrac = (L/b)^2$.

The integral in the left-hand side of \eqref{eq:average-bg-flow-proposition} can now be computed analytically. For example, aligning the polar axis of the spherical coordinate system $(r, \theta, \varphi)$ with the $x_1$-axis, we get, with $\rho = L \sin \theta$,
\begin{align}
    \notag
    \int_\gamma u_{\infty,1}(\bvec{x}) \, \mathrm{d}S
    &= -\frac{C}{4\mu} \int_{0}^{2\pi} \int_{0}^{\pi} (b^2 - L^2 \sin^2 \theta) \, L^2 \sin \theta \, \mathrm{d}\theta \, \mathrm{d}\varphi
    \\
    &=
    -\frac{2\pi C L^2}{4\mu} \left(2 b^2 - \frac{4}{3} L^2 \right)
    =
    A \ubgmax \left( 1 - \frac{2}{3} \areafrac \right).
\end{align}
This proves \cref{eq:average-bg-flow-proposition}.
\end{proof}
\section{Selecting the periodic length $\Lper$}
\label{app:periodic-length}

\newcommand*{\elli}{\ell_i}
\newcommand*{\ella}{\ell_1}
\newcommand*{\ellb}{\ell_2}

This appendix outlines how the periodic length $\Lper$ is
selected in our simulations of pipe-confined flow. As mentioned
at the beginning of \cref{sec:formulation-pipe} and shown in
\cref{fig:pipe-flow-setup}, $\Lper$ is the length of the portion
of the pipe which is discretized and periodically replicated
together with the capsule. In other words, our pipe-confined
simulations model an infinitely long file of capsules with
spacing $\Lper$. In this appendix, we quantify the difference
between this periodic setup and a non-periodic setup with a
\emph{single} capsule in an infinitely long pipe (which is harder
to treat numerically). The difference is due to hydrodynamic
interactions in the periodic case between the capsules in the
file (i.e., between the capsule and its periodic replicas). We
emphasize that this difference is not due to any numerical error,
but due to the choice of mathematical model (periodic or
non-periodic).

We first consider a simplified setting, in which the flow is
generated by a single point force $\bvec{f}$ located at the
origin. In the periodic setting, this point force is replicated
periodically, and the flow field given by \cref{eq:periodic-flow-field}
reduces to
\begin{equation}
  \label{eq:1p-point-flow-field}
  \bvec{u}^\text{1P}(\bvec{x})
  =
  \sum_{n=-\infty}^\infty
  \bmat{G}(\bvec{x} - \bvec{p}_n) \bvec{f},
\end{equation}
where the Stokeslet $\bmat{G}$ is given by \cref{eq:stokeslet},
$\bvec{p}_n = (n \Lper, 0, 0)$, and the force $\bvec{f}$ is a constant
vector. In the non-periodic setting, the flow field is simply
$\bvec{u}^\text{0P}(\bvec{x}) = \bmat{G}(\bvec{x}) \bvec{f}$. For
reasons that will become clear in a moment, we
are interested in the difference between $\bvec{u}^\text{1P}(\bvec{x})$
and $\bvec{u}^\text{0P}(\bvec{x})$ close to the origin. For
simplicity, we will restrict $\bvec{x}$ to lie on the $x_1$-axis
(which is along the periodic direction), i.e., $\bvec{x} = (x_1, 0, 0)$.

To evaluate the series in \cref{eq:1p-point-flow-field}, we use
the Hasimoto decomposition \citep{Hasimoto1959} to write
$\bmat{G}(\bvec{r}) = \bmat{G}^\text{N}(\bvec{r}) +
\bmat{G}^\text{F}(\bvec{r})$, where the near-field part $\bmat{G}^\text{N}(\bvec{r})$
is given by \eqref{eq:hasimoto-near-field-stokeslet}, and the
Fourier transform of the far-field part $\bmat{G}^\text{F}(\bvec{r})$ is given by
\begin{equation}
  \widehat{\bmat{G}}{}^\text{F}(\bvec{k})
  = \frac{1}{8\pi\mu} \left( -\bmat{I} \lVert \bvec{k} \rVert^2
  + \bvec{k} \otimes \bvec{k}
  \right) \left( -\frac{8\pi}{\lVert \bvec{k} \rVert^4} \right)
  \E^{-\lVert \bvec{k} \rVert^2/(2\xi)^2}
  \left(
  1 + \frac{\lVert \bvec{k} \rVert^2}{(2\xi)^2}
  \right).
\end{equation}
Again, $\xi>0$ controls the decay of the two parts of the
decomposition.
Inserting the Hasimoto decomposition of $\bmat{G}$ into
\cref{eq:1p-point-flow-field} leads to a split
$\bvec{u}^\text{1P}(\bvec{x}) =
\bvec{u}^\text{1P,N}(\bvec{x}) +
\bvec{u}^\text{1P,F}(\bvec{x})$.

Let us first consider the case in which the force points along
the $x_1$-axis, i.e., $\bvec{f} = (f_1, 0, 0)$. In this case,
inserting \cref{eq:hasimoto-near-field-stokeslet} into
\cref{eq:1p-point-flow-field} yields the near-field part
\begin{equation}
  \label{eq:pdl-near-field-velocity}
  u_1^\text{1P,N}(\bvec{x})
  =
  \frac{f_1}{4\pi\mu}
  \sum_{n=-\infty}^\infty
  \frac{\erfc(\xi \lvert x_1 - n\Lper \rvert)}{\lvert x_1 - n\Lper \rvert}
  ,
\end{equation}
and $u_i^\text{1P,N}(\bvec{x}) = 0$ for $i=2,3$. For the
far-field part $\bvec{u}^\text{1P,F}$, we follow \cite{Bagge2023}
and use Poisson summation to write it as
\begin{equation}
  \label{eq:pdl-far-field-velocity}
  \bvec{u}^\text{1P,F}(\bvec{x})
  = \sum_{n=-\infty}^\infty
  \bmat{G}^\text{F}(\bvec{x} - \bvec{p}_n) \bvec{f}
  = \frac{1}{\Lper} \frac{1}{(2\pi)^2} \sum_{k_1 \in \mathcal{K}}
  \E^{\I k_1 x_1} \int_{\Reals^2} \widehat{\bmat{G}}{}^\text{F}(\bvec{k})
  \, \mathrm{d}k_2 \mathrm{d}k_3
  \, \bvec{f}
  ,
\end{equation}
where $\mathcal{K} = \frac{2\pi}{\Lper} \Integer$. The integrals
in the right-hand side of \eqref{eq:pdl-far-field-velocity} can
be evaluated analytically as described in \citep[Appendix~C]{Bagge2023}.
The result is
\begin{equation}
  \label{eq:pdl-far-field-velocity-component-1}
  u_1^\text{1P,F}(\bvec{x}) =
  \frac{1}{4\pi\mu} \frac{f_1}{\Lper} \sum_{\substack{k_1 \in \mathcal{K}\\k_1 \neq 0}}
  \E^{\I k_1 x_1}
  E_1\left(
  \frac{k_1^2}{4\xi^2}
  \right)
  +
  \frac{1}{4\pi\mu}
  \frac{f_1}{\Lper}
  (\gamma + \log(\xi^2 \ella^2) - 1)
  ,
\end{equation}
and $u_i^\text{1P,F}(\bvec{x}) = 0$ for $i=2,3$. Here, $E_1(x) = \int_1^\infty
\text{e}^{-xt}t^{-1} \, \text{d}t$ is the exponential integral,
and $\gamma$ is the Euler--Mascheroni constant.
The parameter $\ella$ in the last term of
\eqref{eq:pdl-far-field-velocity-component-1}, corresponding to
$k_1 = 0$, is an arbitrary gauge constant with the physical
dimensions of length.

Summing up \cref{eq:pdl-near-field-velocity,eq:pdl-far-field-velocity-component-1}
gives us the total flow field $\bvec{u}^\text{1P}$ in the
periodic case. Let us now imagine that the point force $\bvec{f}$
is located on the membrane of a spherical capsule of radius $L$. The flow
field should be evaluated on the capsule membrane to compute its
motion. It seems clear that the difference between
$\bvec{u}^\text{1P}$ and $\bvec{u}^\text{0P}$ will be larger
closer to the periodic replicas, so we should compute the
difference at $\bvec{x} = (2L, 0, 0)$. We get
\begin{equation}
  \label{eq:fx-per-field}
  u_1^\text{1P}(2L,0,0) =
  \frac{f_1}{4\pi\mu \Lper} \Bigg(
  \sum_{n=-\infty}^\infty
  \frac{\erfc(a \lvert \lambda - n \rvert)}{\lvert \lambda - n \rvert}
  +
  \sum_{\substack{k \in \Integer\\k \neq 0}}
  \E^{2\pi\I k \lambda}
  E_1\left(
  \frac{\pi^2 k^2}{a^2}
  \right)
  +
  \gamma + \log(b_1^2) - 1
  \Bigg)
  ,
\end{equation}
where we have introduced $\lambda = 2L/\Lper$, $a=\xi\Lper$
and $b_1=\xi\ella$. The expression in \cref{eq:fx-per-field} is
independent of $\xi$, as expected for the Hasimoto decomposition,
which can be shown e.g.\ by noting that its derivative with
respect to $\xi$ is zero. Therefore, we can without loss of
generality set $\xi=\Lper^{-1}$ so that $a=1$ and
$b_1=\ella/\Lper$. Comparing the resulting periodic flow with the
non-periodic flow $u_1^\text{0P}(2L,0,0) = f_1 / (8\pi\mu L)$, we
obtain the relative difference
\begin{multline}
  \label{eq:relative-diff-1}
  \Delta_1 := \frac{\lvert u_1^\text{1P}(2L,0,0) - u_1^\text{0P}(2L, 0, 0) \rvert}{
    \lvert u_1^\text{0P}(2L, 0, 0) \rvert} =
  \\
  =
  \left\lvert
  \lambda \Bigg(
  \sum_{n=-\infty}^\infty
  \frac{\erfc(\lvert \lambda - n \rvert)}{\lvert \lambda - n \rvert}
  +
  \sum_{\substack{k \in \Integer\\k \neq 0}}
  \E^{2\pi\I k \lambda}
  E_1(\pi^2 k^2)
  +
  \gamma + \log(b_1^2) - 1
  \Bigg)
  - 1 \right\rvert
  .
\end{multline}
Here, $\lambda$ is the ratio between the capsule diameter and the
periodic length $\Lper$, while $b_1$ is the ratio between the gauge
constant $\ella$ and $\Lper$. Note that $\lambda < 1$ is a
natural assumption.

If the force instead lies in the $x_2x_3$-plane, e.g.\ $\bvec{f}
= (0, f_2, 0)$, the above derivation can be repeated and we get
\begin{equation}
  \label{eq:relative-diff-2}
  \Delta_2 = \Bigg\lvert
  \lambda
  \Bigg(
  \sum_{n=-\infty}^\infty
  \frac{\erfc(\lvert \lambda - n \rvert)}{\lvert \lambda - n \rvert}
  +
  \sum_{\substack{k \in \Integer \\ k \neq 0}}
  \E^{2\pi\I k \lambda}
  E_1(\pi^2 k^2)
  +
  \gamma + \log (b_2^2) - 3
  \Bigg)
  - 1
  \Bigg\rvert
  ,
\end{equation}
where $b_2 = \ellb/\Lper$ and $\ellb$ is another gauge constant
(the gauge may be different in each coordinate direction).
Due to symmetry, \cref{eq:relative-diff-2} holds for any force in
the $x_2x_3$-plane. Note that the only differences between
\cref{eq:relative-diff-1,eq:relative-diff-2} are the gauge
constants and that a $1$ is replaced by a $3$.

Since $\ella$ and $\ellb$ are arbitrary, they can be set to
minimize $\Delta_1$ and $\Delta_2$ for given $L$ and $\Lper$.
In other words, we can ask the question: given $\lambda$,
which values of $b_i$ ($i=1,2$) minimize
\eqref{eq:relative-diff-1} and \eqref{eq:relative-diff-2}? To
answer this, let us write
\begin{equation}
  \Delta_i = \Bigg\lvert
  \erfc(\lambda)
  +
  \lambda
  \sum_{\substack{n \in \Integer \\ n \neq 0}}
  \frac{\erfc(\lvert \lambda - n \rvert)}{\lvert \lambda - n \rvert}
  +
  2\lambda
  \sum_{k=1}^\infty
  \cos(2\pi k \lambda)
  E_1(\pi^2 k^2)
  +
  \lambda
  (
  \gamma + \log (b_i^2) - \tau_i
  )
  - 1
  \Bigg\rvert
\end{equation}
for $i=1,2$, where $\tau_1 = 1$ and $\tau_2 = 3$. Expanding the
functions involved around $\lambda = 0$, we get
\begin{multline}
  \label{eq:pdl-rel-diff-expansion}
  \Delta_i
  =
  \Bigg\lvert
  \underbrace{\left(
  - \frac{2}{\sqrt{\pi}}
  + 2 (c_1 + c_5)
  + \gamma + \log(b_i^2) - \tau_i
  \right)}_{A_i} \lambda
  + \\
  + \underbrace{2 \left(
  \frac{1}{3 \sqrt{\pi}}
  +
  c_2
  +
  \frac{2}{\sqrt{\pi}} (c_3 + c_4)
  -
  2 \pi^2 c_6
  \right)}_{B} \lambda^3
  + \bigO(\lambda^5)
  \Bigg\rvert,
\end{multline}
where we have introduced
\begin{gather*}
  c_1 = \sum_{n=1}^\infty \frac{\erfc(n)}{n},
  \qquad
  c_2 = \sum_{n=1}^\infty \frac{\erfc(n)}{n^3},
  \qquad
  c_3 = \sum_{n=1}^\infty \text{e}^{-n^2},
  \qquad
  c_4 = \sum_{n=1}^\infty \frac{\text{e}^{-n^2}}{n^2},
  \\
  c_5 = \sum_{n=1}^\infty E_1 ( \pi^2 n^2 ),
  \qquad
  c_6 = \sum_{n=1}^\infty n^2 E_1 ( \pi^2 n^2 )
  .
\end{gather*}
These series converge quickly and can be computed numerically to
arbitrary precision. For small $\lambda$, the expression in
\cref{eq:pdl-rel-diff-expansion} is minimized when the
coefficient $A_i$ in front of $\lambda$ is zero; $\Delta_i$ is then
approximately given by the remaining third-order term, $\Delta_i
\approx B \lambda^3$. This means that the optimal value of $b_i$ is
\begin{equation}
  \label{eq:optimal-bi}
  b_i =
  \exp\left(
  \frac{\tau_i - \gamma}{2}
  + \frac{1}{\sqrt{\pi}}
  - c_1 - c_5
  \right).
\end{equation}
The values are $b_1 = \num{1.851380386386065}$ and $b_2 = \E b_1
= \num{5.032573661878727}$ to 15 decimal places. Furthermore, the
value of the coefficient $B$ above is $B = \num{2.404113806319189}$
to 15 decimal places.

This means that if $b_1$ and $b_2$ are given their optimal values
as stated above, the relative difference is approximately the
same for each force direction, and given by
\begin{equation}
  \label{eq:pdl-rule-of-thumb-1}
  \Delta_1 \approx \Delta_2 \approx 2.5 \lambda^3.
\end{equation}
Numerically, we observe that this approximation is good as long
as $\lambda \leq 0.4$, which means that $\Lper \geq 5L$. In our
simulations, we always select the optimal gauge constants as
given by \cref{eq:optimal-bi}, and assume $\Lper/L \geq 5$, so
that \cref{eq:pdl-rule-of-thumb-1} is a good approximation of the
relative difference between the periodic and non-periodic flow
fields.

To give an example of what the rule-of-thumb \eqref{eq:pdl-rule-of-thumb-1}
means in practice, we provide \cref{tab:pdl-table}.

\begin{table}[htp]
  \centering
  \begin{tabular}{rcccccc}
    \toprule
    $\Lper/L$ & 5 & 7.5 & 10 & 15 & 20 & 40 \\
    \midrule
    $\lambda$ & 0.4 & 0.27 & 0.2 & 0.13 & 0.1 & 0.05 \\
    \midrule
    $\Delta_i$ & 16\% & 4.7\% & 2\% & 0.6\% & 0.25\% & 0.03\% \\
    \bottomrule
  \end{tabular}
  \caption{Values of $\lambda$ and $\Delta_i$ corresponding to
  different $\Lper/L$, using $\lambda=2L/\Lper$ and
  \cref{eq:pdl-rule-of-thumb-1}.}
  \label{tab:pdl-table}
\end{table}

\begin{figure}[htbp]
  \centering
  {\footnotesize\hspace*{10mm}Effect of periodicity, capsule in
  Poiseuille flow} \\
  \includegraphics[width=0.55\textwidth]{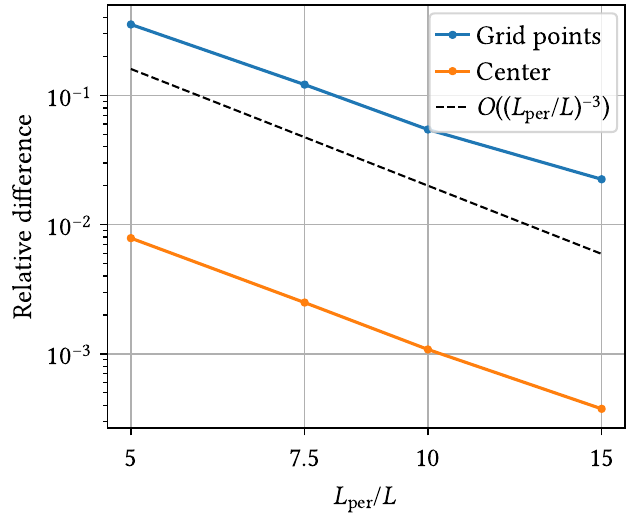}%
  \caption{Relative difference for grid point positions and
  center at time $T=50$ for a single spherical capsule in
  Poiseuille flow ($\Ca=0.6$), periodic with $\Lper$ vs.\ non-periodic.}
  \label{fig:pdl-capsule}
\end{figure}

Finally, we perform a numerical experiment to see the effect of
periodicity in an actual capsule simulation. We consider a single
spherical capsule ($\rvr=1$, $m=48$) in Poiseuille flow
($\Ca=0.6$). The initial location of the capsule center is
$\centercyinit = 0.2$, cf.\ \cref{sec:results-pipe-flow}.
Since there is no pipe in this setup, the simulation
can be run both with and without periodicity. We run with
periodic length $\Lper$ and compute the relative difference in
different quantities against the non-periodic case, at final time
$T=50$. More specifically, we measure the relative difference in
all grid points $\{ \bvec{x}_i \}$,
\begin{equation}
  \frac{\max_i \lVert
  \bvec{x}_i^\text{1P} - \bvec{x}_i^\text{0P}
  \rVert_2}{\max_i \lVert \bvec{x}_i^\text{0P} \rVert_2},
\end{equation}
and in the capsule center, $\lVert \bvec{c}^\text{1P} - \bvec{c}^\text{0P} \rVert_2 / \lVert \bvec{c}^\text{0P} \rVert_2$. The result is shown in
\cref{fig:pdl-capsule} for different periodic lengths.
We observe the cubic dependence on $\lambda = 2L/\Lper$ predicted
by the theory above. Since our focus in
\cref{sec:results-pipe-flow} is on the lateral drift of the
capsule, which is determined through its center position, we have
selected $\Lper/L = 5$ in the simulations, which should give a
relative difference below 1\% compared to a non-periodic setup.

\section{Proofs for stability analysis for a spherical capsule}
\label{app:stability-analysis}

The proofs of the main results of \cref{sec:stability-analysis} are given below.

\paragraph{Proof of \cref{prop:force-linearization}.}
\begin{equation}
    \bmat{F} = \frac{\partial \bvec{x}}{\partial \bvec{x}_\mathrm{r}}.
\end{equation}
With $\bvec{x} = \bvec{x}_\mathrm{r} + \delta \bvec{e}$, we get $\bmat{F} = \bmat{I} + \delta \bmat{E}$, where $\bmat{I}$ is the identity matrix and
\begin{equation}
    \bmat{E} := \frac{\partial \bvec{e}}{\partial \bvec{x}_\mathrm{r}}.
\end{equation}
Thus, $\bmat{F}_\mathrm{S} = \bmat{P}_\mathrm{r} + \delta  \bmat{E} \bmat{P}_\mathrm{r}$. Using that $\bmat{P}_\mathrm{r}$ is symmetric and a projection, we directly get
\begin{equation}
    \label{eq:proof-linearization-V2-value}
    \bmat{V}^2 = \bmat{F}_\mathrm{S} \bmat{F}_\mathrm{S}^\T
    = \bmat{P}_\mathrm{r} + \delta (\bmat{E} \bmat{P}_\mathrm{r}
    + \bmat{P}_\mathrm{r} \bmat{E}^\T)
    + \delta^2 \bmat{E} \bmat{P}_\mathrm{r} \bmat{E}^\T.
\end{equation}
It is known that one eigenvalue of $\bmat{V}^2$ is zero. In order to determine the other two eigenvalues, needed to compute $I_1$, $I_2$ and $J$ in \cref{eq:force-op-s-def,eq:force-op-d-def}, we can use the formula
\begin{equation}
    \det(\bmat{A} + \bmat{B})
    =
    \det(\bmat{A}) + \det(\bmat{B})
    + \tr(\bmat{A}\adj(\bmat{B}))
    + \tr(\bmat{B}\adj(\bmat{A}))
    ,
\end{equation}
which holds for any $3 \times 3$ matrices $\bmat{A}$ and $\bmat{B}$. Here, $\tr(\bcdot)$ denotes the trace, and $\adj(\bcdot)$ denotes the adjugate matrix. Setting $\bmat{A}=\bmat{V}^2$ and $\bmat{B}=-\sigma \bmat{I}$, and using that $\bmat{V}^2$ has determinant zero, we find the characteristic equation
\begin{equation}
\det(\bmat{V}^2 - \sigma \bmat{I}) = -\sigma^3 + \sigma^2 \tr(\bmat{V}^2) - \sigma \tr(\adj(\bmat{V}^2)) = 0.
\end{equation}
Setting $a := \tr(\bmat{V}^2)$ and $b := \tr(\adj(\bmat{V}^2))$, we can write down the two nonzero solutions as
\begin{equation}
    \sigma_1 = \frac{a}{2} + \sqrt{\frac{a^2}{4} - b}, \qquad
    \sigma_2 = \frac{a}{2} - \sqrt{\frac{a^2}{4} - b}.
\end{equation}
Since we are only interested in the quantities $\sigma_1 + \sigma_2$ and $\sigma_1 \sigma_2$, cf.\ \cref{eq:stress-tensor}, terms of $\bigO(\delta^2)$ and higher can be neglected in both $a$ and $b$. Using \cref{eq:proof-linearization-V2-value}, the definition of $\bmat{P}_\mathrm{r}$ and properties of the trace, we get
\begin{equation}
    a = \tr(\bmat{V}^2) = 2 + 2 \delta \tr(\bmat{E}\bmat{P}_\mathrm{r}) + \bigO(\delta^2),
\end{equation}
where $\tr(\bmat{E}\bmat{P}_\mathrm{r})$ can be identified as $\gradsphere \bcdot \bvec{e}$. Furthermore, using the Cayley--Hamilton formula for $3 \times 3$-matrices
\begin{equation}
    \adj(\bmat{A}) = \frac{1}{2} \bmat{I} \Big(
        [\tr(\bmat{A})]^2 - \tr(\bmat{A}^2)
    \Big) - \bmat{A} \tr(\bmat{A}) + \bmat{A}^2,
\end{equation}
we find that
\begin{equation}
    b = \tr(\adj(\bmat{V}^2)) = 1 + 2 \delta \tr(\bmat{E}\bmat{P}_\mathrm{r})
    + \bigO(\delta^2).
\end{equation}
It turns out that $H := \frac{a^2}{4} - b = \bigO(\delta^2)$, so that $\sqrt{H} = \bigO(\delta)$. At this point, we can compute the relevant quantities
\begin{align}
    I_1 &= \sigma_1 + \sigma_2 - 2 = 2 \delta \gradsphere \bcdot \bvec{e} + \bigO(\delta^2), \\
    I_2 &= \sigma_1 \sigma_2 - 1 = 2 \delta \gradsphere \bcdot \bvec{e} + \bigO(\delta^2), \\
    J &= \sqrt{\sigma_1 \sigma_2} = 1 + \delta \gradsphere \bcdot \bvec{e} + \bigO(\delta^2).
\end{align}
Finally, the following lemma is needed before we can evaluate \cref{eq:force-op-s-def,eq:force-op-d-def}.

\begin{lemma}
    When $\bvec{x} = \bvec{x}_\mathrm{r} + \delta \bvec{e}$, the projection matrix $\bmat{P}$ of $\gamma$ is related to the projection matrix $\bmat{P}_\mathrm{r}$ of $\gammar = \Sphere$ via
    \begin{equation}
        \bmat{P} = \bmat{P}_\mathrm{r} + \delta (\bmat{D} + \bmat{D}^\T) + \bigO(\delta^2),
    \end{equation}
    where
    \begin{equation}
        \bmat{D} := \Big(
        \gradsphere (\bvec{n}_r \bcdot \bvec{e}) - \bmat{P}_\mathrm{r} \bvec{e}
        \Big) \otimes \bvec{n}_\mathrm{r}
    \end{equation}
    and $\bvec{n}_\mathrm{r}$ is the outward unit normal of the sphere.
\end{lemma}
\begin{proof}
    The difference between the projection matrices is $\bmat{P} - \bmat{P}_\mathrm{r} = -\bvec{n} \bvec{n}^\T + \bvec{n}_\mathrm{r} \bvec{n}_\mathrm{r}^\T$, where $\bvec{n}$ is the unit normal vector of $\gamma$. The tangent vectors, cf.\ \cref{eq:surface-deformation-gradient-1,eq:surface-deformation-gradient-2,eq:surface-deformation-gradient-3}, are related via
    \begin{equation}
        \bvec{a}_i = \bvec{a}_i^\mathrm{r} + \delta \frac{\partial \bvec{e}}{\partial u_i},
        \qquad i=1,2,
    \end{equation}
    where $(u_1, u_2) = (\theta, \varphi)$ are the (polar and azimuthal) angles in a spherical coordinate system. Defining the unnormalized normal vector $\tilde{\bvec{n}} = \bvec{a}_1 \times \bvec{a}_2$ and the unit normal $\bvec{n} = \tilde{\bvec{n}} / \lVert \tilde{\bvec{n}} \rVert$, we get
    \begin{equation}
        \label{eq:proof-linearization-normal-vector}
        \bvec{n} = \bvec{n}_\mathrm{r} + \delta \bmat{P}_\mathrm{r} \bvec{\eta} + \bigO(\delta^2),
        \qquad
        \bvec{\eta} := \frac{1}{\lVert \tilde{\bvec{n}}_\mathrm{r} \rVert}
        \left(
        \bvec{a}_1^\mathrm{r} \times \frac{\partial \bvec{e}}{\partial \varphi}
        + \frac{\partial \bvec{e}}{\partial \theta} \times \bvec{a}_2^\mathrm{r}
        \right).
    \end{equation}
    Working with the expression for $\bvec{\eta}$, it can be simplified to
    $\bvec{\eta} = \bvec{n}_\mathrm{r} (\gradsphere \bcdot \bvec{e}) - \gradsphere(\bvec{n}_\mathrm{r} \bcdot \bvec{e}) + \bmat{P}_\mathrm{r} \bvec{e}$. Plugging this into the left equation of \eqref{eq:proof-linearization-normal-vector}, we can compute $\bvec{n} \bvec{n}^\T = \bvec{n}_\mathrm{r} \bvec{n}_\mathrm{r}^\T - \delta (\bmat{D} + \bmat{D}^\T) + \bigO(\delta^2)$, from which the lemma follows.
\end{proof}

As a corollary to the lemma, $\gradgamma = \gradsphere + \bigO(\delta)$. This lets us replace $\gradgamma$ by $\gradsphere$ in \cref{eq:force-op-s-def,eq:force-op-d-def} since the stress tensors themselves are already $\bigO(\delta)$. For $\forceopd$, \cref{eq:force-op-d-def}, we now get
\begin{equation}
    \forceopd[\delta \bvec{e}] = \gradsphere \bcdot
    [(\gradsphere \bcdot \delta\bvec{e}) \bmat{P}_\mathrm{r}]
    + \bigO(\delta^2)
    = (\gradsphere - 2 \bvec{n}_\mathrm{r}) (\gradsphere \bcdot \delta \bvec{e}) + \bigO(\delta^2),
\end{equation}
as claimed by \cref{eq:prop-linear-force-fd}. For $\forceops$, \cref{eq:force-op-s-def}, we get
\begin{align}
    \notag
    \forceops[\delta \bvec{e}] &= \frac{1}{2}
        \gradgamma \bcdot \left(
       (1 + \gradsphere \bcdot \delta \bvec{e}) \bmat{P}_\mathrm{r} + \delta (\bmat{E} \bmat{P}_\mathrm{r} + \bmat{P}_\mathrm{r} \bmat{E}^\T)
      -  (1 + \gradsphere \bcdot \delta \bvec{e})\bmat{P}_\mathrm{r} - \delta (\bmat{D} + \bmat{D}^\T)
    \right)
    + \bigO(\delta^2)
    \\[5pt]
    \label{eq:proof-linearization-fs-step}
    &= \frac{\delta}{2}
        \gradsphere \bcdot \left(
       (\bmat{E} \bmat{P}_\mathrm{r} + \bmat{P}_\mathrm{r} \bmat{E}^\T)
       - (\bmat{D} + \bmat{D}^\T)
    \right)
    + \bigO(\delta^2).
\end{align}
Applying the surface divergence to each term, one arrives, after some work, at
\begin{align}
    \gradsphere \bcdot \bmat{E} \bmat{P}_\mathrm{r} &=
    (\gradsphere - \bvec{n}_\mathrm{r})(\gradsphere \bcdot \bvec{e}) + \gradsphere (\bvec{n}_\mathrm{r} \bcdot \bvec{e}) - \bmat{P}_\mathrm{r} \bvec{e}, \\
    \gradsphere \bcdot \bmat{P}_\mathrm{r} \bmat{E}^\T &=
    \Deltasphere \bvec{e}, \\
    \gradsphere \bcdot \bmat{D} &=
    \bvec{n}_\mathrm{r} (\gradsphere \bcdot \bvec{e}) + \bvec{n}_\mathrm{r} \bvec{n}_\mathrm{r}^\T
    \Deltasphere \bvec{e} + \gradsphere(\bvec{n}_\mathrm{r} \bcdot \bvec{e}) - \bmat{P}_\mathrm{r} \bvec{e}, \\
    \gradsphere \bcdot \bmat{D}^\T &=
    2 \gradsphere(\bvec{n}_\mathrm{r} \bcdot \bvec{e}) - 2 \bmat{P}_\mathrm{r} \bvec{e}.
\end{align}
Combining these into \cref{eq:proof-linearization-fs-step} yields \cref{eq:prop-linear-force-fs}, which completes the proof of the proposition. \qed

\paragraph{Proof of \cref{thm:force-eigenfunctions}.}
To show \cref{eq:eig-fd-a,eq:eig-fd-b}, use linearity of $\forceopdlin$ and the relations
\begin{align}
    \gradsphere \bcdot \vsphv_\spha^\sphb &= -(\spha+1)(\spha+2) \sph_\spha^\sphb, \\
    \gradsphere \bcdot \vsphw_\spha^\sphb &= -\spha(\spha-1) \sph_\spha^\sphb.
\end{align}
It is straightforward to check that
\begin{equation}
    \forceopdlin[\bvec{A}_\spha^\sphb] = -(\spha^2 + \spha + 4)(2\spha+1)(\gradsphere - 2 \bvec{n}_\mathrm{r}) \sph_\spha^\sphb = -(\spha^2 + \spha + 4) \bvec{A}_\spha^\sphb,
\end{equation}
and that $\forceopdlin[\bvec{B}_\spha^\sphb] = \bvec{0}$. \Cref{eq:eig-fd-x} follows from the fact that $\gradsphere \bcdot \vsphx_\spha^\sphb = 0$.

To show \cref{eq:eig-fs-c,eq:eig-fs-d}, the same method is applied, now also using the relations
\begin{align}
    \Deltasphere \vsphv_\spha^\sphb &= -(\spha+1)(\spha+2) \vsphv_\spha^\sphb, \\
    \Deltasphere \vsphw_\spha^\sphb &= -\spha(\spha-1) \vsphw_\spha^\sphb.
\end{align}
It can be checked that
\begin{align}
    \forceops[\vsphv_\spha^\sphb] &= (\spha+2) \Big(
    -\spha \gradsphere \sph_\spha^\sphb + (\spha+1) \sph_\spha^\sphb \bvec{n}_\mathrm{r}
    \Big)
    = -\frac{r+2}{2r+1} \Big(
    (\spha^2+\spha+1) \vsphv_\spha^\sphb + (\spha-1)(\spha+1) \vsphw_\spha^\sphb
    \Big), \\[5pt]
    \forceops[\vsphw_\spha^\sphb] &= (\spha-1) \Big(
    -(\spha+1) \gradsphere \sph_\spha^\sphb + \spha \sph_\spha^\sphb \bvec{n}_\mathrm{r}
    \Big)
    = -\frac{r-1}{2r+1} \Big(
    \spha (\spha+2) \vsphv_\spha^\sphb + (\spha^2+\spha+1) \vsphw_\spha^\sphb
    \Big).
\end{align}
Using orthogonality between $\vsphv_\spha^\sphb$ and $\vsphw_\spha^\sphb$, it can be checked that $\forceops[\bvec{C}_\spha^\sphb] = 2\spha(\spha-1)(\spha+2) \forceops[\vsphv_\spha^\sphb] + K_{+} \forceops[\vsphw_\spha^\sphb]$ satisfies \cref{eq:eig-fs-c}, and similarly for $\forceops[\bvec{D}_\spha^\sphb]$ and \cref{eq:eig-fs-d}. Finally, \cref{eq:eig-fs-x} follows by using the relation $\Deltasphere \vsphx_\spha^\sphb = -\spha(\spha+1)\vsphx_\spha^\sphb$, and noting that $\gradsphere \bcdot \vsphx_\spha^\sphb = 0$, and furthermore, that $\vsphx_\spha^\sphb$ is orthogonal to $\bvec{n}_\mathrm{r}$. \qed
\section{Definition and properties of the discrete spherical harmonic transform (DSHT)}\label{appendix:DSHT}

Define an inner product $\langle\cdot,\cdot\rangle_{G_q}: \mathbb{C}^{2q(q + 1)}\times \mathbb{C}^{2q(q + 1)} \to \mathbb{C}$, by
\begin{equation}
\langle\boldsymbol{x},\boldsymbol{y}\rangle_{G_q} = \frac{\pi}{q} \sum_{j = 0}^q\sum_{k = 0}^{2q - 1}w_j x_{jk} \overline{y_{jk}},
\end{equation}
where $\{w_j\}_{j = 0}^{q}$ are Gauss-Legendre weights. 

Define a linear map $\bmat{Q}^*_{qp}: (\mathbb{C}^{2q(q + 1)}, \langle\cdot,\cdot\rangle_{G_q}) \to (\mathbb{C}^{p(p + 2)}, \langle\cdot,\cdot\rangle)$, where $p \leq q$, by
\begin{equation}
\begin{aligned}
\hat{\boldsymbol{x}} & = \bmat{Q}^*_{qp}\boldsymbol{x},\\
\hat{x}_{ns} &= \langle\boldsymbol{x},\boldsymbol{Y}^{ns}\rangle_{G_q},\ -p \leq s \leq p - 1,\ \vert s \vert \leq n \leq p,
\end{aligned}
\end{equation}
where $Y_{jk}^{ns} = Y_n^s(u_j, v_k)$, with $Y_n^s$ being the spherical harmonic function of degree $n$ and order $s$, $u_j = \cos^{-1}t_j$, and $v_k = \frac{2\pi k}{2q}$. $\{t_j\}_{j = 0}^{q}$ are Gauss-Legendre nodes.
We refer to $\bmat{Q}^*_{qp}$ as the discrete spherical harmonic transform (DSHT) on a spherical harmonic grid with $2q(q + 1)$ points, used to compute the spherical harmonic coefficients up to degree $p$.
We can easily verify that the adjoint operator of $\bmat{Q}^*_{qp}$ is $\bmat{Q}_{qp}:(\mathbb{C}^{p(p + 2)}, \langle\cdot,\cdot\rangle) \to (\mathbb{C}^{2q(q + 1)}, \langle\cdot,\cdot\rangle_{G_q})$, defined by 
\begin{equation}
\begin{aligned}
\boldsymbol{{x}} & = \bmat{Q}_{qp}\boldsymbol{\hat{x}}\\
{x}_{jk} & = \sum_{s = -p}^{p - 1}\sum_{n = |s|}^{p} \hat{x}_{ns}{Y}^{ns}_{jk},\ 0\leq j \leq q,\ 0\leq k \leq 2q - 1.
\end{aligned}
\end{equation}
We refer to $\bmat{Q}_{qp}$ as the inverse discrete spherical harmonic transform (iDSHT).

\begin{proposition} 
\begin{equation}
\Vert I - \bmat{Q}_{qp}^*\bmat{Q}_{qp} \Vert_2 \leq (p + 1) \epsilon_{G},
\end{equation}
where $\epsilon_{G}$ is the maximum quadrature error for $\int_{\mathbb{S}^2}Y_n^s\overline{Y_{n'}^{s'}}$ with $n, n'\leq p$.
\end{proposition}

\begin{proof}
Let $\hat{\boldsymbol{y}} = \bmat{Q}_{qp}^*\bmat{Q}_{qp}\hat{\boldsymbol{x}}$, then
\begin{equation}
\begin{aligned}
\hat{x}_{ns} - \hat{y}_{ns} & =  \hat{x}_{ns} - \langle \sum_{s' = {-p}}^{p - 1}\sum_{n' = \vert s'\vert}^p \hat{x}_{n's'}\boldsymbol{Y}^{n's'}, \boldsymbol{Y}^{ns} \rangle_{G_q}\\
& =  \sum_{s' = {-p}}^{p - 1}\sum_{n' = \vert s'\vert}^p  \left(\delta_{nn'}\delta_{ss'}  - \langle \boldsymbol{Y}^{n's'}, \boldsymbol{Y}^{ns} \rangle_{G_q}\right)\hat{x}_{n's'}.
\end{aligned}
\end{equation}
Thus,
\begin{equation}
\begin{aligned}
\Vert I - \bmat{Q}_{qp}^*\bmat{Q}_{qp}\Vert_2 & = \left\Vert \left\{ \delta_{nn'}\delta_{ss'}  - \langle \boldsymbol{Y}^{n's'}, \boldsymbol{Y}^{ns} \rangle_{G_q} \right\}_{n,n',s,s'} \right\Vert_2 \\
& \leq \sqrt{p(p + 2)} \max_{n,n',s,s'} \left\vert \delta_{nn'}\delta_{ss'}  - \langle \boldsymbol{Y}^{n's'}, \boldsymbol{Y}^{ns} \rangle_{G_q} \right\vert\\
& \leq (p + 1) \epsilon_G.
\end{aligned}
\end{equation}

\end{proof}

In the following analysis we neglect $\epsilon_{G}$ since it is small compared to other parts of error. Note that $\Vert I - \bmat{Q}_{qp}\bmat{Q}_{qp}^* \Vert_{G_q}$ is not small.  

If we consider a bandlimited $\boldsymbol{x}$, meaning there exists $\hat{\boldsymbol{x}}\in\mathbb{C}^{2p(p + 1)}$ such that $\boldsymbol{x} = \bmat{Q}_{qp}\hat{\boldsymbol{x}}$ ($q > p$), then we have $\bmat{Q}_{qp}\bmat{Q}_{qp}^*\boldsymbol{x} =\bmat{Q}_{qp}\bmat{Q}_{qp}^* \bmat{Q}_{qp}\hat{\boldsymbol{x}} = \bmat{Q}_{qp}\hat{\boldsymbol{x}} = \boldsymbol{x}$ (i.e., $\boldsymbol{x} = \text{iDSHT}\circ \text{DSHT}\,\boldsymbol{x}$ neglecting $\epsilon_G$). However, for a random $\boldsymbol{x}$, this property does not hold. (This is unlike the discrete Fourier transform (DFT), where $\text{iDFT}\circ \text{DFT}$ is always identity.)

\section{Interpolation error analysis and parameter selection for spectral filtering}\label{appendix:parameter_filtering}

We select the parameters $p_{\max}$ and $\text{tol}_{\min}$ by analyzing the interpolation error of the surface function on the POU grid.

\subsection{The interpolation error}\label{appendix:parameter_filtering-interpolation_err}

Let $\boldsymbol{x}_m\in \mathbb{R}^{6(m - 1)^2}$ be the $x$-axis of the surface positions on the POU grid. (For simplicity, we consider only one spatial component.)
Let $x\in C^{\infty} (\mathbb{S}^2)$ be a smooth function such that its restriction to the POU grid matches the discrete values: $x\big|_{\text{POU}_m} = \boldsymbol{x}_m$.
We define $x_m\in L^2(\mathbb{S}^2)$ to be the continuous function obtained by interpolating the discrete data $\boldsymbol{x}_m$ to $\mathbb{S}^2$, i.e., $x_m = I_m \boldsymbol{x}_m$, where $I_m$ is the interpolation operator from the discretized POU parameterization space to the sphere.

We are interested in the interpolation error $\Vert x - x_m\Vert_{L^2}$. We want to smooth $\boldsymbol{x}_m$ so that $\Vert x - x_m\Vert_{L^2}$ remains small in the simulation.

We can approximate $\Vert x - x_m\Vert_{L^2}$ by considering the expansion of the interpolated spherical harmonic functions. Let $\widetilde{Y}_m^{ns}\in L^2(\mathbb{S}^2)$ be defined as $\widetilde{Y}^{ns}_m = I_m\left(Y_n^{s}\big|_{\text{POU}_m}\right)$, and
let the expansion of $\widetilde{Y}_m^{ns}$ be
\begin{equation}
\widetilde{Y}_m^{ns} = \sum_{j = 0}^{\infty}\sum_{k = -j}^j {y}^{nsjk}_{m} {Y_{j}^k}.
\end{equation}
Notice that ${y}_{m}^{nsjk} \to \delta_{nj} \delta_{sk}$ as $m \to \infty$.
Let the spherical harmonic expansion of $x$ be
\begin{equation}
x = \sum_{n = 0}^{\infty}\sum_{s = -n}^n a_{ns} Y_n^{s}.
\end{equation}
Then the spherical harmonic expansion of $x_m$ can be written as 
\begin{equation}
\begin{aligned}
x_m = \sum_{n = 0}^{\infty}\sum_{s = -n}^n a_{ns} \widetilde{Y}^{ns}_m
= \sum_{j = 0}^{\infty}\sum_{k = -j}^j \left(\sum_{n = 0}^{\infty}\sum_{s = -n}^n {y}^{nsjk}_m a_{ns} \right)  {Y_j^{k}}.
\end{aligned}
\end{equation}
Now we have
\begin{equation}
\begin{aligned}
& \Vert x - x_m \Vert_{L^2}^2\\
= & \sum_{j = 0}^{\infty} \sum_{k = -j}^j \left\vert a_{jk} - \sum_{n = 0}^{\infty}\sum_{s = -n}^{n} y^{nsjk}_m a_{ns} \right\vert^2 \\
\leq & \sum_{n = 0}^{\infty} \sum_{j = 0}^{\infty} \sum_{k = -j}^j \left\vert \sum_{s = -n}^{n} (\delta_{nj}\delta_{sk} - y_{m}^{nsjk}) a_{ns} \right\vert^2\\
\leq &  \sum_{n = 0}^{\infty} 
\left\Vert \left\{\delta_{nj}\delta_{sk} -  y_{m}^{nsjk}\right\}_{(jk),s} \right\Vert_2^2
\sum_{s = -n}^{n} \left\vert a_{ns} \right\vert^2 \\
= & \sum_{n = 0}^{\infty} \left\Vert  \Delta \boldsymbol{Y}_n^m \left\Vert_2^2  \right\Vert \boldsymbol{a}_{n} \right\Vert_2^2 .
\end{aligned}
\end{equation}
Therefore, the relative error
\begin{equation}\label{eq:eps_m_interp}
\begin{aligned}
(\epsilon_m^\text{interp})^2 = & \frac{\Vert x - x_m \Vert_{L^2}^2}{\Vert x \Vert_{L^2}^2} \\
\leq & \sum_{n = 0}^{\infty} \left\Vert  \Delta \boldsymbol{Y}_n^m \right\Vert_2^2  \frac{\left\Vert \boldsymbol{a}_{n} \right\Vert_2^2}{\sum_{j = 0}^{\infty}\left\Vert \boldsymbol{a}_{j} \right\Vert_2^2} \\
= & \sum_{n = 0}^{\infty} \left\Vert  \Delta \boldsymbol{Y}_n^m \right\Vert_2^2  c_n^2 ,
\end{aligned}
\end{equation}
where $c_n$ is the normalized $n_{th}$ mode of $x$.

\begin{figure}[htbp]
\centering
\begin{subfigure}[b]{0.45\textwidth}
  \centering
  \includegraphics[width=\textwidth]{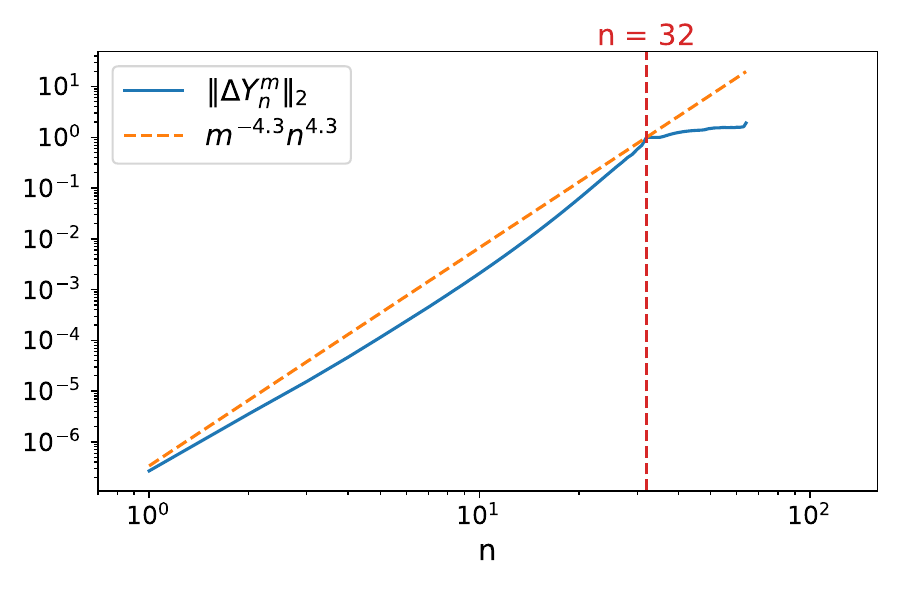}
  \caption{$m = 32$}
\end{subfigure}
\hfill
\begin{subfigure}[b]{0.45\textwidth}
  \centering
  \includegraphics[width=\textwidth]{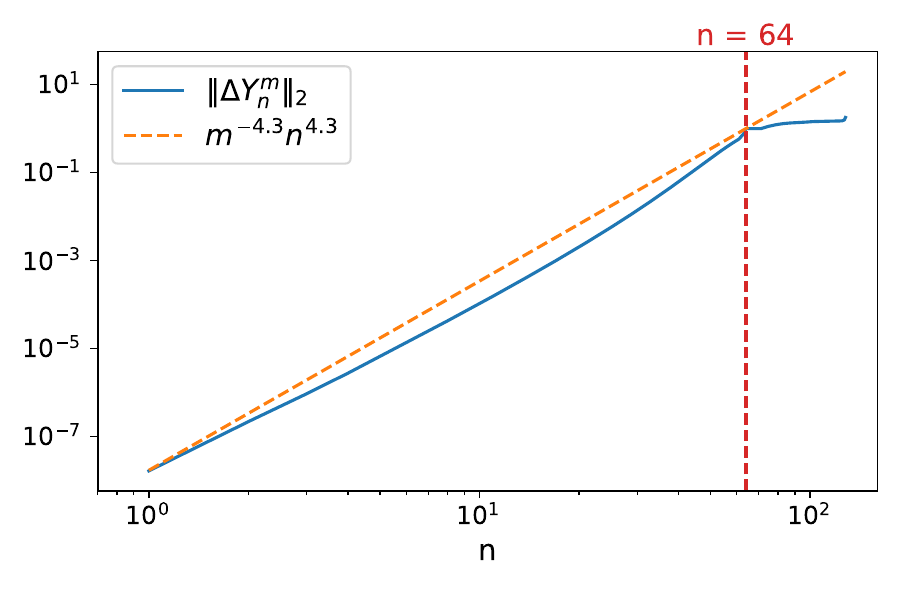}
  \caption{$m = 64$}
\end{subfigure}
\caption{The value of $\left\Vert  \Delta \boldsymbol{Y}_n^m \right\Vert_2$ for $0\leq n \leq 2m$, for $m = 32$ and $m = 64$, with $I_m$ being the cubic spline interpolation.}
\label{fig:deltyn2_spline}
\end{figure}

\begin{figure}[htbp]
\centering
\begin{subfigure}[b]{0.45\textwidth}
  \centering
  \includegraphics[width=\textwidth]{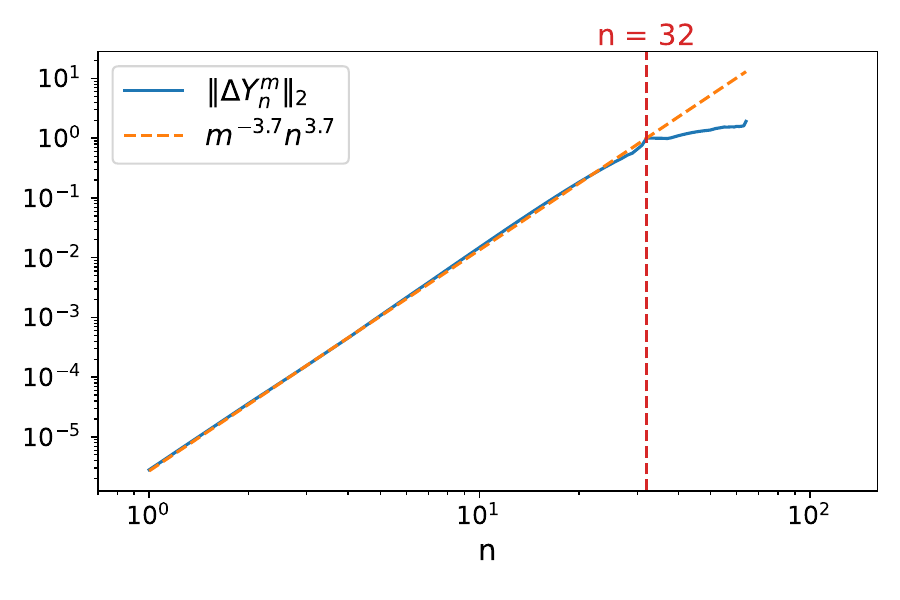}
  \caption{$m = 32$}
\end{subfigure}
\hfill
\begin{subfigure}[b]{0.45\textwidth}
  \centering
  \includegraphics[width=\textwidth]{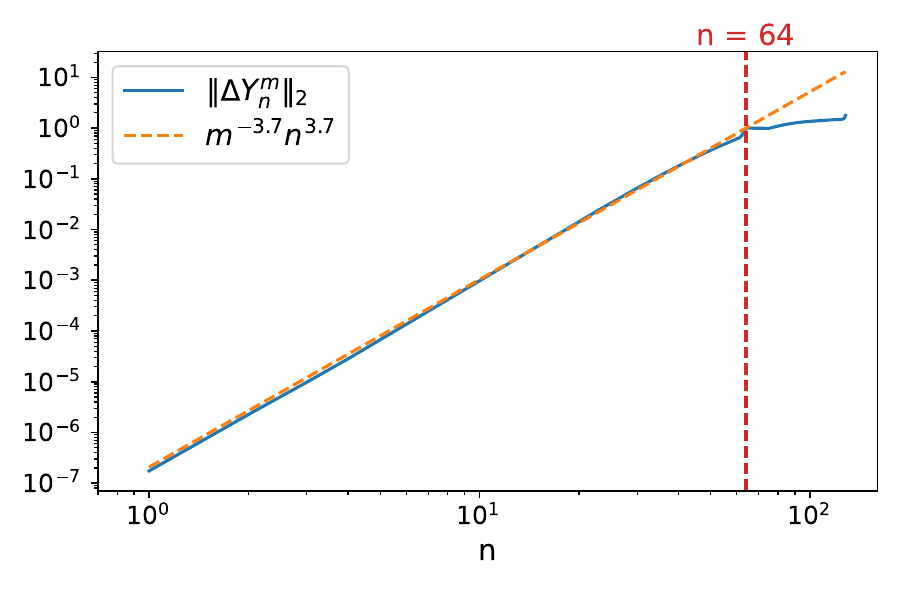}
  \caption{$m = 64$}
\end{subfigure}
\caption{The value of $\left\Vert  \Delta \boldsymbol{Y}_n^m \right\Vert_2$ for $0\leq n \leq 2m$, for $m = 32$ and $m = 64$, with $I_m$ being the cubic polynomial interpolation.}
\label{fig:deltyn2_cubic}
\end{figure}

Figure \ref{fig:deltyn2_spline} shows the value of $\left\Vert  \Delta \boldsymbol{Y}_n^m \right\Vert_2$ for $0\leq n \leq 2m$, for $m = 32$ and $m = 64$, with $I_m$ being the cubic spline interpolation. 
We can observe
\begin{equation}
\begin{aligned}
\left\Vert  \Delta \boldsymbol{Y}_n^m \right\Vert_2 & \approx m^{-k_0}n^{k_0}, \ k_0 = 4.3, \text{ for } 0\leq n \leq m, \\
\left\Vert \Delta \boldsymbol{Y}_n^m \right\Vert_2 & > 1, \text{ for } n > m.
\end{aligned}
\end{equation}
Figure \ref{fig:deltyn2_cubic} shows the value of $\left\Vert  \Delta \boldsymbol{Y}_n^m \right\Vert_2$ with $I_m$ being the cubic polynomial interpolation, in which case $k_0 = 3.7$.

\subsection{Selection of \texorpdfstring{$p_{\max}$}{} and \texorpdfstring{$\mathrm{tol}_{\min}$}{}}\label{appendix:parameter_filtering-parameter_selection}

When selecting the filtering parameters, we consider two key errors. 
The first is the interpolation error of the filtered surface, denoted by $\epsilon_{m,\, p_{\max}}^\text{interp}$.
The second is the filtering error $\epsilon^\text{tail}_{p_{\max}}$, defined in section \ref{sec:spectral_filtering}, which measures how far we deviate from the original physical problem. 
If we truncate at $p_{\max}$, the errors can be estimated as
\begin{equation}
\begin{aligned}
& \epsilon^\text{interp}_{m,\, p_{\max}} \leq \sqrt{ \sum_{n = 0}^{p_{\max}}  \left\Vert  \Delta \boldsymbol{Y}_n^m \right\Vert_2^2c_n^2},\\
& \epsilon^\text{tail}_{p_{\max}} \leq \sqrt{ \sum_{n = p_{\max} + 1}^{p}  c_n^2}.
\end{aligned}
\end{equation}
We always set 
\begin{equation}
p_{\max} \leq m,
\end{equation}
since $\left\Vert  \Delta \boldsymbol{Y}_n^m \right\Vert_2 > 1$ when $n > m$ and we want to avoid amplifying the influence of high frequencies in the interpolation error.

\begin{figure}[htbp]
\centering
\includegraphics[width=0.5\textwidth]{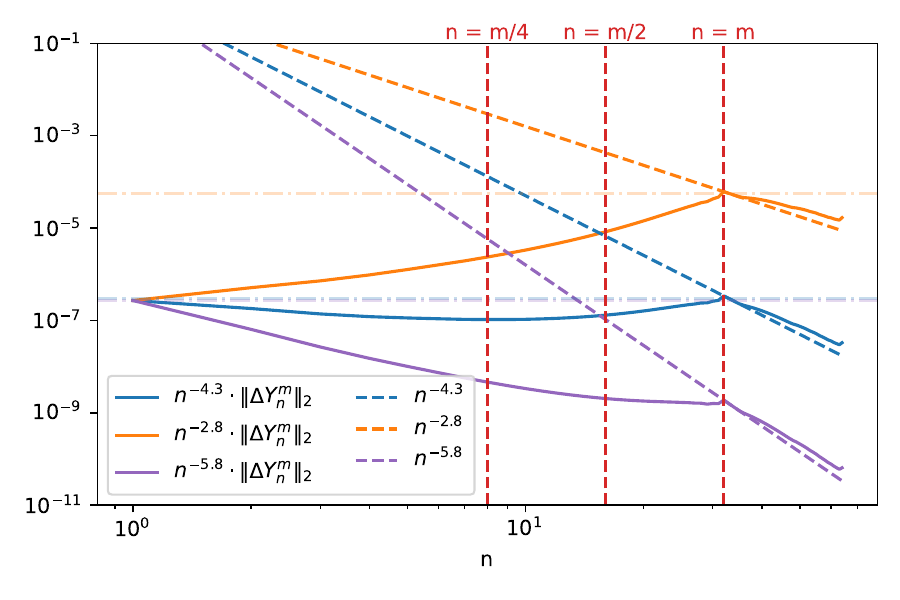}
\caption{The plot of $\left\Vert  \Delta \boldsymbol{Y}_n^m \right\Vert_2 n^{-k}$ (solid lines) and $n^{-k}$ (dashed lines), for $k<k_0$, $k = k_0$, and $k > k_0$, where $m = 32$, $k_0 = 4.3$, and $I_m$ is cubic spline interpolation.}
\label{fig:deltyn2_cn_lge_spline}
\end{figure}

Now we further analyze $\epsilon^\text{interp}_{m,\, p_{\max}}$ and $\epsilon^\text{tail}_{p_{\max}}$ by assuming a specific decay in $c_n$.
Assume
\begin{equation}
c_n\propto n^{-k}, \ k\geq 0.
\end{equation}
We can then approximate
\begin{equation}
\begin{aligned}
& \epsilon^\text{interp}_{m,\, p_{\max}} \sim  \max_{n \leq p_{\max}} m^{-k_0} n^{k_0 - k}\\
& \epsilon^\text{tail}_{p_{\max}} \sim p_{\max}^{-k}.
\end{aligned}
\end{equation}
Figure \ref{fig:deltyn2_cn_lge_spline} shows the plot of $\left\Vert  \Delta \boldsymbol{Y}_n^m \right\Vert_2n^{-k}$ and $n^{-k}$, for $k<k_0$, $k = k_0$, and $k > k_0$.

We observe that for $k\geq k_0$, $\epsilon^\text{interp}_{m,\, p_{\max}} \sim \left\Vert  \Delta \boldsymbol{Y}_1^m \right\Vert_2 \approx  m^{-k_0}$, which is independent of the value of $p_{\max}$. However, $\epsilon^\text{tail}_{p_{\max}} \sim p_{\max}^{-k}$, which increases as $p_{\max}$ decreases. In this case, truncation is not advantageous.

For $k < k_0$, $\epsilon^\text{interp}_{m,\,p_{\max}} \sim m^{-k_0} p_{\max}^{k_0 - k}$ and $\epsilon^\text{tail}_{p_{\max}} \sim p_{\max}^{-k}$. If we truncate at $p_{\max} = m$, both $\epsilon^\text{interp}_{m,\,p_{\max}}$ and $\epsilon^\text{tail}_{p_{\max}}$ are $\mathcal{O}(m^{-k})$. When we further reduce $p_{\max}$, $\epsilon^\text{interp}_{m,\,p_{\max}}$ decreases while $\epsilon^\text{tail}_{p_{\max}}$ increases. 

Therefore, for any $p_{\max} \leq m$ we select, we aim to avoid cases where $c_n$ decays faster than $n^{-k_0}$, based on which we set 
\begin{equation}
\text{tol}_{\min} = \frac{1}{\sqrt{2k_0 - 1}}{p_{\max}}^{-\frac{2k_0 - 1}{2}}, 
\end{equation}
which approximates $\epsilon^\text{tail}_{p_{\max}}$ for $c_n \approx n^{-k_0}$.

For the selection of $p_{\max}$, we can set 
\begin{equation}
p_{\max} = m,
\end{equation}
to balance $\epsilon^\text{tail}_{p_{\max}} \sim \epsilon^\text{interp}_{m,\,p_{\max}}$, meaning the deviation from the original problem is no greater than the interpolation error.
Alternatively, more aggressive truncation ($p_{\max} < m$) can be used to further reduce $\epsilon^\text{interp}_{m,\,p_{\max}}$, at the cost of increasing the deviation from the original problem.

\subsection{Aggressive filtering}\label{appendix:parameter_filtering-aggressive_filtering}

As discussed above, we typically set $p_{\max} = m$ when applying the spectral filtering method to smooth the surface function. However, when accurate representation of surface derivatives is required, especially in the presence of high-curvature regions during capsule deformation, more aggressive truncation can be necessary.

Similarly, we can estimate the error of the $r_\text{th}$ derivative of the interpolated surface function by 
\begin{equation}\label{eq:eps_m_interp_deriv}
(\epsilon_m^{\text{interp}, (r)})^2 
=  \frac{\Vert D^{(r)} x - D^{(r)}  x_m \Vert_{L^2}^2}{\Vert D^{(r)} x \Vert_{L^2}^2}
\leq  \sum_{n = 0}^{\infty} \left\Vert  \Delta \boldsymbol{Y}_n^{m,(r)} \right\Vert_2^2  c_n^2, 
\end{equation}
in which 
\begin{equation}
 \Delta \boldsymbol{Y}_n^{m,(r)}  =
 \left\{y^{nsjk, (r)} -  y_{m}^{nsjk, (r)}\right\}_{(jk),s}, 
\end{equation}
where $y^{nsjk, (r)}$ denotes the $(j, k)$ mode of the $r_\text{th}$ derivative $D^{(r)}Y_n^{s}$, and $y_m^{nsjk, (r)}$ denotes the $(j, k)$ mode of $D^{(r)}\widetilde{Y}_m^{ns} = D^{(r)}(Y_n^{s}\big|_{\text{POU}_m})$. And $c_n$ is the normalized $n_{th}$ mode of $x$.

In particular, for second derivatives, we observe that
\begin{equation}
\left\Vert \Delta \boldsymbol{Y}_n^{m,(2)} \right\Vert_2\approx m^{-4.3}n^{6.3}, \text{ for } 0\leq n \leq m,
\end{equation}
with cubic spline interpolation.
To control the interpolation error $\epsilon_m^{\text{interp}, (2)}$, we choose a more aggressive truncation threshold, setting
\begin{equation}
p_{\max} = m^{0.68}.
\end{equation}

\section{Strain energy functional and its properties}\label{appendix:strain_energy_functional}

\subsection{Functional form of the strain energy and relation to interfacial force}\label{appendix:strain_energy_functional-force}

Let 
\begin{equation}
    e(\boldsymbol{x}) = 
    \frac{\Es}{4} \left[
    \frac{1}{2}I_1(\boldsymbol{x})^2 + I_1(\boldsymbol{x}) - I_2(\boldsymbol{x})
    \right]
    + \frac{\Ed}{8} I_2(\boldsymbol{x})^2,
\end{equation}
then the strain energy can be written as
\begin{equation}
    \mathcal{E} = \int_{\gamma} \frac{1}{J(\boldsymbol{x})} e(\boldsymbol{x})\, \mathrm{d} S 
    = \int_{[0, \pi]^2}  \frac{1}{J\left(\boldsymbol{x}(\boldsymbol{u})\right)} e\left(\boldsymbol{x}(\boldsymbol{u})\right) W(\boldsymbol{u}) \, \mathrm{d} \boldsymbol{u},
\end{equation}
where $\boldsymbol{u} = (u, v)$ is the parameterization of $\gamma$, and $W(\boldsymbol{u})$ is the surface area element.
$\mathcal{E}$ can be considered as a functional
\begin{equation}
    \mathcal{E}[\boldsymbol{x}] = \int_{[0, \pi]^2}  f_{\mathcal{E}}(\boldsymbol{u}, \boldsymbol{x}(\boldsymbol{u}), \bnabla_{\boldsymbol{u}}\boldsymbol{x}(\boldsymbol{u})) \, \mathrm{d} \boldsymbol{u},
\end{equation}
where $\bnabla_{\boldsymbol{u}}\boldsymbol{x} = (\boldsymbol{x}_u, \boldsymbol{x}_v)$, and
$f_{\mathcal{E}}(\boldsymbol{u}, \boldsymbol{x}(\boldsymbol{u}), \bnabla_{\boldsymbol{u}}\boldsymbol{x}(\boldsymbol{u})) = \frac{1}{J\left(\boldsymbol{x}(\boldsymbol{u})\right)} e\left(\boldsymbol{x}(\boldsymbol{u})\right) W(\boldsymbol{u})$.
The interfacial force is given by 
\begin{equation}
    \boldsymbol{f} = \bnabla_{\gamma} \bcdot \bLambda (\boldsymbol{x}),
\end{equation}
in which $\bLambda$ is defined in equation \eqref{eq:stress-tensor-nondim}

\begin{proposition}
\begin{equation}
\boldsymbol{f} = - \frac{\delta \mathcal{E}}{\delta \boldsymbol{x}} \,\frac{1}{W},
\end{equation}
i.e.,
\begin{equation}
\delta \mathcal{E} =  - \int_{[0, \pi]^2} \boldsymbol{f} \cdot {\delta \boldsymbol{x}}\, W \mathrm{d} \boldsymbol{u} =  - \int_{\gamma} \boldsymbol{f} \cdot {\delta \boldsymbol{x}} \, \mathrm{d} S
\end{equation}
i.e., $\boldsymbol{f}$ acts to minimize the strain energy $\mathcal{E}$.
\end{proposition}

\begin{proof}
\begin{equation}
J = \frac{W}{W_r},
\end{equation}
where $W_r$ is the surface area element of the reference configuration $\gamma_r$. Then
\begin{equation}
f_{\mathcal{E}} = \frac{1}{J} e W = e W_r,
\end{equation}
and
\begin{equation}
\begin{aligned}
\frac{\delta \mathcal{E}}{\delta \boldsymbol{x}} =  \frac{\partial f_{\mathcal{E}}}{\partial \boldsymbol{x}} - \bnabla_{\boldsymbol{u}} \bcdot \frac{\partial f_{\mathcal{E}}}{\partial \bnabla_{\boldsymbol{u}} \boldsymbol{x}} =  W_r  \frac{\partial e}{\partial \boldsymbol{x}} - \bnabla_{\boldsymbol{u}} \bcdot \left( W_r \frac{\partial e}{\partial \bnabla_{\boldsymbol{u}} \boldsymbol{x}} \right).
\end{aligned}
\end{equation}
We have
\begin{equation}
\begin{aligned}
I_1 & = \mathrm{tr} (\bmat{F}_\mathrm{S}\bmat{F}_\mathrm{S}^\T) - 2
= \frac{1}{W_r^2}(G_r E - 2F_r F + E_r G) - 2, \\
I_2 & = \det (\bmat{F}_\mathrm{S}\bmat{F}_\mathrm{S}^\T) - 1
= \frac{W^2}{W_r^2} - 1,
\end{aligned}
\end{equation}
where $E$, $F$, $G$ are coefficients of the first fundamental form of $\gamma$, while $E_r$, $F_r$, $G_r$ are those of $\gamma_r$. This gives
\begin{equation}
\frac{\partial e}{\partial \boldsymbol{x}}  = 0,
\end{equation}
and
\begin{equation}
\begin{aligned}
\frac{\partial e}{\partial \bnabla_{\boldsymbol{u}} \boldsymbol{x}} & =
\frac{\partial e}{\partial I_1} \frac{I_1}{\partial \bnabla_{\boldsymbol{u}} \boldsymbol{x}} + \frac{\partial e}{\partial I_2} \frac{I_2}{\partial \bnabla_{\boldsymbol{u}} \boldsymbol{x}} \\
& = \frac{E_s(I_1 + 1)}{2} \frac{1}{W_r^2} \bnabla_{\boldsymbol{u}} \boldsymbol{x}\begin{bmatrix}G_r & -F_r \\- F_r& E_r\end{bmatrix}
+ \frac{E_d I_2 - E_s}{2} \frac{1}{W_r^2} \bnabla_{\boldsymbol{u}} \boldsymbol{x} \begin{bmatrix}G & -F \\- F& E\end{bmatrix}. \\
& = \frac{E_s(I_1 + 1)}{2W}  \Fs\Fs^\T \frac{1}{W} \bnabla_{\boldsymbol{u}} \boldsymbol{x} \begin{bmatrix}G & -F \\- F& E\end{bmatrix}
+ \frac{(E_d I_2 - E_s)W}{2W_r^2} \frac{1}{W} \bnabla_{\boldsymbol{u}} \boldsymbol{x} \begin{bmatrix}G & -F \\- F& E\end{bmatrix}.
\end{aligned}
\end{equation}
Let 
\begin{equation}
\bmat{P} = \frac{1}{W} \bnabla_{\boldsymbol{u}} \boldsymbol{x} \begin{bmatrix}G & -F \\- F& E\end{bmatrix},
\end{equation}
and we can derive
\begin{equation}
\begin{aligned}
\bnabla_{\boldsymbol{u}} \bcdot \bmat{P} & = W \Delta_{\gamma} \boldsymbol{x},\\
\bnabla_{\boldsymbol{u}} \bmat{A} \bvec{\colon} \bmat{P} & = W \bnabla_{\gamma} \bcdot \bmat{A}, \ \forall \bmat{A} \in \mathbb{R}^{3\times 3}, \\
\bmat{P} (\bnabla_{\boldsymbol{u}} a ) & = W \bnabla_{\gamma} a, \ \forall a\in \mathbb{R}.
\end{aligned}
\end{equation}
Therefore,
\begin{equation}
\begin{aligned}
- \frac{\delta \mathcal{E}}{\delta \boldsymbol{x}} = \bnabla_{\boldsymbol{u}} \bcdot\left( W_r \frac{\partial e}{\partial \bnabla_{\boldsymbol{u}} \boldsymbol{x}}\right) & =  \frac{E_s(I_1 + 1)W}{2J}  \Fs\Fs^\T \Delta_{\gamma} \boldsymbol{x} + W \bnabla_{\gamma}\bcdot \left(\frac{E_s(I_1 + 1)}{2J} \Fs\Fs^T \right)  \\
& \quad + \frac{(E_d I_2 - E_s)JW}{2} \Delta_{\gamma} \boldsymbol{x} + W \bnabla_{\gamma} \left(\frac{(E_d I_2 - E_s)J}{2}  \right) \\
& = W \bnabla_{\gamma} \bcdot \left( \frac{E_s(I_1 + 1)}{2J} \Fs\Fs^T + \frac{(E_d I_2 - E_s)J}{2} (\mathbf{I} - \boldsymbol{n}\boldsymbol{n}^T) \right)\\
& = W \boldsymbol{f}.
\end{aligned}
\end{equation}

\end{proof}

\subsection{Conditions for zero strain energy}\label{appendix:strain_energy_functional-zero_condition}

\begin{proposition}
\begin{equation}
\mathcal{E} \geq 0
\end{equation}
and
\begin{equation}
\mathcal{E} = 0 \iff I_1(\boldsymbol{x}) = I_2(\boldsymbol{x}) = 0, \text{ a.e. on } \gamma.
\end{equation}
\end{proposition}

\begin{proof}
$\impliedby$ is trivial.

$\implies$: We have
\begin{equation}
e = (I_1 + 1)^2 + 10(I_2 - \frac{1}{10})^2 - \frac{11}{10},
\end{equation}
and
\begin{equation}
I_2 + 1 \leq \frac{(I_1 + 2)^2}{4}.
\end{equation}
Consider the ellipse 
\begin{equation}
\frac{(I_1 + 1)^2}{\frac{11}{10}} + \frac{(I_2 - \frac{1}{10})^2}{\frac{11}{100}} = 1,
\end{equation}
and the parabola 
\begin{equation}
I_2 = \frac{(I_1 + 2)^2}{4} - 1.
\end{equation}
Their unique intersection is a tangency at $(I_1, I_2) = (0, 0)$. Therefore, for any feasible $(I_1, I_2)$ with $I_1 \neq 0$ or $I_2 \neq 0$, we have $e > 0$.

\end{proof}

\subsection{Non-convexity and non-unique minimizers}\label{appendix:strain_energy_functional-non_convexity}

\begin{proposition}
$\mathcal{E}[\boldsymbol{x}]$ is not convex.
\end{proposition}

\begin{proof}
Let $\boldsymbol{x}_0$ be the reference configuration, and $\bmat{Q}$ a rigid transform, then 
\begin{equation}
\mathcal{E}[\boldsymbol{x}_0] = \mathcal{E}[\bmat{Q}\boldsymbol{x}_0] = 0.
\end{equation}
It is easy to find a case in which $\mathcal{E}[\frac{1}{2} \boldsymbol{x}_0 + \frac{1}{2} \bmat{Q}\boldsymbol{x}_0] > 0$. Then we have
\begin{equation}
\mathcal{E}[\frac{1}{2} \boldsymbol{x}_0 + \frac{1}{2} \bmat{Q}\boldsymbol{x}_0] > 0 = \frac{1}{2} \mathcal{E}[\boldsymbol{x}_0] + \frac{1}{2} \mathcal{E}[\bmat{Q}\boldsymbol{x}_0].
\end{equation}
This violates the convexity inequality, proving $\mathcal{E}$ is non-convex.
\end{proof}

\begin{proposition}
If $\boldsymbol{x}_{\min}$ is a (global or local) minimizer of $\mathcal{E}[\boldsymbol{x}]$, $\boldsymbol{x}_{\min}\in {C}^{\infty}$, then there exists $\boldsymbol{x}_{\min}'$, $\boldsymbol{x}_{\min}' \in {C}^{1}$, $\mathcal{E}[\boldsymbol{x}_{\min}] = \mathcal{E}[\boldsymbol{x}_{\min}']$, and $\boldsymbol{x}_{\min}$ cannot be transformed to $\boldsymbol{x}_{\min}'$ by a rigid transform.
\end{proposition} 
\begin{proof}
$\boldsymbol{x}_{\min}: \gamma_{\min} \to \mathbb{R}^3$ can be considered as a smooth isometric embedding from a $2$-dimensional Riemannian manifold $(\gamma_{\min},\, g)$ to Euclidean space $\mathbb{R}^3$. Let 
\begin{equation}
\boldsymbol{y} = (1 - \epsilon) \boldsymbol{x}_{\min},
\end{equation}
where $0 < \epsilon < 1$, then $\boldsymbol{y}: \gamma_{\min} \to \mathbb{R}^3$ is a smooth \emph{short} embedding (i.e., a smooth embedding that never increases distances) into $\mathbb{R}^3$. Let $\gamma_{\min} = \partial \Omega$, then we can assume the origin $O \in \Omega$ and select $\epsilon$ properly to make $\boldsymbol{y}\subset \Omega$

By the Nash--Kuiper theorem \citep{MR75640,nash1954c}, there exists a sequence of continuously differentiable isometric embeddings $\left\{\boldsymbol{y}_n\right\}_n$ which uniformly converges to $\boldsymbol{y}$. Therefore, there exists $N > 0$, s.t. 
\begin{equation}
\left\Vert \boldsymbol{y} - \boldsymbol{y}_N \right\Vert_{L^{\infty}} < \frac{1}{2} \epsilon \left\Vert \boldsymbol{x}_{\min} \right\Vert_{L^{\infty}}, \text{ and } \boldsymbol{y}_N \subset \Omega
\end{equation} 
Let $\boldsymbol{x}_{\min}' = \boldsymbol{y}_N$, then 
\begin{equation}
\mathcal{E}[\boldsymbol{x}_{\min}] = \mathcal{E}[\boldsymbol{x}_{\min}'], 
\end{equation}
and 
\begin{equation}
\begin{aligned}
\left\Vert \boldsymbol{x}_{\min} - \boldsymbol{x}_{\min}' \right\Vert_{L^{\infty}}
> &  \left\Vert \boldsymbol{x}_{\min} - \boldsymbol{y} \right\Vert_{L^{\infty}}
- \left\Vert \boldsymbol{y} - \boldsymbol{x}_{\min}' \right\Vert_{L^{\infty}} \\
> & \epsilon\left\Vert \boldsymbol{x}_{\min} \right\Vert_{L^{\infty}} -  \frac{1}{2}\epsilon\left\Vert \boldsymbol{x}_{\min} \right\Vert_{L^{\infty}}\\
= & \frac{1}{2}\epsilon\left\Vert \boldsymbol{x}_{\min} \right\Vert_{L^{\infty}}.
\end{aligned}
\end{equation} 
Combined with $\boldsymbol{x}_{\min}' \subset \Omega$, this shows $\boldsymbol{x}_{\min}$ cannot be transformed to $\boldsymbol{x}_{\min}'$ by a rigid transform.
\end{proof}
Although we have shown that the strain energy functional admits multiple global minimizers that are not related by rigid transforms, we have not established whether the minimizer remains non-unique under the additional constraint of volume preservation. Our numerical simulations enforce incompressible flow, and therefore the enclosed volume is always preserved. Whether multiple global minimizers can coexist under this constraint remains an open question.

\section{Supplementary stability and convergence tables}
\label{appendix:stable_conv_tbls}

Tables \ref{tab:relax_stability}--\ref{tab:relax_vloss} show the stability and convergence results for unconfined relaxation flow, while Tables \ref{tab:pois_stability}--\ref{tab:pois_vloss} show those for  Poiseuille flow through a pipe. 

\begin{table}[htbp]
\centering
\setlength{\aboverulesep}{1pt}
\setlength{\belowrulesep}{1pt}
\setlength{\tabcolsep}{4pt}
\small
\begin{tabular}{rrrrrrrrrrrrrr}
\toprule
\multirow{2}{*}{$\rvr$} & \multirow{2}{*}{method} & \multicolumn{4}{c}{$m = 24$} & \multicolumn{4}{c}{$m = 48$} & \multicolumn{4}{c}{$m = 96$} \\
\cmidrule(lr){3-6} \cmidrule(lr){7-10} \cmidrule(lr){11-14}
 & & $\Ca = 0.125$ & $0.25$ & $0.5$ & $1$ & $\Ca = 0.125$ & $0.25$ & $0.5$ & $1$ & $\Ca = 0.125$ & $0.25$ & $0.5$ & $1$ \\
\midrule
$1$&none&51&51&51&51&51&51&51&51&51&51&51&51\\
    &filter&51&51&51&51&51&51&51&51&51&51&51&51\\
    &reparam&51&51&51&51&51&51&51&51&51&51&51&51\\
\midrule
$0.89$&none&51&51&51&51&51&51&51&1&51&51&51&0\\
    &filter&51&51&51&51&51&51&51&51&51&51&51&51\\
    &reparam&51&51&51&2&51&51&51&1&51&51&18&0\\
\midrule
$0.78$&none&51&51&51&51&51&51&51&1&51&51&51&0\\
    &filter&51&51&51&51&51&51&51&51&51&51&51&51\\
    &reparam&51&51&51&2&51&51&51&1&51&51&2&0\\
\midrule
$0.64$&none&51&51&51&12&51&51&51&1&51&51&51&0\\
    &filter&51&51&51&51&51&51&51&51&51&51&51&51\\
    &reparam&51&51&51&1&51&51&51&0&51&51&51&0\\
\bottomrule
\end{tabular}
\caption{Maximum time the simulation remains stable (up to final time $T = 51$) for single biconcave capsule in relaxation flow over the interval $[1, 51]$, following shear flow in $[0, 1]$ with varying reduced volume $\rvr$, capillary number $\Ca$, capsule resolution $m$, and simulation method. Three methods are compared: unfiltered (none), spectral filtering (filter), and reparametrization (reparam). A value of 51 indicates full stability throughout the entire simulation window.}\label{tab:relax_stability}
\end{table}

\begin{table}[htbp]
\centering
\setlength{\aboverulesep}{1pt}
\setlength{\belowrulesep}{1pt}
\setlength{\tabcolsep}{2.5pt}
\small
\begin{tabular}{cccc>{\columncolor[gray]{0.9}}c>{\columncolor[gray]{0.9}}ccc>{\columncolor[gray]{0.9}}c>{\columncolor[gray]{0.9}}c}
\toprule
\multirow{2}{*}{$\rvr$} & \multirow{2}{*}{QoI} & \multicolumn{2}{c}{$\Ca = 0.125$} & \multicolumn{2}{c}{$\Ca = 0.25$} & \multicolumn{2}{c}{$\Ca = 0.5$} & \multicolumn{2}{c}{$\Ca = 1$} \\
\cmidrule(lr){3-4} \cmidrule(lr){5-6} \cmidrule(lr){7-8} \cmidrule(lr){9-10}
&& $m = 24$ & $m = 48$ & $m = 24$ & $m = 48$ & $m = 24$ & $m = 48$ & $m = 24$ & $m = 48$ \\
\midrule
$1$&Pos&$1.7\times 10^{-4}$&$1.2\times 10^{-4}$&$1.7\times 10^{-4}$&$1.2\times 10^{-4}$&$1.7\times 10^{-4}$&$1.0\times 10^{-4}$&$1.9\times 10^{-4}$&$8.3\times 10^{-5}$\\
    &Asp&$6.9\times 10^{-7}$&$6.4\times 10^{-7}$&$1.3\times 10^{-6}$&$1.3\times 10^{-6}$&$2.3\times 10^{-6}$&$1.4\times 10^{-6}$&$3.2\times 10^{-6}$&$1.4\times 10^{-6}$\\
    &Inc&$7.1\times 10^{-3}$&$1.5\times 10^{-3}$&$3.9\times 10^{-3}$&$8.2\times 10^{-4}$&$2.5\times 10^{-3}$&$5.7\times 10^{-4}$&$1.9\times 10^{-3}$&$3.6\times 10^{-4}$\\
\midrule
$0.89$&Pos&$2.1\times 10^{-3}$&$2.7\times 10^{-4}$&$2.1\times 10^{-3}$&$2.7\times 10^{-4}$&$2.1\times 10^{-3}$&$9.7\times 10^{-4}$&$7.9\times 10^{-3}$&$7.9\times 10^{-3}$\\
    &Asp&$1.8\times 10^{-4}$&$2.5\times 10^{-5}$&$1.8\times 10^{-4}$&$2.6\times 10^{-5}$&$1.9\times 10^{-4}$&$2.6\times 10^{-5}$&$2.2\times 10^{-4}$&$6.6\times 10^{-5}$\\
    &Inc&$2.9\times 10^{-2}$&$6.0\times 10^{-3}$&$1.5\times 10^{-2}$&$3.0\times 10^{-3}$&$9.1\times 10^{-3}$&$2.0\times 10^{-3}$&$1.2\times 10^{-2}$&$3.6\times 10^{-3}$\\
\midrule
$0.78$&Pos&$1.8\times 10^{-3}$&$2.7\times 10^{-4}$&$1.7\times 10^{-3}$&$3.0\times 10^{-4}$&$3.1\times 10^{-3}$&$1.4\times 10^{-3}$&$2.0\times 10^{-2}$&$1.3\times 10^{-2}$\\
    &Asp&$2.0\times 10^{-4}$&$2.9\times 10^{-5}$&$1.9\times 10^{-4}$&$2.8\times 10^{-5}$&$1.8\times 10^{-4}$&$2.9\times 10^{-5}$&$2.8\times 10^{-4}$&$6.0\times 10^{-5}$\\
    &Inc&$1.2\times 10^{-2}$&$2.4\times 10^{-3}$&$5.5\times 10^{-3}$&$1.1\times 10^{-3}$&$2.3\times 10^{-3}$&$5.7\times 10^{-4}$&$4.6\times 10^{-3}$&$1.8\times 10^{-3}$\\
\midrule
$0.64$&Pos&$2.8\times 10^{-3}$&$4.9\times 10^{-4}$&$4.0\times 10^{-3}$&$6.4\times 10^{-4}$&$1.2\times 10^{-2}$&$3.4\times 10^{-3}$&$3.8\times 10^{-2}$&$1.6\times 10^{-2}$\\
    &Asp&$2.2\times 10^{-4}$&$3.2\times 10^{-5}$&$2.2\times 10^{-4}$&$3.0\times 10^{-5}$&$3.3\times 10^{-4}$&$6.3\times 10^{-5}$&$2.5\times 10^{-4}$&$2.7\times 10^{-4}$\\
    &Inc&$8.7\times 10^{-3}$&$1.6\times 10^{-3}$&$3.8\times 10^{-3}$&$7.0\times 10^{-4}$&$1.5\times 10^{-3}$&$6.2\times 10^{-4}$&$7.2\times 10^{-4}$&$3.3\times 10^{-4}$\\
\bottomrule
\end{tabular}
\caption{Relative errors at final time $T = 51$ for single biconcave capsule in relaxation flow with spectral filtering. Relative errors are reported for position (Pos), Taylor asphericity (Asp), and inclination angle (Inc), for different reduced volumes $\rvr$ and capillary numbers $\Ca$. The reference solution is computed with resolution $m = 96$.}\label{tab:relax_conv}
\end{table}

\begin{table}[htbp]
\centering
\setlength{\aboverulesep}{1pt}
\setlength{\belowrulesep}{1pt}
\setlength{\tabcolsep}{4pt}
\small
\begin{tabular}{cccc>{\columncolor[gray]{0.9}}c>{\columncolor[gray]{0.9}}c>{\columncolor[gray]{0.9}}c}
\toprule
\multirow{2}{*}{$\rvr$} & \multicolumn{3}{c}{$\Ca = 0.125$} & \multicolumn{3}{c}{$\Ca = 0.25$} \\
\cmidrule(lr){2-4} \cmidrule(lr){5-7} 
 & $m = 24$ & $m = 48$ & $m = 96$ & $m = 24$ & $m = 48$ & $m = 96$ \\
\midrule
$1$&$2.1\times 10^{-4}$&$3.0\times 10^{-4}$&$3.9\times 10^{-6}$&$2.2\times 10^{-4}$&$3.0\times 10^{-4}$&$4.0\times 10^{-6}$\\
\midrule
$0.89$&$1.1\times 10^{-3}$&$1.6\times 10^{-4}$&$2.0\times 10^{-5}$&$1.1\times 10^{-3}$&$1.6\times 10^{-4}$&$2.1\times 10^{-5}$\\
\midrule
$0.78$&$2.0\times 10^{-3}$&$2.9\times 10^{-4}$&$3.8\times 10^{-5}$&$1.9\times 10^{-3}$&$2.8\times 10^{-4}$&$3.7\times 10^{-5}$\\
\midrule
$0.64$&$3.3\times 10^{-3}$&$4.8\times 10^{-4}$&$6.5\times 10^{-5}$&$3.2\times 10^{-3}$&$4.5\times 10^{-4}$&$6.2\times 10^{-5}$\\
\bottomrule
\end{tabular}
\begin{tabular}{cccc>{\columncolor[gray]{0.9}}c>{\columncolor[gray]{0.9}}c>{\columncolor[gray]{0.9}}c}
\toprule
\multirow{2}{*}{$\rvr$}  & \multicolumn{3}{c}{$\Ca = 0.5$} & \multicolumn{3}{c}{$\Ca = 1$} \\
\cmidrule(lr){2-4} \cmidrule(lr){5-7} 
 & $m = 24$ & $m = 48$ & $m = 96$ & $m = 24$ & $m = 48$ & $m = 96$ \\
\midrule
$1$&$2.3\times 10^{-4}$&$2.7\times 10^{-4}$&$4.2\times 10^{-6}$&$2.5\times 10^{-4}$&$2.0\times 10^{-4}$&$4.5\times 10^{-6}$\\
\midrule
$0.89$&$1.1\times 10^{-3}$&$1.6\times 10^{-4}$&$2.8\times 10^{-5}$&$1.2\times 10^{-3}$&$3.3\times 10^{-4}$&$2.9\times 10^{-5}$\\
\midrule
$0.78$&$1.8\times 10^{-3}$&$2.5\times 10^{-4}$&$3.9\times 10^{-5}$&$2.8\times 10^{-3}$&$4.7\times 10^{-4}$&$6.4\times 10^{-5}$\\
\midrule
$0.64$&$3.7\times 10^{-3}$&$5.2\times 10^{-4}$&$5.3\times 10^{-5}$&$4.7\times 10^{-3}$&$6.7\times 10^{-4}$&$1.2\times 10^{-4}$\\
\bottomrule
\end{tabular}
\caption{Relative volume loss at final time $T = 51$ for single biconcave capsule in relaxation flow with spectral filtering, for different reduced volumes $\rvr$ and capillary numbers $\Ca$.}\label{tab:relax_vloss}
\end{table}

\begin{table}[htbp]
\centering
\setlength{\aboverulesep}{1pt}
\setlength{\belowrulesep}{1pt}
\setlength{\tabcolsep}{4pt}
\small
\begin{tabular}{rrrrrrrrrrr}
\toprule
\multirow{2}{*}{$\rvr$} & \multirow{2}{*}{method} & \multicolumn{3}{c}{$m = 24$} & \multicolumn{3}{c}{$m = 48$} & \multicolumn{3}{c}{$m = 96$} \\
\cmidrule(lr){3-5} \cmidrule(lr){6-8} \cmidrule(lr){9-11}
 & & $\Ca = 0.15$ & $0.3$ & $0.6$  & $\Ca = 0.15$ & $0.3$ & $0.6$ & $\Ca = 0.15$ & $0.3$ & $0.6$ \\
\midrule
$1$&none&100&100&100&100&100&100&100&100&100\\
    &filter&100&100&100&100&100&100&100&100&100\\
    &agg-filter&100&100&100&100&100&100&100&100&100\\
\midrule
$0.89$&none&100&100&69&100&100&100&100&100&100\\
    &filter&100&100&74&100&100&100&100&100&100\\
    &agg-filter&100&100&100&100&100&100&100&100&100\\
\midrule
$0.78$&none&71&35&23&100&60&31&100&100&5\\
    &filter&64&37&22&87&54&33&100&88&46\\
    &agg-filter&100&81&59&100&100&100&100&100&100\\
\bottomrule
\end{tabular}
\caption{Maximum time the simulation remains stable (up to final time $T = 100$) for single biconcave capsule in Poiseuille flow through a pipe, with varying reduced volume $\rvr$, capillary number $\Ca$, capsule resolution $m$, and simulation method. Three methods are compared: unfiltered (none), spectral filtering (filter), and aggressive spectral filtering (agg-filter). A value of 100 indicates full stability throughout the entire simulation window.}\label{tab:pois_stability}
\end{table}

\begin{table}[htbp]
\centering
\setlength{\aboverulesep}{1pt}
\setlength{\belowrulesep}{1pt}
\setlength{\tabcolsep}{4pt}
\small
\begin{tabular}{cc>{\columncolor[gray]{0.9}}c>{\columncolor[gray]{0.9}}ccc>{\columncolor[gray]{0.9}}c>{\columncolor[gray]{0.9}}c}
\toprule
\multirow{2}{*}{$\rvr$} & \multirow{2}{*}{QoI} & \multicolumn{2}{c}{$\Ca = 0.15$} & \multicolumn{2}{c}{$\Ca = 0.3$} & \multicolumn{2}{c}{$\Ca = 0.6$}  \\
\cmidrule(lr){3-4} \cmidrule(lr){5-6} \cmidrule(lr){7-8} 
&& $m = 24$ & $m = 48$ & $m = 24$ & $m = 48$ & $m = 24$ & $m = 48$  \\
\midrule
$1$&Pos&$6.1\times 10^{-2}$&$5.3\times 10^{-3}$&$8.4\times 10^{-2}$&$6.2\times 10^{-3}$&$2.7\times 10^{-1}$&$1.7\times 10^{-2}$\\
    &Asp&$2.1\times 10^{-3}$&$2.4\times 10^{-4}$&$2.2\times 10^{-3}$&$1.9\times 10^{-4}$&$2.4\times 10^{-3}$&$9.6\times 10^{-5}$\\
\midrule
$0.89$&Pos&$8.1\times 10^{-2}$&$3.5\times 10^{-2}$&$1.2\times 10^{-1}$&$3.9\times 10^{-2}$&$2.9\times 10^{-1}$&$7.7\times 10^{-2}$\\
    &Asp&$1.2\times 10^{-3}$&$3.5\times 10^{-4}$&$2.5\times 10^{-3}$&$2.3\times 10^{-4}$&$9.8\times 10^{-3}$&$4.8\times 10^{-4}$\\
\midrule
$0.78$&Pos&$2.6\times 10^{-1}$&$7.0\times 10^{-2}$&$4.5\times 10^{-1}$&$9.4\times 10^{-2}$&$3.3\times 10^{0}$&$1.1\times 10^{-1}$\\
    &Asp&$2.2\times 10^{-3}$&$1.1\times 10^{-3}$&$3.6\times 10^{-2}$&$1.4\times 10^{-3}$&$1.6\times 10^{-1}$&$1.3\times 10^{-3}$\\
\bottomrule
\end{tabular}
\caption{Relative errors at final time $T = 100$ for single biconcave capsule in Poiseuille flow through a pipe, with aggressive spectral filtering. Relative errors are reported for position (Pos) and Taylor asphericity (Asp), for different reduced volumes $\rvr$ and capillary numbers $\Ca$. The reference solution is computed with resolution $m = 96$.}\label{tab:pois_conv}
\end{table}

\begin{table}[htbp]
\centering
\setlength{\aboverulesep}{1pt}
\setlength{\belowrulesep}{1pt}
\setlength{\tabcolsep}{1.5pt}
\small
\begin{tabular}{c>{\columncolor[gray]{0.9}}c>{\columncolor[gray]{0.9}}c>{\columncolor[gray]{0.9}}cccc>{\columncolor[gray]{0.9}}c>{\columncolor[gray]{0.9}}c>{\columncolor[gray]{0.9}}c}
\toprule
\multirow{2}{*}{$\rvr$} & \multicolumn{3}{c}{$\Ca = 0.15$} & \multicolumn{3}{c}{$\Ca = 0.3$} & \multicolumn{3}{c}{$\Ca = 0.6$} \\
\cmidrule(lr){2-4} \cmidrule(lr){5-7}  \cmidrule(lr){8-10} 
 & $m = 24$ & $m = 48$ & $m = 96$ & $m = 24$ & $m = 48$ & $m = 96$  & $m = 24$ & $m = 48$ & $m = 96$ \\
\midrule
$1$&$1.1\times 10^{-2}$&$1.4\times 10^{-3}$&$1.7\times 10^{-4}$&$2.0\times 10^{-2}$&$2.3\times 10^{-3}$&$2.6\times 10^{-4}$&$6.7\times 10^{-2}$&$6.9\times 10^{-3}$&$7.8\times 10^{-4}$\\
\midrule
$0.89$&$3.1\times 10^{-2}$&$4.4\times 10^{-3}$&$7.2\times 10^{-4}$&$6.9\times 10^{-2}$&$9.4\times 10^{-3}$&$1.7\times 10^{-3}$&$2.2\times 10^{-1}$&$2.9\times 10^{-2}$&$5.1\times 10^{-3}$\\
\midrule
$0.78$&$7.8\times 10^{-2}$&$1.6\times 10^{-2}$&$4.1\times 10^{-3}$&$2.2\times 10^{-1}$&$3.6\times 10^{-2}$&$9.3\times 10^{-3}$&$5.4\times 10^{-1}$&$8.3\times 10^{-2}$&$2.0\times 10^{-2}$\\
\bottomrule
\end{tabular}
\caption{Relative volume loss at final time $T = 100$ for single biconcave capsule in Poiseuille flow through a pipe with aggressive spectral filtering, for different reduced volumes $\rvr$ and capillary numbers $\Ca$.}\label{tab:pois_vloss}
\end{table}

\begin{table}[htbp]
\centering
\setlength{\aboverulesep}{1pt}
\setlength{\belowrulesep}{1pt}
\setlength{\tabcolsep}{6pt}
\small
\begin{tabular}{cccc}
\toprule
\multirow{2}{*}{$\rvr$} & \multicolumn{3}{c}{$\Ca$} \\
\cmidrule(lr){2-4}
 & $0.15$ & $0.3$ & $0.6$ \\
\midrule
$1$ & $1.3\times 10^{-6}$ & $2.4\times 10^{-6}$ & $3.2\times 10^{-6}$ \\
\midrule
$0.89$ & $1.4\times 10^{-5}$ & $1.1\times 10^{-5}$ & $1.1\times 10^{-5}$ \\
\midrule
$0.78$ & $4.3\times 10^{-5}$ & $2.8\times 10^{-5}$ & $2.2\times 10^{-5}$ \\
\bottomrule
\end{tabular}
\caption{Residual $\|\dot{\boldsymbol{x}} - \boldsymbol{u}(\boldsymbol{x})\|$ at final time $T = 100$ for single biconcave capsule in
Poiseuille flow through a pipe with aggressive spectral filtering, for different reduced volumes $\rvr$ and capillary numbers $\Ca$ (simulated at resolution $m=48$). The residual is evaluated by upsampling the surface points $\boldsymbol{x}$ to a finer grid of $m = 96$. $\dot{\boldsymbol{x}}$ is approximated via central difference as $\dot{\boldsymbol{x}}(t+h) \approx (\boldsymbol{x}(t+2h) - \boldsymbol{x}(t))/(2h)$, and compared against the fluid velocity $\boldsymbol{u}(\boldsymbol{x}(t+h))$.}\label{tab:pois_res}
\end{table}
\section{Supplementary capsule evolution figures}
\label{app:capsule-evolution}

\begin{figure}[h]
  \centering
  Capsule shape evolution for case B (\iconBsmall)\\
  {\footnotesize%
  \begin{minipage}[t]{0.5\textwidth}\vskip0pt\centering
    \hspace*{30pt}(a) Times $0 \leq t \leq 5$\\
    \includegraphics[width=\textwidth]{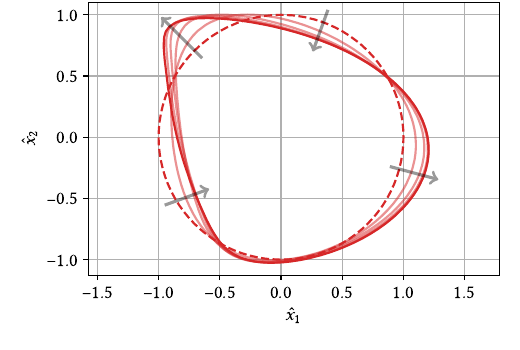}%
  \end{minipage}%
  \begin{minipage}[t]{0.5\textwidth}\vskip0pt\centering
    \hspace*{30pt}(b) Times $5 \leq t \leq 100$\\
    \includegraphics[width=\textwidth]{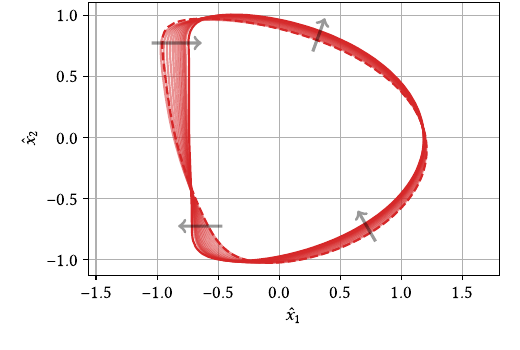}%
  \end{minipage}%
  }
  \caption{Capsule outline over time for case~B. (a) $t=0, 1, 2, \ldots, 5$.
  (b) $t=5, 10, 15, \ldots, 70$ and $100$.}
  \label{fig:arbitrary-evolution-B}
\end{figure}

\begin{figure}[h]
  \centering
  Capsule shape evolution for case E (\iconEsmall)\\
  {\footnotesize%
  \begin{minipage}[t]{0.5\textwidth}\vskip0pt\centering
    \hspace*{30pt}(a) Times $0 \leq t \leq 5$\\
    \includegraphics[width=\textwidth]{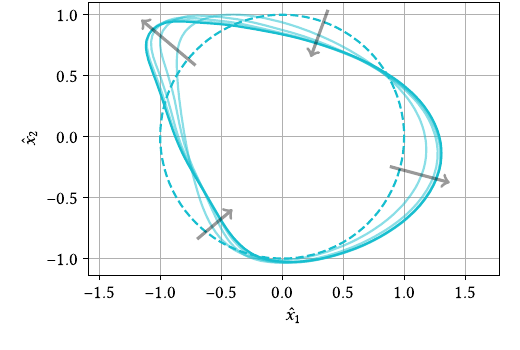}%
  \end{minipage}%
  \begin{minipage}[t]{0.5\textwidth}\vskip0pt\centering
    \hspace*{30pt}(b) Times $5 \leq t \leq 100$\\
    \includegraphics[width=\textwidth]{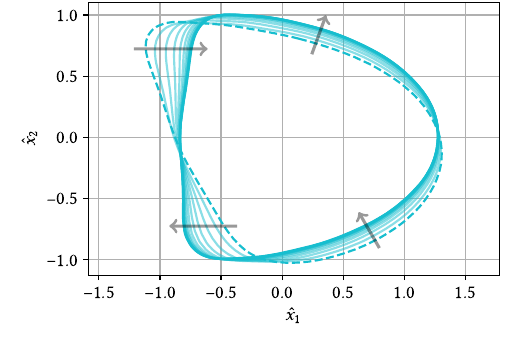}%
  \end{minipage}%
  }
  \caption{Capsule outline over time for case~E. (a) $t=0, 1, 2, \ldots, 5$.
  (b) $t=5, 10, 15, \ldots, 70$ and $100$.}
  \label{fig:arbitrary-evolution-E}
\end{figure}

The capsule evolutions for cases B and E are shown in \cref{fig:arbitrary-evolution-B,fig:arbitrary-evolution-E}, respectively. These complete the set of \cref{fig:circle-evolution-y02,fig:arbitrary-evolution-C,fig:arbitrary-evolution-D} found in \cref{sec:results-pipe-flow}.

\end{document}